\documentclass[showpacs,preprintnumbers,amsmath,amssymb,prb,twocolumn,superscriptaddress,floatfix]{revtex4}
\usepackage{graphicx}
\usepackage{bm}

\newcommand{\f}[1]{^{(#1)}} 
\newcommand{\ve}[1]{\mathbf{#1}}

\newcommand{\im}{{\rm i}}
\newcommand{\eq}[1]{Eq. (\ref{#1})}
\newcommand{\xy}{3D$xy$\ }
\newcommand{\xxy}{2D$xy$\ }
\newcommand{\fig}[1]{Fig. \ref{#1}}

\newcommand{\tab}[1]{Tab. \ref{#1}}
\newcommand{\beg}{\begin{eqnarray}}
\newcommand{\eee}{\end{eqnarray}}
\def\comment#1{}

\begin{document}

\title{Field- and temperature induced topological phase transitions in the 
       three-dimensional $N$-component London superconductor}

\author{J. Smiseth}
\affiliation{Department of Physics, Norwegian University of
Science and Technology, N-7491 Trondheim, Norway}

\author{E. Sm{\o}rgrav}
\affiliation{Department of Physics, Norwegian University of
Science and Technology, N-7491 Trondheim, Norway}

\author{E. Babaev}
\affiliation{Laboratory of Atomic and Solid State Physics, Cornell University, Ithaca, NY 14853-2501, USA}
\affiliation{Department of Physics, Norwegian University of Science and Technology, N-7491 Trondheim, Norway}

\author{A. Sudb{\o}}
\affiliation{Department of Physics, Norwegian University of
Science and Technology, N-7491 Trondheim, Norway}

\date{Received \today}

\begin{abstract}
The phase diagram and critical properties of the $N$-component London superconductor are studied 
both analytically and through large-scale Monte-Carlo simulations in $d=2+1$ dimensions 
(components here refer to different replicas of the complex scalar field). Examples are given 
of physical systems to which this model is applicable. 
The model with different bare phase stiffnesses for each component, is a model of superconductivity 
which should arise out of metallic phases of light atoms under extreme pressure. A projected 
mixture of electronic and protonic condensates in liquid metallic hydrogen under extreme 
pressure is the simplest example, corresponding to $N=2$. These are such that Josephson coupling 
between different matter field components {\it is precisely zero on symmetry grounds}. 
The $N$-component London model is dualized to a theory involving $N$ vortex 
fields with highly nontrivial interactions. We compute critical
exponents $\alpha$ and $\nu$ for $N=2$ and $N=3$. Direct and dual gauge field correlators
for general $N$ are given and the $N=2$ case is studied in detail.  
The model with $N=2$ shows two anomalies in 
the specific heat when the bare phase stiffnesses of each matter field species are 
different. One anomaly corresponds to an {\it inverted} \xy  
fixed point, while the other corresponds to a  \xy fixed point. Correspondingly, for $N=3$, we 
demonstrate the existence of two neutral \xy fixed points and one inverted charged \xy fixed point. 
For the general case, there are  $N$ 
fixed points, namely one charged inverted \xy fixed point, and $N-1$ neutral \xy fixed points. We explicitly identify one charged vortex mode and $N-1$ neutral vortex 
modes. The model for $N=2$ and equal bare phase 
stiffnesses corresponds to a field theoretical description of an easy-plane quantum
antiferromagnet. In this case, the critical exponents are computed and found to be non \xy values. 
The  $N$-component London superconductor model in 
an external magnetic field, with no inter-species Josephson coupling, will be shown to have a novel feature, namely 
$N-1$ superfluid phases  arising 
out of $N$ charged condensates. In particular, for $N=2$ we point out the possibility of  two 
novel types of field-induced phase transitions in ordered quantum fluids: {\it i)} A phase 
transition  from a superconductor to a superfluid  or vice versa,
driven by tuning an external magnetic field. {\it This identifies the superconducting phase of liquid metallic hydrogen as a novel
quantum fluid.}
{\it ii)} A phase transition corresponding to a 
quantum fluid analogue of sublattice melting, where a composite field-induced Abrikosov 
vortex lattice is decomposed and disorders the phases of the constituent condensate  
with lowest bare phase stiffness. Both transitions belong to the \xy universality class. 
For $N \geq 3$, there is a new feature not present in the cases $N=1$ and $N=2$, namely 
a partial decomposition of composite field-induced vortices driven by thermal fluctuations. 
A ``color electric charge" concept, useful for establishing the character of these phase transitions,
is introduced. 
\end{abstract}


\maketitle

\section{\label{Intro} Introduction}
Ginzburg-Landau (GL) theories with several complex scalar matter fields minimally 
coupled to one gauge field are of interest in a wide variety of condensed matter 
systems and beyond. This includes such apparently disparate systems as the two-Higgs 
doublet model \cite{lee1973}, superconducting low temperature phases of light atoms 
such as hydrogen \cite{Ashcroft1999,Neilnew} under extreme enough pressures to 
produce liquid metallic states, and effective theories for easy-plane quantum 
antiferromagnets  \cite{senthil2003,motrunich2004,sachdev2004}. Well known cases 
of  multicomponent systems are represented by  multiband  superconductors \cite{suhl} 
like $MgB_2$ where there are two order parameters corresponding to Cooper pairs
made up of electrons living on different sheets of Fermi surface. In that case however 
condensates are not independently conserved and the $U(1)\times U(1)$ symmetry is broken 
to $U(1)$, so the main results of this paper do not apply to multiband superconductors.
In contrast, in the projected liquid metallic state of  hydrogen \cite{Ashcroft1999,Neilnew}, 
which appears being close to a realization in high pressure experiments \cite{Datchi2000,Bonev},
the scalar fields represent Cooper pairs of electrons and protons. This excludes, on symmetry 
grounds, the possibility of inter-flavor pair tunneling, i.e. there is no intrinsic  Josephson 
coupling between  different species of the condensate. This sets it apart from systems with 
multi-flavor {\it electronic} condensates arising out of superconducting order parameters 
originating on multiple-sheet Fermi surfaces, such as is the case in $MgB_2$. For the latter 
system, Josephson coupling in internal order parameter space {\it cannot} be ruled out on 
symmetry grounds, and must therefore be included in the description. This is so because the 
Josephson coupling represents a singular perturbation and can never be ignored on sufficiently 
long length scales. This is otherwise well known from studies of extremely layered 
superconductors \cite{clem}, where the critical sector is that of the \xxy model in the 
absence of Josephson coupling, while any amount of interlayer phase-coupling (in an extended 
system) produces a critical sector belonging to the \xy universality class.  {\it It is 
precisely the lack of Josephson coupling in certain, but by no means
all systems with multiple 
flavor order parameters, that opens up the possibility of novel and interesting critical 
phenomena.} However, even in inter-band Josephson-coupled condensates, interesting physics 
arises at finite length scales \cite{egor2002,frac}.

A two-component action with no Josephson coupling in $(2+1)$ dimensions, with matter fields 
originating in a bosonic representation of spin operators, is also claimed to be the critical 
sector of a field theory separating a N{\'e}el state and a paramagnetic (valence bond ordered) 
state of a two dimensional quantum antiferromagnet at zero temperature with easy-plane 
anisotropy\cite{senthil2003,sachdev2004}. This happens because, although the 
effective description of the antiferromagnet involves an  {\it a priori} compact gauge field, 
it must be supplemented by Berry-phase terms in order to properly describe $S=1/2$ spin 
systems \cite{haldane1988,read_sachdev}. Berry-phase terms in turn cancel the effects of 
monopoles at the critical point \cite{senthil2003,sachdev2004}. Hence, an effective 
description in terms of two complex scalar matter fields coupled to one {\it non-compact} gauge 
field suffices to describe the non-trivial quantum critical point separating a state with 
broken {\it internal} $SU(2)$ {\it symmetry} and a paramagnetic $SU(2)$-symmetric state with 
broken {\it external symmetry} (lattice translational invariance). The latter state is the 
valence-bond ordered state. Critical behavior separating states differing in this manner is 
not captured by the Landau-Wilson-Ginzburg paradigm \cite{senthil2003,senthil_science2004}, 
and requires a description of a phase transition without a local order parameter. An example 
of such a description is the well known Kosterlitz-Thouless phase transition taking place in 
the \xxy model \cite{KT1974}. The difference from the Kosterlitz-Thouless case and the quantum 
critical behavior described above is that while the low temperature phase of the \xxy model is 
a Gaussian fixed line, this is not so for either side of the quantum critical point of the 
easy-plane quantum antiferromagnet \cite{senthil2003,motrunich2004,senthil_science2004,sachdev2004}. 
We also mention that another example of a multicomponent system with no inter-component Josephson 
effect are spin-triplet superconductors which are well known to allow a variety of topological 
defects and phase transitions \cite{volovik}. Some of the topics we discuss below are 
related to the models of spin-triplet paired electrons \cite{ssf}.

Since the condensates described above in the context of light atoms and easy-plane 
quantum antiferromagnets are gauge-charged condensates, the order parameter flavors 
are all coupled to each other via a non-compact gauge field. This 
coupling is vastly different from the Josephson coupling in the sense that 
while an $N$-flavor order parameter condensate with no coupling 
between different species in general will have $N$ phase transitions, a 
Josephson coupling between a pair of order parameter species will 
collapse the two independent phase transitions they undergo with no coupling, 
down to one. Josephson coupling between all pairs of order parameter species 
will collapse all $N$ phase transitions down to a single one, namely an 
inverted \xy transition. On the other hand, $N$ order parameter species
coupled to one and the same gauge field will still undergo in general 
$N$ phase transitions, namely one {\it inverted} \xy transition where 
a Higgs phenomenon takes place, followed by $N-1$ \xy transitions as the coupling 
constants are increased beyond the Higgs/\xy critical point \cite{sachdev2004,smiseth2004}.

A special feature is presented by the important case $N=2$. Here, it turns out 
that the {\it dual} description of the theory is isomorphic to  the starting 
point \cite{motrunich2004,sachdev2004,smiseth2004}. Normally, in $d=2+1$, a gauge 
theory dualizes into a global theory and vice versa. In contrast a $U(1)\times U(1)$ gauge theory 
dualizes into another $U(1)\times U(1)$ gauge theory, i.e. the theories are {\it self-dual}. 
In general the theory has two separate critical points, one inverted \xy and one \xy 
critical point \cite{smiseth2004}. For the special case where the bare phase stiffnesses
of the two matter fields are equal, as they naturally are in the case of easy-plane 
quantum antiferromagnets in the absence of an external magnetic field 
\cite{motrunich2004,sachdev2004}, another interesting feature appears. 
In this case, there is only one critical point separating two phases described by 
{\it self-dual field theories}. This cannot be either an inverted \xy or a \xy 
fixed point. Self-duality also precludes the possibility of a $Z(2)$ universality 
class although the exponent $\nu$ that we find for this case appears to be close to 
the Ising value (while $\alpha$ is not). This phase transition therefore defines a 
new universality class, namely that of the $d=2+1$ $U(1)\times U(1)$ self-dual gauge theory. 

What happens to such multi-component charged condensates in three dimensions in the 
absence of Josephson coupling between the order parameter components, but in the 
presence of an external magnetic field, has been recently studied in
Refs. \onlinecite{BSA},\onlinecite{SSBS}  for the case $N=2$, with particular emphasis 
on applications to liquid metallic hydrogen. In this paper, we extend on this and consider 
in detail the effects of tuning the external magnetic field and temperature when also 
$N \geq 3$. New features appear compared to the $N=2$ case, because composite vortices 
consisting of non-trivial windings in all order parameter components can now undergo partial
decompositions by tearing vortices  of individual order parameter components off the
composite vortices, one after the other. We provide a dual picture of these processes: 
{\it i)} as a vortex loop proliferation in the background of a composite vortex lattice, 
and {\it ii)} as a  metal-insulator transition  in a system 
consisting of several ``colors of electric charges'' in a multi color dielectric 
background. The new concept of ``color charge''  will be introduced and 
explained in  detail in this paper. {\it It allows us 
to determine the universality class, and the partially broken
symmetries of the partial decomposition
transitions taking place in multi flavor superconductors in an external magnetic field. }
 We also show that the  number of colors $N_{\rm color}$ of dual charges exceeds
the number of field components (flavors) for $N>3$.

The outline of the paper is as follows. The first six sections of the paper deal with results
in zero external magnetic field. In Sections VII and VIII  we present results 
in finite magnetic field. Readers who wish to consult results on finite magnetic
field may proceed directly to Section VII. 

In Section \ref{model}, we introduce the model and the main approximation we will use to 
study the model, as well as the duality transform that will be used extensively, along 
with the explicit vortex representation of the model. In Section \ref{Neutral_modes}, we 
explicitly transform the action for the $N=2$ case into an action consisting of two parts: 
{\textit{i})} one charged vortex mode with vortex interactions mediated by a massive vector 
field, and {\textit{ii})} one neutral vortex mode with vortex interactions mediated by a 
gauge field. In Section \ref{gauge_corr}, we compute gauge field correlators and dual gauge 
field correlators in terms of vortex correlators. {\it This explicitly identifies the mechanism 
by which a thermally driven vortex loop proliferation destroys the Higgs phase (Meissner effect) 
and dual Higgs phase} \cite{hove2000,smiseth2004}. Gauge field correlators are useful in 
characterizing the charged fixed point of the $N$-flavor London model \cite{hove2000,smiseth2004}, 
while dual gauge field correlators are also useful in characterizing the $N-1$ neutral fixed points 
\cite{smiseth2004}. In Section \ref{mc_simsN=2}, we present large-scale Monte Carlo (MC) simulations 
for the case $N=2$, computing critical exponents at the neutral and charged fixed points, as well 
as the mass of the gauge field as
a function of temperature. The neutral fixed point is found to be in the \xy universality class, 
while the charged fixed point is shown to be in the inverted \xy universality class. We also 
consider in detail the case when the two bare phase stiffnesses of the model are identical, 
showing that the resulting one fixed point is in a new universality class distinct from
the \xy and inverted \xy universality classes. In Section \ref{mc_simsN=3}, we  present 
corresponding results for the case $N=3$. In Section \ref{N=2_extfield}, we outline the phases 
to expect for the case $N=2$ when an external magnetic field is applied. We also present
results from large-scale Monte-Carlo simulations revealing a novel phase transition
in the \xy universality class inside the Abrikosov vortex lattice phase at low magnetic
fields when temperature is increased. In Section \ref{N>2_extfield}, 
we do the same when $N > 2$, emphasizing the qualitatively new features compared to the case 
$N=2$. We also introduce  a useful ``color charge" picture of the various partial
decomposition transitions of the composite vortex lattice that we encounter
for the case when $N \geq 3$. In Section \ref{Concl}, we summarize our results. In 
Appendix \ref{App:charged_neutral}, we identify charged an neutral vortex modes for 
general $N$. In Appendix \ref{App:vortex}, we derive the vortex representation
for the general-$N$ case. In Appendices \ref{App:gauge} and \ref{App:dualgauge}, 
we derive expressions for gauge field correlators and dual gauge field correlators,
respectively. In Appendix \ref{App:josephson} we generalize our dual representation
for arbitrary $N$ to also include inter-flavor Josephson coupling.  In Appendix 
\ref{App:bkt}, we consider Kosterlitz-Thouless transitions for the 
general-$N$ case in two spatial dimensions at finite temperature.

\section{\label{model} Model and dual action}
For an  analysis of the possible phase transitions in a  GL model of $N$ individually 
conserved bosonic  matter fields, each coupled to one and the same  $U(1)$ non-compact 
gauge field, we study a version of the  $N$-flavor GL theory in $(2+1)$ dimensions 
{\it with no Josephson coupling terms} between order parameter components. Moreover,
we ignore mixed gradient terms, such that there is  no Andreev-Bashkin effect \cite{AB}.
The model  is defined by $N$ complex scalar fields 
$\{\Psi_0^{(\alpha)}(\ve r) \ | \ \alpha=1,\dots,N\}$ coupled 
through the charge $e$ to a fluctuating gauge field $\ve{A}(\ve r)$, with the action
\begin{equation}
\begin{split}
\label{gl_action}
S =& \int d^3\ve r \left[\sum_{\alpha=1}^N \frac{|(\nabla -\im e\ve{A}(\ve r))\Psi_0^{(\alpha)}(\ve r)|^2}{2 M^{(\alpha)}}\right. \\
&+ \left.V(\{\Psi_0^{(\alpha)}(\ve r)\}) + \frac{1}{2}(\nabla\times\ve{A}(\ve r))^2\right],
\end{split}
\end{equation}
where $M^{(\alpha)}$ is the mass of the condensate species
$\alpha$. 
Assuming that the individual condensates are conserved, the potential 
$V(\{\Psi_0^{(\alpha)}(\ve r)\})$ must be function of 
$|\Psi_0^{(\alpha)}(\ve r)|^2$ only. 
In this paper, we focus on  the critical phenomena and phase diagram of Eq. (\ref{gl_action})
in zero as well as finite external magnetic field, and for these purposes the model 
in Eq. (\ref{gl_action}) will be studied in the phase-only approximation 
$\Psi_0^{(\alpha)}(\ve r) = |\Psi^{(\alpha)}_0|\exp [ \im\theta^{(\alpha)}(\ve r) ]$ 
where $|\Psi^{(\alpha)}_0|$ is a constant, i.e. we freeze out amplitude fluctuations 
of {\it each individual matter  field}. The model we study is therefore the generalization 
to arbitrary $N$ of the frozen-amplitude one gap lattice superconductor model also known 
as the London superconductor model \cite{dasgupta1981}. 

One may well ask what confidence one should put in the phase only approximation
for all fields when the bare phase stiffness of each individual condensate is 
very different, such as is the case in LMH. The answer is that one can be quite 
confident that this is a useful and reasonable approximation. Consider first
the case $N=2$. We use the phase only approximation with confidence for considering 
the criticality here. It certainly works at the lowest critical temperature.  After 
that point, we are left with a one-component superconductor. What the field with the 
lowest phase stiffness does above the lowest critical temperature is not of interest, 
it is only the remaining field with criticality at higher temperature that matters. 
Hence, significantly above the lowest critical temperature, we may still apply the 
phase only approximation for the remaining one-component case. For this field, we may 
use the phase only approximation up to the highest critical temperature with the same 
confidence as we can use the phase only approximation for the field with the lowest 
phase stiffness up to and slightly above the lowest critical temperature. The same 
argument can  be repeated for arbitrary $N$:  We can use the phase only approximation for 
the fields up to and slightly above their respective critical temperatures. After 
that it is immaterial what they do, it is only the remaining components that matter. 

\subsection{Basic properties of the model}
Varying \eq{gl_action} with respect to ${\bf A}$, we obtain the equation for the 
supercurrent
\beg 
{\bf J} 
&=&  
 \sum_{\alpha=1}^N
\frac{\im e  }{2 M^{(\alpha)}}
\left\{{\Psi_0^{(\alpha)}}^* \nabla \Psi_0^{(\alpha)}-
\Psi_0^{(\alpha)} \nabla {\Psi_0^{(\alpha)}}^* \right\}
\nonumber \\&&
-2 e^2 \left( \frac{|\Psi_0^{(\alpha)}|^2}{M^{(\alpha)}} 
\right) {\bf A}. \ 
\label{AA}
\eee
 Vortex excitations in such an $N$-flavor GL model carry fractional flux.
 Consider a vortex
 where the phase  $\theta^{(\eta)}(\ve r)$  has a $2\pi$
 winding around a vortex core, while other phases do
 not have nontrivial windings. Expressing ${\bf A}$
from  \eq{AA}, and integrating 
along a path around the vortex core at a distance larger than the
 magnetic penetration length, 
 we obtain an expression for the magnetic 
 flux encompassed by the path given by 
 \beg
 \Phi^{(\eta)}=\oint {\bf A} d {\bf l} =
\Phi_0 
\frac{|\Psi_0^{(\eta)}|^2}{M^{(\eta)}}
\left[ \sum_{\alpha=1}^N\frac{
|\Psi_0^{(\alpha)}|^2}{M^{(\alpha)}}
\right]^{-1},
\label{Phi}
 \eee
 where $\Phi_0 = 2.07 \cdot 10^{-15} {\rm Tm}^2$ is the flux quantum. As it will be clear
 from a discussion following \eq{potential_0} (see \eq{potential}), such a vortex has 
 a logarithmically divergent energy\cite{frac,smiseth2004}. Only a {\it composite}
 vortex where all phases $\theta^{(\alpha)}$ have $2\pi n$ winding around the core 
 carries integer flux and has finite energy. As detailed below,  the composite
 vortices are responsible for the magnetic properties of the
 system at low temperatures while thermal excitations
in the form of loops of  individual fractional-flux vortices are
responsible for the critical properties of the system in the
absence of an external field.

{\it Note that since each individual amplitude is frozen, this model will 
be different from the case where only the sums of the squares of the 
amplitudes are frozen} \cite{Nogueira1}. The latter is usually referred to 
as the $N$-component scalar QED $(NSQED)$ \cite{Coleman_Weinberg,HLM}, or 
the CP${}^{N-1}$  model \cite{Hikami}. (As far as critical properties are 
concerned, the NSQED model and the CP${}^{N-1}$ model
have been shown to belong to the same universality class \cite{Hikami}). 
We strongly emphasize that we must distinguish our model from NSQED 
and CP${}^{(N-1)}$, and will consequently be referring to it
as the $N$-flavor London superconductor (NLS) model.   
The NLS is in fact the natural model to consider for the physical
systems mentioned in the introduction, in particular
pertaining to the superconducting mixtures of metallic
phases of light atoms. As we shall see, the NLS model
has physics which sets it distinctly apart from the
NSQED and the CP${}^{N-1}$ 
models, and it does not have critical properties in the same
universality class as they do. This becomes particularly
apparent in the large-$N$ limit, as we shall see in section \ref{vort_rep}. 

\subsection{Separation of variables}
Before we proceed further, it is useful to give another form of the action. 
For brevity we introduce the bare phase stiffness of the matter field with
flavor index $\alpha$ defined as $|\psi^{(\alpha)}|^2 =
|\Psi^{(\alpha)}_0|^2/M^{(\alpha)}$. 
Then  \eq{gl_action} may be rewritten
in terms of {\it one} charged and $N-1$ neutral modes as follows (details of this 
are found in Appendix \ref{App:charged_neutral}). We have 
$S = \int d^3 \ve r {\cal{L}}$, with 
\begin{eqnarray}
{\cal L} & = & \frac{1}{2 \Psi^2}  \left( 
\sum_{\alpha = 1}^N |\psi^{(\alpha)}|^2 \nabla \theta^{(\alpha)}  - e \Psi^2 \ve A 
\right)^2  + \frac{1}{2}(\nabla\times\ve{A})^2 \nonumber \\
& &+  \frac{1}{4 \Psi^2}  \sum_{\alpha,\beta=1}^N |\psi^{(\alpha)}|^2  |\psi^{(\beta)}|^2 
 \left( \nabla (\theta^{(\alpha)}- \theta^{(\beta)}) \right)^2,
 \label{charge_neutral}
\end{eqnarray}
where
\beg
\label{defpsi2}
 \Psi^2 \equiv \sum_{\alpha=1}^N |\psi^{(\alpha)}|^2. 
 \eee
The first term in \eq{charge_neutral} represents the charged mode
 coupling to the gauge 
field $\ve A$, and the remaining terms are the $N-1$  neutral modes which do not couple 
to $\ve A$. This means that they have gauge charge equal to zero. We will come back to 
this in Section \ref{Neutral_modes}. This form \eq{charge_neutral} will be useful later 
when we discuss finite field effects in section \ref{N>2_extfield}. We 
also stress that $\Psi$ in the above expression should not be confused with 
$\Psi_0^{(\alpha)}$ defined in \eq{gl_action}.

Counting degrees of freedom  in \eq{charge_neutral} requires
care.  The case $N=1$  yields the well known answer that 
a phase variable (which is not a gauge invariant quantity) is higgsed into 
a massive vector field by coupling to the vector potential. In the case $N=2$,  the 
situation is different in the  sense that one can form a gauge invariant 
quantity by subtracting phase gradients. Thus the $U(1)\times U(1)$ system may 
be viewed as possessing {\it i}) a local $U(1)$ gauge symmetry associated with the phase 
sum which is  coupled to the vector potential and thus yields  a massive vector field, 
and {\it ii}) a global  $U(1)$ symmetry which is associated with a phase difference where 
there is no coupling to the vector potential. These charged and neutral modes 
are naturally described by the first and third terms in \eq{charge_neutral}, respectively. 
For  $N=3$, the situation is  principally different from both the $N=1$ and $N=2$ cases. That 
is, in \eq{charge_neutral} for $N=3$, we find one term describing the charged mode  (the 
first term) and {\it three} terms describing gauge-invariant neutral phase combinations.

The two neutral modes in Eq. (\ref{charge_neutral}), in the $N=3$ case,
cannot be properly described by only two terms, for
topological reasons. A vortex excitation produces a zero in the 
order parameter space, thus making the superconductor  multiply connected.
A vortex with a non-trivial phase winding in any of the three components
would result in non-trivial contributions to 
two of three phase-difference terms in Eq. (\ref{charge_neutral}).
{\it Hence, for $N=3$ an elementary vortex {\it i.e.} with nontrivial
winding only in one of the phases excites two neutral modes}.
In general, when all $|\psi^{(\alpha)}|$ differ, the bare phase stiffnesses of two neutral modes
excited by each of the three possible elementary vortices, 
are different. Thus, the neutral modes in the system are described by  
three  phase-difference terms in Eq. (\ref{charge_neutral}).
{\it These three 
terms are not independent when the condition of single-valuedness of each of
the $N$ order parameter components is enforced}, namely that individual phases
may change only by integer multiples of $2\pi$ around zeroes of the order parameters.

Using  Eq. (\ref{charge_neutral}) as opposed to Eq.  (\ref{gl_action}), has 
advantages, because the neutral and charged modes are explicitly identified.
This facilitates a discussion of the critical properties of the $N$-flavor system.
Moreover, Eq. (\ref{charge_neutral}) will allow us to identify various 
states of {\it partially} broken symmetry which emerge if an $N$-flavor system is 
subjected to external magnetic field \cite{BSA}. We will come back to these 
points in detail in Sections \ref{N=2_extfield} and \ref{N>2_extfield}.

\subsection{The Villain approximation}
The theory Eq. (\ref{gl_action}) is discretized on a $d=3$ dimensional
cubic lattice with spacing
$a=1$ and size $L^3$, and in the phase only approximation the action reads
\begin{eqnarray}
S &=& \sum_{\ve r}\left[-
  \beta\sum_{\alpha=1}^N|\psi^{(\alpha)}|^2\sum_{\mu=1}^3\cos(\Delta^\mu\theta^{(\alpha)}(\ve
  r) - eA^\mu(\ve r)) \right. \nonumber\\
& &+ \left. \frac{\beta}{2}(\nabla\times\ve{A}(\ve r))^2\right]. 
\end{eqnarray}
Here, we have included the inverse temperature coupling $\beta = 1/T$. The symbol $\Delta^\mu$ denotes the lattice difference
operator in direction $\mu$ in Euclidean space and the position vector $\ve r$ runs over
all points on the lattice. The partition function in the Villain approximation is 
\begin{equation}
\begin{split}
  \label{villain}
Z =& \int_{-\infty}^{\infty}\mathcal{D}\ve A\prod_{\gamma=1}^N\int_{-\pi}^{\pi}\mathcal{D}\theta^{(\gamma)}\prod_{\eta=1}^N\sum_{\ \ve n^{(\eta)}}\exp(-S) \\
S =& \sum_{\ve r}\left[\sum_{\alpha=1}^N\frac{\beta|\psi^{(\alpha)}|^2}{2}(\Delta\theta^{(\alpha)} - e\ve{A} +2\pi \ve n^{(\alpha)})^2\right. \\
&\left.+\frac{\beta}{2}(\Delta\times\ve{A})^2\right],
\end{split}
\end{equation}
where  $\ve n^{(\alpha)}(\ve r)$ are integer vector fields
ensuring $2\pi$ periodicity, and the lattice position index vector
$\ve r$ is suppressed. {\it Here, we stress the importance of keeping 
track of the $2\pi$ periodicity of the individual phases}. For $N=1$ it has been
shown that thermal fluctuations in this model excite topological
defects in form of closed vortex loops. At the critical temperature
the system undergoes a vortex loop
proliferation phase transition\cite{kleinert_book,tesanovic1999,nguyen1999}. 

\subsection{\label{vort_rep}Vortex representation}

In the following, we transform the model Eq. (\ref{villain}) into a theory of interacting vortex 
loops of different flavors. The procedure is described in detail in Appendix \ref{App:vortex}. 
The kinetic energy terms are linearized by introducing $N$ auxiliary
fields $\ve v^{(\alpha)}(\ve r)$. 
Applying the Poisson summation formula and integrating over $\ve n^{(\alpha)}(\ve r)$ constrains the 
fields $\ve v^{(\alpha)}(\ve r)$ to take only integer values $\ve{\hat v}^{(\alpha)}(\ve r)$. Integration over 
all $\theta^{(\alpha)}(\ve r)$ produces the local constraints $\Delta\cdot\ve{\hat v}^{(\alpha)}(\ve r)=0$, 
which are fulfilled by replacing 
$\ve{\hat v}^{(\alpha)}(\ve r)$ with $\Delta\times\ve{\hat h}^{(\alpha)}(\ve r)$ 
where $\ve{\hat h}^{(\alpha)}(\ve r)$ are integer-valued fields. 
By applying the Poisson summation once more and summing over 
all $\ve{\hat h}^{(\alpha)}(\ve r)$, the fields $\ve{\hat  h}^{(\alpha)}(\ve r)$ 
take continuous values $\ve h^{(\alpha)}(\ve r)$ and the integer-valued vortex 
fields $\ve m^{(\alpha)}(\ve r)$ are introduced. We recognize $\ve h^{(\alpha)}(\ve r)$ as 
the dual gauge fields of the theory. To preserve the gauge symmetry of
$\ve h^{(\alpha)}(\ve r)$ each vortex field of flavor index $\alpha$
is constrained by the condition 
\begin{equation}
\label{closedloops}
\Delta\cdot\ve m^{(\alpha)}(\ve r)=0. 
\end{equation}
Hence, the vortex fields form closed loops. At this stage, the action reads
\begin{equation}
\begin{split}
  \label{dual1}
S =&\sum_{\ve r}\left[\sum_{\alpha=1}^N\frac{(\Delta\times\ve
    h^{(\alpha)})^2}{2\beta|\psi^{(\alpha)}|^2} - \im e\ve
  A\cdot\left(\sum_{\alpha=1}^N\Delta\times\ve h^{(\alpha)}\right) \right.\\
&+ \left.2\pi\im\sum_{\alpha=1}^N \ve m^{(\alpha)}\cdot\ve h^{(\alpha)} + \frac{\beta}{2}(\Delta\times\ve A)^2\right],
\end{split}
\end{equation}
where the vortex fields $\ve m^{(\alpha)}(\ve r)$ are constrained by Eq. (\ref{closedloops}). We integrate 
out the gauge field $\ve A(\ve r)$ and get a theory in the dual gauge fields  $\ve h^{(\alpha)}(\ve r)$ 
and the vortex fields $\ve m^{(\alpha)}(\ve r)$ \cite{smiseth2004}
\begin{equation}
\begin{split}  
\label{dual2}
S = \sum_{\ve r}&\left[2\pi\im\sum_{\alpha=1}^N \ve m^{(\alpha)}\cdot\ve h^{(\alpha)}+ 
\sum_{\alpha=1}^N\frac{(\Delta\times\ve h^{(\alpha)})^2}{2\beta|\psi^{(\alpha)}|^2}\right.\\
+&\left. \frac{e^2}{2\beta}\left(\sum_{\alpha=1}^N\ve h^{(\alpha)}\right)^2\right].
\end{split}
\end{equation}
This  generalizes to arbitrary $N$ the results of Peskin \cite{Peskin}, and Thomas and 
Stone \cite{Thomas_Stone}. In Appendix \ref{App:josephson} we generalize this result even 
further by including inter-flavor Josephson coupling.

When $N \geq 2$ there is an important difference from the $N=1$ case, which gives 
rise to entirely novel physics. Note how it is the {\it algebraic sum} of the dual photon fields 
in \eq{dual2} that is massive. This differs from the case $N=1$, where $e$ produces one massive dual 
photon with bare mass $e^2/2$, and the model describes a vortex field $\ve m(\ve r)$ interacting through a 
\textit{massive} dual vector field $\ve h(\ve r)$. However, when $N\ge 2$,  since $\Delta\cdot\ve m^{(\alpha)}(\ve r)=0$, 
a gauge transformation $\ve h^{(\alpha)}(\ve r)\to \tilde{\ve h}^{(\alpha)}(\ve r) = \ve h^{(\alpha)}(\ve r) + \Delta
g^{(\alpha)}(\ve r)$ for $\alpha\in [1,\dots ,N]$ 
leaves the action in \eq {dual2} invariant if one of the gauge fields, say $\tilde{\ve h}^{(\eta)}(\ve r) $ 
compensates the sum in the last term in the action with $\Delta g^{(\eta)}(\ve r) =
-\sum_{\gamma\neq\eta}\Delta  g^{(\gamma)}(\ve r)$. 
Thus, even in the presence of a gauge charge $e$, such that the direct model is a gauge 
theory, the dual description is such that the individual dual photon fields are also 
gauge fields.

Integrating out the dual gauge fields we get a generalized theory
of vortex fields of $N$ flavors interacting through the potential $D^{(\alpha,\eta)}(\ve r)$
\begin{equation}
  \label{vortex_action}
\begin{split}
Z &= \prod_{\alpha=1}^N ~ \sum_{\ve m^{(\gamma)}} ~\delta_{\Delta\cdot\ve m^{(\gamma)},0} ~ {\rm e}^{-S_{\ve V}} \\
S_{\rm V} &= \pi^2 \sum_{\ve r,\ve r^\prime}\sum_{\alpha,\eta}\ve m^{(\alpha)}(\ve r)D^{(\alpha,\eta)}(\ve r - \ve r^\prime)\ve m^{(\eta)}(\ve r^\prime),
\end{split}
\end{equation}
where $\delta_{x,y}$ is the Kronecker-delta, and the discrete Fourier
transform of the vortex interaction potential is
$\widetilde{D}^{(\alpha,\eta)}(\ve q)$, given by \cite{smiseth2004}
\begin{equation}
\label{potential}
\frac{\widetilde{D}^{(\alpha,\eta)}(\ve q)}{2 \beta|\psi^{(\alpha)}|^2} = 
\frac{\lambda^{(\eta)}}{|\ve Q_{\ve q}|^2 + m_0^2} + \frac{\delta_{\alpha,\eta} 
- \lambda^{(\eta)}}{|\ve Q_{\ve q}|^2},
\end{equation}
where $\lambda^{(\alpha)} \equiv |\psi^{(\alpha)}|^2/\Psi^2$, and
$\Psi^2$ is given by Eq. (\ref{defpsi2}). Here, the bare mass $m_0$ is
the inverse bare screening length given by $m^2_0 = e^2 \Psi^2$, and $|\ve Q_{\ve q}|^2 =
\sum_{\mu=1}^3 (2\sin(q^\mu/2))^2$ is the 
Fourier representation of the lattice Laplace operator, where
$q^\mu=2\pi n^\mu/L$ with $n^\mu\in [1,\dots,L]$. Note that
$\sum_{\alpha} \lambda^{(\alpha)} =1$. Note also that when $e^2=0$,
the interaction matrix reduces to 
\begin{equation}
\label{potential_0}
\widetilde{D}^{(\alpha,\eta)}(\ve q) = 2 \beta|\psi^{(\alpha)}|^2
 \frac{\delta_{\alpha,\eta}}{|\ve Q_{\ve q}|^2}.
\end{equation}
This means that when there is no charge coupling the matter fields to 
a {\it fluctuating} gauge field, there is no interaction between
vortices of different flavors. 
This simple case corresponds to Eq. (\ref{gl_action}) 
representing a system of $N$ decoupled \xy models. Also note that for vortices
of different flavors, $\eta \neq \alpha$, when $e \neq 0$,  the interaction matrix
tends to vanish when the inter-vortex distance is much smaller
then the effective penetration length $\lambda = 1/m_0$. It follows from
the fact that when the inter-vortex distance is much smaller than
$\lambda$, the vortices interact as if $\ve A$ does not screen, i.e.
as if $\ve A$  does not fluctuate. In this case, it is clear that the 
action we describe is simply that of $N$ decoupled \xy models, i.e. 
inter-flavor interactions vanish, cf. Eq. (\ref{potential_0}).
For instance, for the case $N=2$,
there will be no interactions between vortices of condensate
$\Psi_0^{(1)}$ and 
vortices of the condensate $\Psi_0^{(2)}$ unless we allow the gauge field to 
fluctuate. In the extreme type-II limit where $\lambda \to \infty$ only intra-flavor interactions between 
vortices will exist (see also Ref. \onlinecite{npb}). 



The first term of the vortex interaction potential Eq. (\ref{potential}) is a Yukawa 
screened potential, {\it while the second term mediates long range Coulomb 
interactions between vortex fields}. If $N=1$ 
the latter cancels out exactly and we are left with the well studied vortex theory of the GL model 
which has a charged fixed point for $e \neq 0$ \cite{Herbut_Tesanovic1996,hove2000}. For $N\ge2$ 
we find a theory of vortex loops of $N$ flavors interacting through long range Coulomb with an 
additive screened part. If the number of species $N$ grows to infinity and $\Psi^2 \to \infty$, 
the vortex interaction receives the 
dominant contribution from a diagonal unscreened $N \times N$ Coulomb matrix. 
But there are physical situations where off-diagonal interactions play an
important role even in the large-$N$ limit (to be discussed below).
One can also observe from Eq. (\ref{Phi}) that in the $N \to \infty$ limit when all components 
have similar stiffness the magnetic flux enclosed by elementary vortices also tends to zero. 
Thus, for $N \to \infty$ the physics of the model is governed by
neutral modes only.

The energy density of one straight vortex line of flavor $\alpha$ in a distance $\ve r$ 
larger than the effective penetration depth $\lambda$ is found by integrating along the 
line using the last term in the potential \eq{potential} only\cite{Fossheim_Sudbo_book}. 
This produces an energy term  of the form $D({\ve{r}}) \sim \ln( |\ve r|)$, and shows that 
such a vortex has logarithmically divergent energy. 

The large-$N$ limit of the NLS serves to illustrate how different the physics is from the 
large-$N$ limit of the NSQED model and the CP${}^{N-1}$ model \cite{HLM,Hikami}. In the 
large-$N$ expansion of the NSQED model, only one charged fixed point is found (which is 
infrared stable provided $2N > 365$), with critical exponent $1/\nu = 1+48/N+\dots$ in $D=3$ 
\cite{HLM}. This is consistent with the results found in the large-$N$ limit of the
CP${}^{(N-1)}$ model \cite{Hikami}. The origin of the difference between these results and 
the results we find for the NLS model is easily traced  to the following fact. The treatment 
of the NSQED model in Ref. \onlinecite{HLM} is strictly speaking correct only in the case 
of type-I superconductivity, since they find that for physical values of $N$, only a first 
order phase transition from a superconductor to a normal metal takes place (no infrared stable 
{\it fixed point} is found for physical values of $N$). This is correct only for values of the 
Ginzburg-Landau parameter $\kappa < 0.8/\sqrt{2}$, as has been shown
in recent large-scale MC simulations  \cite{Mo2002} and in earlier
analytical treatments \cite{kleinert_book}.   
The transitions discussed below where neutral modes appear do not 
significantly  depend on whether the system is type-I or type-II. Our results are therefore best 
thought of as generalizations to arbitrary $N$ of the problem studied many years ago by Dasgupta 
and Halperin on the frozen-amplitude $N=1$ lattice superconductor model \cite{dasgupta1981}. It 
is this fact that in the present model the modulus of each component is fixed, along with the 
{ precise absence of internal Josephson coupling} between matter field species, that brings out 
the novel physics we shall describe, namely the {\it charge-neutral superfluid modes arising out 
of $N$ charged condensate fields}. 

\subsection{Dual field theory}
Starting from \eq{dual2} the above vortex system may be formulated as a field theory, introducing $N$ complex 
matter fields $\phi^{(\alpha)}$ for each vortex species, minimally coupled to the dual 
gauge fields $\bf{h}^{(\alpha)}$. This generalizes the dual theory for $N=1$ 
in Refs. \onlinecite{Thomas_Stone,kleinert_book}. The theory reads \cite{smiseth2004} 
(for a comment on the case of general $N$, see also bottom of page 42, 
Ref. \onlinecite{sachdev2004})
\begin{equation}
\begin{split}
S_{\rm{dual}}  =& \sum_{\ve r} \left[ \sum_{\alpha=1}^N \left( m_\alpha^2 |\phi^{(\alpha)}|^2 + 
|(\Delta - \im \ve h^{(\alpha)})\phi^{(\alpha)}|^2 \right.\right.\\
&+ \left.\frac{(\Delta\times\ve h^{(\alpha)})^2}{2\beta|\psi^{(\alpha)}|^2} \right) 
 +  \frac{e^2}{2\beta} \left( \sum_{\alpha=1}^N\ve h^{(\alpha)}
 \right)^2 \\
&+ \left.\sum_{\alpha,\eta} g^{(\alpha, \eta)} |\phi^{(\alpha)}|^2 |\phi^{(\eta)}|^2 \right].
\end{split}
\label{S_dual}
\end{equation}
Here, we have added chemical potential (core-energy) terms for the vortices, as well 
as steric  short-range repulsion interactions between vortex elements. In the $N=1$ 
case, a RG treatment of the term $\frac{e^2}{2\beta}\ve h^2$ yields 
\begin{equation}
\label{rg_flow}
\frac{\partial e^2}{\partial \ln l} = e^2,
\end{equation}
and hence this term
scales up, suppressing the dual vector field $\ve h$. The charged 
theory in $d=2+1$ therefore dualizes into a $|\phi|^4$ theory and vice versa \cite{hove2000}.
Correspondingly, for $N \geq 2$, \eq{rg_flow} suppresses $\sum_{\alpha=1}^N\ve h^{(\alpha)}$, but
not each individual dual gauge field. For the particular case $N=2$, assuming the same 
to hold, we end up with a gauge theory of two complex matter fields coupled minimally to 
one gauge field, which was also precisely the starting point. Thus the theory is 
self-dual for $N=2$ \cite{motrunich2004,sachdev2004}. 

\section{\label{Neutral_modes} Charged and neutral vortex modes}
In this section, we present a straightforward method of 
identifying charged and neutral vortex modes for the
model Eq. (\ref{gl_action}). Consider first the case $N=2$,
when the action Eq. (\ref{dual2}) reads
\begin{widetext}
\begin{eqnarray}
S = \sum_{\ve r} \left \{ 2 \pi \im \left[ \ve m^{(1)}\cdot \ve h^{(1)} +  \ve m^{(2)}\cdot \ve h^{(2)} \right]
+\frac{1}{2 \beta} 
\left[ \frac{(\ve \nabla \times \ve h^{(1)})^2}{|\psi^{(1)}|^2} 
     + \frac{(\ve \nabla \times \ve h^{(2)})^2}{|\psi^{(2)}|^2} \right]
 + \frac{e^2}{2 \beta} \left( \ve h^{(1)} + \ve h^{(2)} \right)^2 \right \}.
\label{dual2_N=2} 
\end{eqnarray}
\end{widetext}
From this we identify the massive linear combination of the dual gauge
fields $\ve h^{(\alpha)}$, 
namely ${\cal H} = \ve h^{(1)} + \ve h^{(2)}$. If a neutral vortex mode exists
in the system, this implies the existence also of a gauge field in the 
problem, which we will denote by $\ve{\cal A}$. We therefore write $\ve h^{(\alpha)}$ 
as linear combinations of ${\cal H}$ and ${\cal A}$ as follows
\begin{eqnarray}
\ve h^{(\alpha)} & = & \Gamma^{(\alpha)} {\mathcal H} + \Lambda^{(\alpha)} {\mathcal A}.
\end{eqnarray} 
We insert this into Eq. (\ref{dual2_N=2}) and demand that cross-terms between
${\cal H}$ and ${\cal A}$ vanish, thus obtaining the following set of equations
determining the coefficients $ (\Gamma^{(\alpha)}, \Lambda^{(\alpha)} )$
\begin{eqnarray}
\Gamma^{(1)} + \Gamma^{(2)} &=&  1, \nonumber \\
\Lambda^{(1)} + \Lambda^{(2)} & = & 0, \nonumber \\
\Gamma^{(1)} \Lambda^{(1)}/|\psi^{(1)}|^2 + 
\Gamma^{(2)} \Lambda^{(2)}/|\psi^{(2)}|^2 & = & 0.
\end{eqnarray}
Thus, we have $\Gamma^{(\alpha)} = |\psi^{(\alpha)}|^2/\Psi^2$,
where $\Psi^2 =  |\psi^{(1)}|^2 + |\psi^{(2)}|^2$, which yields the following expression for the gauge field ${\cal A}$
\begin{eqnarray}
{\cal A} = \frac{1}{\Lambda^{(1)}} 
\frac{ |\psi^{(2)}|^2  \ve h^{(1)} - |\psi^{(1)}|^2 \ve h^{(2)}}{\Psi^2}.
\end{eqnarray}
Since we have three equations and four unknowns, we may choose $\Lambda^{(1)}$ freely, 
and determine it by simplifying the prefactor in ${\cal A}$ to get
$\Lambda^{(1)} = 1/\Psi^2 = - \Lambda^{(2)}$,
whence we have
\begin{eqnarray}
\label{neutralgaugefield}
{\cal A} =  |\psi^{(2)}|^2  \ve h^{(1)} - |\psi^{(1)}|^2  \ve h^{(2)}.
\end{eqnarray}
Inverting the relations for ${\cal H}$ and ${\cal A}$, we have
\begin{eqnarray}
\ve h^{(1)} & =  & 
(|\psi^{(1)}|^2 {\cal H} + {\cal A})/\Psi^2, \nonumber \\
\ve h^{(2)} & =  & 
(|\psi^{(2)}|^2 {\cal H} - {\cal A})/\Psi^2.
\end{eqnarray}
Inserting this back into Eq. (\ref{dual2_N=2}), collecting terms, and redefining
the fields ${\cal H}/\Psi^2 \to {\cal H}$ and ${\cal A}/\Psi^2 \to  {\cal A}$,
we have the action $S =  S_{{\cal H}} + S_{{\cal A}}$ where
\begin{eqnarray}
S_{{\cal H}}  & = & \sum_{\ve r} \left \{ 2 \pi \im  {\cal H} \cdot \ve m^{(+)}  
 +  \frac{1}{2 \beta_{{\cal H}}}  \left[(\nabla \times {\cal H})^2 + m_0^2  {\cal H}^2 \right] \right \}, \nonumber \\
S_{{\cal A}} & = & \sum_{\ve r} \left \{ 2 \pi \im  {\cal A} \cdot \ve m^{(-)} 
                   + \frac{1}{2 \beta_{{\cal A}}}  (\nabla \times {\cal A})^2 \right \}, 
\label{Sdual_AH}
\end{eqnarray}
where 
\beg
\ve m^{(+)}  &=&    |\psi^{(1)}|^2  \ve m^{(1)} + |\psi^{(2)}|^2  \ve m^{(2)}, \nonumber \\
\ve m^{(-)}  &=&  \ve m^{(1)} - \ve m^{(2)}, \nonumber \\
\frac{1}{2 \beta_{{\cal H}}}  &=&    \frac{ (|\psi^{(1)}|^2 + |\psi^{(2)}|^2)}{2 \beta}, \nonumber \\
\frac{1}{2 \beta_{{\cal A}}}  &=&  \frac{(1/|\psi^{(1)}|^2 + 1/|\psi^{(2)}|^2 )}{2\beta}, 
\label{defs_neutral}
\eee
and $ m_0^2 = e^2\Psi^2$. The action in Eq. (\ref{Sdual_AH}), which is equivalent to Eq. (\ref{dual2_N=2}), therefore describes a
 vortex mode $\ve m^{(+)}$ interacting with itself via a screened anti Biot-Savart interaction mediated by the 
 massive vector field ${\cal H}$, and the
vortex mode $\ve m^{(-)}$ interacting with itself via an 
unscreened anti Biot-Savart interaction mediated by  the gauge field
${\cal A}$. Hence, the former vortex mode is charged, the latter is
neutral. In Appendix  
\ref{App:charged_neutral}, we present an alternative method of identifying
charged and neutral modes for general $N$.

\section{\label{gauge_corr} Gauge field correlators}
Gauge field correlation functions are useful objects to study when considering 
the critical properties of gauge theories. The main reason is that they 
provide non-local gauge invariant order parameters for the theories,
which in turn enable reliable determination of critical exponents, including anomalous 
scaling dimensions. {\it Moreover, these correlators explicitly identify the 
mechanism by which the Meissner effect is destroyed in type-II superconductors:
The  mass of the gauge field $\ve A$, and hence the Higgs phase (equivalently 
the Meissner phase) is destroyed by a thermally driven vortex loop proliferation 
of the charged vortex mode} \cite{tesanovic1999,nguyen1999,hove2000,smiseth2004}.

In this section, we study in detail the direct gauge field 
correlation function, as well as various combinations of dual gauge field correlation 
functions, in order to gain insights into the nature of the critical points 
Eq. (\ref{gl_action}) can exhibit.

\subsection{$\ve A$-field correlator and Higgs mass}
We first consider the propagator for the gauge field $\ve A$, which provides information 
about at which of the critical points the Higgs phenomenon takes place, and where the 
remaining (neutral) fixed points appear. We present  compact expressions for 
the general-$N$ case, in later sections we present explicit numerical results for 
the cases  $N=2$ and $N=3$. 

We compute the correlation function $\langle \ve A(\ve r)\cdot\ve
A(0)\rangle$ in terms of vortex correlators in the standard way by starting from the action 
Eq. (\ref{dual1}), {\it prior to integrating out the gauge field $\ve A$},
adding source terms containing currents $\ve J$ minimally coupled to $\ve A$, 
and performing functional derivations with respect to the currents that are subject 
to the constraint $\ve \nabla \cdot \ve J = 0$, after which the currents are 
set to zero.  The details of the computations required to compute the 
$\ve A$-field correlator are given in Appendix \ref{App:gauge}. The
discrete Fourier transform of the gauge field propagator is
$\mathcal{G}_{\ve A}(\ve q) = \langle\ve A_{\ve q}\cdot\ve A_{-\ve q}\rangle$. We find 
\begin{equation}
  \mathcal{G}_{\ve A}(\ve q)  = \frac{2/\beta}{|\ve Q_{\ve q}|^2 + m_0^2}\left(1 +\frac{2\pi^2\beta
  e^2}{|\ve Q_{\ve q}|^2 }\frac{G^{(+)}(\ve q)}{|\ve Q_{\ve  q}|^2 + m_0^2}\right),
 \label{A_gaugeprop}
\end{equation}
where we have defined the correlation function of the charged vortex
mode as
\begin{eqnarray}
G^{(+)}(\ve q) & = & 
\langle (\sum_{\alpha=1}^N|\psi^{(\alpha)}|^2\ve m^{(\alpha)}_{\ve q})
\cdot(\sum_{\eta=1}^N|\psi^{(\eta)}|^2\ve m^{(\eta)}_{-\ve q})\rangle.
\label{Gpluss_corr}
\end{eqnarray}
Notice in Eq. (\ref{A_gaugeprop}), that the $\ve A$-field  correlator is only affected 
by the gauge-charged vortex mode $\sum_{\alpha=1}^N
|\psi^{(\alpha)}|^2\ve m^{(\alpha)}_{\ve q}$ via the coupling constant $m_0^2 \propto e^2$.

Eq. (\ref{A_gaugeprop}) is useful in MC simulations, 
in conjunction with scaling forms to be presented below, for extracting 
the gauge field mass and the anomalous scaling dimension of the gauge
field. The correlation length  
$\xi_{\ve A}$ that appears in a scaling Ansatz for the $\ve A$-field correlator
\begin{eqnarray}
{\cal{G}}_{\ve A}(\ve x) = \frac{1}{|\ve x|^{D-2+\eta_{\ve A}}} 
{\cal{G}}_{\pm} \left( \frac{|\ve x|}{\xi_{\ve A}} \right),
\end{eqnarray} 
is related to the mass of the gauge field via 
$m_{\ve A} = \xi_{\ve A}^{-1}$.
Here, $\eta_{\ve A}$ is the anomalous scaling dimension of 
the gauge field $\ve A$. 
Consequently, the gauge field
propagator \eq{A_gaugeprop} has the general structure\cite{kajantie2004}
\begin{equation}
  \label{def_gaugemass}
  \mathcal{G}_{\ve A}(\ve q) \sim \frac{1}{|\ve Q_{\ve q}|^2 + \Sigma_{\ve A}(\ve q)},
\end{equation}
where, close to the critical point 
\begin{equation}
  \label{ansatz_gaugemass}
  \Sigma_{\ve A}(\ve q) = m_{\ve A}^2 + C|\ve q|^{2-\eta_{\ve A}} + \mathcal{O}(|\ve q|^\delta),
\end{equation}
$C$ is a constant and $\delta > 2-\eta_{\ve A}$. By taking the $\ve q\to 0$ limit of
the Eqs. (\ref{def_gaugemass}) and (\ref{ansatz_gaugemass}) we may
extract the gauge mass from MC simulations. From the relation $\ve
B = \Delta\times\ve A$ the gauge mass is identified as the inverse magnetic
penetration depth $\lambda$. The masses of dual gauge fields are
defined in a similar fashion.

Let us make a remark concerning how  a charged fixed
point ($\eta_{\ve A} = 1$) could be distinguished from a neutral fixed
point ($\eta_{\ve A} = 0$) by gauge mass measurements. The magnetic
penetration length  
is related to the {\it superconducting} coherence length $\xi$
via \cite{Herbut_Tesanovic1996,hove2000}
\begin{eqnarray}
\lambda^{-1} \sim \xi^{\frac{2-d}{2-\eta_{\ve A}}} \sim |T-T_c|^{\frac{\nu(d-2)}{2-\eta_{\ve A}}},
\end{eqnarray}
where  $\nu$ is the critical exponent of the coherence length in the 
superconductor, i.e. $\nu = 0.67155(3)$\cite{Hasenbusch}, and $d$ is dimensionality. Therefore, 
we see that when $\eta_{\ve A}=0$, we have \cite{Herbut_Tesanovic1996,hove2000}
\begin{eqnarray}
\lambda \sim \sqrt{\xi}  \sim |T-T_c|^{-\frac{\nu}{2}},
\end{eqnarray}
while when $\eta_{\ve A}=1$, we have
\begin{eqnarray}
\lambda \sim \xi \sim |T-T_c|^{-\nu}.
\end{eqnarray}
Hence, the gauge mass $m_{\ve A}=\lambda^{-1}$ plotted as a function
of temperature in the critical regime should for $\eta_{\ve A}=1$ give a curve with {\it
  positive curvature},
while for $\eta_{\ve A}=0$ it should give a curve with {\it negative curvature}.

The compact expression Eq. (\ref{A_gaugeprop}) is valid for arbitrary
number of matter field 
flavors $N$, and generalizes the expression obtained in
\cite{hove2000}. 
Note that if $e^2=0$, we have trivially that \eq{A_gaugeprop} reduces to
\begin{equation}
  \mathcal{G}_{\ve A}(\ve q)  = \frac{2/\beta}{|\ve Q_{\ve q}|^2},
\end{equation}
which is always massless.
In Sections \ref{mc_simsN=3} 
and \ref{mc_simsN=3} we will use large-scale MC simulations to study in detail the case 
$N=2$ and $N=3$, respectively. The main feature of Eq. (\ref{A_gaugeprop}) is that at low temperatures, 
we may in the very simplest approximation entirely ignore the vortex
correlation function $ G^{(+)}(\ve q)$
such that $\mathcal{G}_{\ve A}(\ve q)$ is obviously massive with photon mass given by the bare 
mass $m_0$ of the problem. Actually, in the low-temperature regime, we have  
$G^{(+)}(\ve q) \sim \ve q^2$ which in the long-wavelength limit exactly cancels the factor 
$1/|\ve Q_{\ve q}|^2$, rendering the propagator massive. 

However, at the superconducting critical temperature, vortex loops proliferate \cite{tesanovic1999,nguyen1999,hove2000,loops_transition,qed3} 
resulting in vortex condensation and hence $\lim_{\ve q \to 0}
G^{(+)}(\ve q) \sim \rm{const}$. 
Now, the term inside the  brackets in Eq. (\ref{A_gaugeprop}) will diverge, dominating 
the behavior of the $\ve A$-field correlator, such that
$\mathcal{G}_{\ve A}(\ve q) \sim 1/\ve q^2$.
Thus, the Higgs mass is destroyed. {\it Note that the amplitudes
of the matter fields play no role in this, since they are entirely
frozen in the present London approximation. It is the condensation of 
topological defects of the matter fields, i.e. vortex loops, that are responsible 
for bringing the Higgs mass to zero, not the vanishing of the amplitudes} \cite{loops_transition}.
Therefore, we may view the divergence of the penetration length (the correlation length in the 
$\ve A$-field propagator), as a manifestation of the vortex loop blowout in the 
system. {\it Vortex loops} have dual counterparts in the current loops of the 
matter fields $\Psi_0^{(\alpha)}(\ve r)$ in Eq. (\ref{gl_action}). Conversely therefore, 
we  may also  view the Higgs-mass, i.e. the Meissner effect in the superconductor, 
as a manifestation of blowout of super-current loops upon entering the low-temperature 
phase. Again, the amplitudes of the matter fields
$\Psi_0^{(\alpha)}(\ve r)$ play no special role
here, other than that they have to be non-zero across the Higgs transition 
\cite{tesanovic1999,nguyen1999,hove2000,loops_transition,qed3}.

\begin{widetext}
\subsection{Dual gauge field correlators}
The details of the computations required for finding the dual gauge field
correlation functions in terms of vortex fields are found in Appendix \ref{App:dualgauge}. We find
the following "Dysons's equation" for the gauge field correlator
\begin{eqnarray}
\langle  \ve h^{(\alpha)}_{\ve q}  \cdot \ve h^{(\beta)}_{-\ve q} \rangle
 = \widetilde{D}^{(\alpha,\beta)} (\ve q) 
- \pi^2 \widetilde{D}^{(\alpha,\eta)} (\ve q) \widetilde{D}^{(\beta,\kappa)} (\ve q) 
\langle \ve m^{(\eta)}_{\ve q} \cdot  \ve m^{(\kappa)}_{-\ve q} \rangle, 
\label{dual_h_individual}
\end{eqnarray}
\end{widetext}
where we have used the fact that the trace of the transverse projection operator is 
given by  $\rm{Tr} \left[ P_{T}^{\mu\nu} \right] = 2$,  the matrix elements 
$\widetilde{D}^{(\alpha,\eta)} (\ve q) $ are defined
in Eq. (\ref{potential}), and a summation over the indices $(\eta,\kappa) \in [1,\dots,N]$
is understood. These results are valid for all $N$. 

To obtain more explicit expressions, we will work out in detail what we obtain for $N=2$. As we have 
seen above, in this case it is natural to use  Eq. (\ref{dual_h_individual}) to 
form correlation functions of the combination $\ve h^{(1)}+ \ve h^{(2)}$. We will, for completeness
also consider the combination and 
$\ve h^{(1)}- \ve h^{(2)}$ and $|\psi^{(2)}|^2 \ve h^{(1)} - |\psi^{(1)}|^2 \ve h^{(2)}$. We also 
use the fact that the interaction matrix $\widetilde{D}^{(\alpha,\beta)}(\ve q)$ is symmetric, and 
introduce the definitions
\begin{eqnarray}
\ve h^{(\pm)}_{\ve q} & \equiv & \ve h^{(1)}_{\ve q} \pm  \ve h^{(2)}_{\ve q}                  \nonumber \\
a^{(\pm)}         & \equiv & \widetilde{D}^{(1,1)}(\ve q) \pm \widetilde{D}^{(1,2)}(\ve q) \nonumber \\
b^{(\pm)}         & \equiv & \widetilde{D}^{(2,2)}(\ve q) \pm \widetilde{D}^{(1,2)}(\ve q). 
\label{ab1_eqs}
\end{eqnarray}
It is enlightening at this stage to introduce the expressions for $\widetilde{D}^{(\alpha,\beta)}(\ve q)$, 
as follows
\begin{eqnarray}
\frac{\widetilde{D}^{(1,1)}(\ve q)\Psi^2}{2 \beta |\psi^{(1)}|^2  |\psi^{(2)}|^2 } & = & 
\frac{1}{|\ve Q_{\ve q}|^2} + \frac{|\psi^{(1)}|^2 }{|\psi^{(2)}|^2}
\frac{1}{|\ve Q_{\ve q}|^2 + m_0^2}, \nonumber \\
\frac{\widetilde{D}^{(2,2)}(\ve q)\Psi^2}{2 \beta |\psi^{(1)}|^2  |\psi^{(2)}|^2 } & = & 
 \frac{1}{|\ve Q_{\ve q}|^2} + \frac{|\psi^{(2)}|^2 }{|\psi^{(1)}|^2}
\frac{1}{|\ve Q_{\ve q}|^2 + m_0^2}, \nonumber \\
\frac{\widetilde{D}^{(1,2)}(\ve q)\Psi^2}{2 \beta |\psi^{(1)}|^2  |\psi^{(2)}|^2 } & = & 
-\frac{1}{|\ve Q_{\ve q}|^2} + \frac{1}{|\ve Q_{\ve q}|^2 + m_0^2},
\label{D_eqs}
\end{eqnarray}
where $\Psi^2 = |\psi^{(1)}|^2 + |\psi^{(2)}|^2$. Using Eqs. (\ref{D_eqs}) in Eqs. (\ref{ab1_eqs}), we find
\begin{eqnarray}
a^{(+)} & \equiv & \frac{2 \beta |\psi^{(1)}|^2}{|\ve Q_{\ve q}|^2 + m_0^2}, \nonumber \\ 
b^{(+)} & \equiv & \frac{2 \beta |\psi^{(2)}|^2}{|\ve Q_{\ve q}|^2 + m_0^2},
\label{ab_pluss}
\end{eqnarray}
and $a^{(-)}$ and $b^{(-)}$ given by
\begin{eqnarray}
\frac{a^{(-)}\Psi^2}{2 \beta |\psi^{(1)}|^2 |\psi^{(2)}|^2} & \equiv & \frac{2}{|\ve Q_{\ve q}|^2} +  \frac{|\psi^{(1)}|^2/|\psi^{(2)}|^2-1}{|\ve Q_{\ve q}|^2 + m_0^2}, \nonumber \\ 
\frac{b^{(-)}\Psi^2}{2 \beta |\psi^{(1)}|^2 |\psi^{(2)}|^2} & \equiv & \frac{2}{|\ve Q_{\ve q}|^2} +  \frac{|\psi^{(2)}|^2/|\psi^{(1)}|^2-1}{|\ve Q_{\ve q}|^2 + m_0^2}, 
\label{ab_minus}
\end{eqnarray}
where $m_0^2 = e^2\Psi^2$. 
{\it Notice how the  unscreened part of the interactions cancel out in $(a^{(+)}, b^{(+)})$
but not in $(a^{(-)}, b^{(-)})$}. This is the origin of the qualitatively different behavior
we will find for the $\ve h^{(+)}_{\ve q}$ and   $\ve h^{(-)}_{\ve q}$ correlators.
Notice also how the expressions 
simplify when $|\psi^{(1)}|^2 = |\psi^{(2)}|^2  $, when the screened part of the interactions 
appearing in $a^{(-)}, b^{(-)}$ vanish,  such  that $a^{(-)}= b^{(-)}$.

We may now write the correlation functions of the two relevant linear combinations
of dual gauge fields as follows
\begin{widetext}
\begin{eqnarray}
\mathcal{G}^{(\pm)}_{\ve h}(\ve q) & \equiv & \langle  \ve h^{(\pm)}_{\ve q} \cdot \ve h^{(\pm)}_{-\ve q}  \rangle
= a^{(\pm)} + b^{(\pm)} 
 -  \pi^2 \langle (a^{(\pm)} \ve m^{(1)}_{\ve q} \pm  b^{(\pm)} \ve m^{(2)}_{\ve q}) \cdot
(a^{(\pm)}\ve m^{(1)}_{-\ve q} \pm  b^{(\pm)} \ve m^{(2)}_{-\ve q}) \rangle.
\label{hpm_corr_eq}
\end{eqnarray}
\end{widetext}
Using Eqs. (\ref{ab_pluss}), and (\ref{hpm_corr_eq}),
we find the surprisingly compact expression, valid for all $N$
\begin{eqnarray}
\mathcal{G}^{(+)}_{\ve h}(\ve q) = \frac{2\beta\psi^2}{|\ve Q_{\ve q}|^2 + m_0^2}
\left(1  -  \frac{2\pi^2\beta}{\psi^2}\frac{G^{(+)}(\ve q)}{|\ve Q_{\ve q}|^2 + m_0^2}
  \right),
\label{hp_corr}
\end{eqnarray}
where we have again introduced $G^{(+)}(\ve q)$ appearing in Eq. (\ref{Gpluss_corr}).
In fact, this result could have been written down using the known result for
the charged case for $N=1$ \cite{hove2000}, in combination with Eq. (\ref{Sdual_AH}),
considering the part of Eq. (\ref{Sdual_AH}) only pertaining to the massive vector
field ${\cal H}$. This provides a nice consistency check on the general expression
for the dual gauge field correlators, as well as on the interaction
matrix $\widetilde D^{(\alpha,\eta)}(\ve q)$. In the low- and
high-temperature phase, the vortex correlator $G^{(+)}(\ve q)$
behaves as $\sim \ve q^2$ and $\sim c(T)$, respectively. In either case, the 
dual gauge field correlator $\mathcal{G}^{(+)}_{\ve h}(\ve q)$ is always massive.

Consider the correlation function of the combination of dual gauge fields
$\mathcal{A} = |\psi^{(2)}|^2 \ve h^{(1)} - |\psi^{(1)}|^2 \ve
h^{(2)}$ which couples to the gauge-neutral vortex mode in \eq{Sdual_AH}. In principle 
we may follow the routes used in the above
calculations, but by now we realize that a quick way of obtaining the
results is to use \eq{Sdual_AH} in combination with the
known results for the case $N=1$ in the {\it neutral} case \cite{hove2000}.
We define
\begin{eqnarray}
{\mathcal G}_{{\mathcal A}}(\ve q) \equiv  \langle {\mathcal A}_{\ve q}  {\mathcal A}_{-\ve q}  \rangle,
\end{eqnarray}
and find immediately, using the results of Ref. \onlinecite{hove2000} along 
with the definitions in \eq{defs_neutral}
\begin{eqnarray}
{\mathcal G}_{\mathcal A}(\ve q) = \frac{2 \beta_{{\mathcal A}}}{|\ve Q_{\ve q}|^2} 
\left(1 - \frac{2 \pi^2 \beta_{{\mathcal A}} G^{(-)}(\ve q)}{|\ve Q_{\ve q}|^2} \right),
\label{dual_corr3}
\end{eqnarray}
where
\begin{equation}
G^{(-)}(\ve q) =  
\langle (\ve m^{(1)}_{\ve q}  - \ve m^{(2)}_{\ve q}) \cdot (\ve
m^{(1)}_{-\ve q}  -
\ve m^{(2)}_{-\ve q}) \rangle
\end{equation}
is the correlation function of the gauge-neutral vortex mode.

In the long wave length limit the behavior of $G^{(-)}(\ve q)$ gives rise
to a dual Higgs mechanism. This comes about because  
the $G^{(-)}(\ve q)$ correlation function is always $\sim \ve q^2$ at long 
wavelengths, but has a non-analytic coefficient in front of the $\ve q^2$ 
term given by the helicity modulus of the gauge-neutral mode 
$\ve m^{(1)} - \ve m^{(2)}$. This serves to cancel the $1/\ve q^2$ term 
in the $ \mathcal{G}_{\mathcal A}(\ve q)$ correlation function exactly.
This cancellation, originating in the vanishing of the helicity
modulus of the gauge-neutral mode, is responsible for producing a dual 
Higgs mass $m_\mathcal{A}$ in $\mathcal{G}_{\mathcal A}(\ve q) $. Higher order terms
determine the actual value of the dual Higgs mass.  
Thus, we see that while $\ve h^{(1)} + \ve h^{(2)}$ is always massive,
$|\psi^{(2)}|^2\ve h^{(1)} - |\psi^{(1)}|^2\ve h^{(2)}$ plays the role of a gauge degree of freedom
which provides a dual counterpart to $\ve A$ in \eq{gl_action}.
This is a manifestation of the self-duality of the theory which we have 
alluded to above \cite{motrunich2004,sachdev2004,smiseth2004}.

Notice that the existence of a dual Meissner effect arising out of
\eq{dual_corr3} is a substantially more subtle effect than 
the direct Meissner effect coming out of \eq{A_gaugeprop}.  
The correlator of the gauge-neutral mode
has the property  
\begin{equation}
 \label{g_min}
 G^{(-)}(\ve q) =  C_2 \ve q^2 + C_4 \ve q^4 + \mathcal{O}(\ve q^6),
\end{equation}
 for all 
temperatures, in analogy with the vortex correlator of the \xy model
for the case $N=1$. It is the non-analytic behavior of the
coefficient $C_2$, involving the {\it helicity modulus} of the
gauge-neutral mode, which is responsible for {\it producing} 
a dual Higgs mass as the gauge-neutral mode proliferates. To obtain a 
dual Meissner effect, a subtle cancellation is required, namely 
that at some critical temperature $T_{\rm c1}$, we must have 
\begin{eqnarray}
\label{criterion2}
1-\frac{2\pi^2 \beta |\psi^{(1)}|^2|\psi^{(2)}|^2C_2(T_{c1})}{|\psi^{(1)}|^2
  + |\psi^{(2)}|^2} = 0,
\end{eqnarray}
where we have used the expression for $\beta_{{\mathcal A}}$ from \eq{defs_neutral}.
It is important to note that while the actual value of the dual Higgs mass 
is influenced by the higher order terms in \eq{g_min}, the {\it criterion} for obtaining a dual Higgs 
phenomenon is only determined by the cancellation among the terms of
order $1/\ve q^2$ terms in \eq{dual_corr3}. 
This differs from the mechanism that destroys the Higgs mass in the $\ve A$ 
correlator, since there no such subtle cancellations are required, it suffices 
that the correlator $G^{(+)}(\ve q) $ changes behavior from a constant to 
$\sim \ve q^2$ in the long-wavelength limit.

We finally consider the correlation function of $\ve h^{(-)}$. Applying the results 
from \eq{hpm_corr_eq}, we find \cite{smiseth2004} 
\begin{widetext}
\begin{equation}
\mathcal{G}_{\ve h}^{(-)}(\ve q) =
\frac{8 \beta \lambda^{(1)}\lambda^{(2)}\Psi^2}{|\ve Q_{\ve q}|^2 }
 \left\{ 1 - \frac{2\pi^2\beta\lambda^{(1)}\lambda^{(2)}\Psi^2
   G^{(-)}(\ve q)}{|\ve Q_{\ve q}|^2}
- \frac{2\pi^2\beta(\lambda^{(1)} - \lambda^{(2)}) G^{(m)}(\ve q)}{|\ve Q_{\ve q}|^2 + m_0^2} 
\right\}  + (\lambda^{(1)} - \lambda^{(2)})^2\mathcal{G}_{\ve h}^{(+)}(\ve q),
\label{hm_dual}
\end{equation}
\end{widetext}
where $\lambda^{(\alpha)} = |\psi^{(\alpha)}|^2/\Psi^2$, $\Psi^2 =
  |\psi^{(1)}|^2 + |\psi^{(2)}|^2$, and the mixed gauge-neutral and gauge-charged vortex field
correlator is given by
\begin{equation}
G^{(\rm m)}(\ve q) = 
\langle (\ve m^{(1)}_{\ve q}  - \ve m^{(2)}_{\ve q}) \cdot
(\sum_{\alpha=1}^2|\psi^{(\alpha)}|^2\ve m^{(\alpha)}_{-\ve q}) \rangle.
\end{equation}
Note that for the case $N=1$, such that either $\lambda^{(1)}$ or $\lambda^{(2)}$
vanishes, then the remaining $\lambda^{(\eta)}=1$, only the last term in \eq{hm_dual} 
survives, and $\mathcal{G}^{(-)}_{\ve h}(\ve q)$ correctly reduces to
$\mathcal{G}^{(+)}_{\ve h}(\ve q)$
in \eq{hp_corr}.
In the long wave length limit, it is the second term in the curly
brackets in \eq{hm_dual} that 
dominates, giving rise to a dual Higgs mechanism. Notice again how it
is the vortex correlator $G^{(-)}(\ve q)$ which determines the fate of the
massless dual gauge field $\ve h^{(1)} - \ve h^{(2)}$, just like in
\eq{dual_corr3}. This is particularly evident for the case $|\psi^{(1)}|=|\psi^{(2)}|$,
when \eq{hm_dual} reduces to
\begin{eqnarray}
 \mathcal{G}^{(-)}_{\ve h}(\ve q) 
= \frac{4 \beta |\psi^{(1)}|^2 }{|\ve Q_{\ve q}|^2}\left(1  
- \pi^2\beta |\psi^{(1)}|^2  \frac{G^{(-)}(\ve q)}{|\ve Q_{\ve q}|^2}
  \right).
\label{H_corr_iso}
\end{eqnarray}
This correlator for $N=2$, $e \neq 0$ has precisely the same form
as the dual gauge field correlator for the case $N=1, e=0$, which
exhibits a dual Higgs phenomenon \cite{hove2000}.

Substituting $\lambda^{(\alpha)} =
|\psi^{(\alpha)}|^2/\Psi^2$ in \eq{hm_dual}, we see that the criterion
for destroying the dual Higgs mass is precisely the same as the
criterion we arrived at in \eq{criterion2}. 
Thus, whether we compute the correlator 
in \eq{hm_dual} or that in \eq{dual_corr3} to establish the 
existence of a dual Higgs phase does not matter. 
Furthermore, for $N=2, e \neq 0$, 
$\ve m^{(1)} - \ve m^{(2)}$ behaves  as vortices for $N=1, e=0$, {\it i.e. it is 
a superfluid mode arising out of superconducting condensates.} A nonzero  $m_{\cal A}$ 
for the dual gauge field ${\cal A}$ is {\it produced} by disordering $\theta^{(1)}$ 
at a critical temperature $T_{\rm c1}$ while a nonzero  $m_{\ve A}$
for the gauge 
field $\ve A$ is {\it destroyed} by disordering $\theta^{(2)}$ at a critical temperature 
$T_{\rm c2}$.

\section{\label{mc_simsN=2}Monte Carlo simulations, $N=2$}
Since the bare interaction between vortices is dominated at
long distances by an unscreened part, it is of interest to study the 
character of the phase transition associated with the generation of a Higgs 
mass for the gauge field ${\bf{A}}$. For the $N=1$ case, it is 
known that the vortex tangle of the \xy model is incompressible and the dual 
theory is a gauge theory such that $\langle \phi \rangle \neq 0$ is 
prohibited. For the charged case, the vortex tangle is compressible, the dual 
theory only has global symmetry,  and hence vortex condensation and 
$\langle \phi \rangle \neq 0$ is possible. The introduction of charge 
destabilizes the \xy fixed point. 

To investigate what happens for the case $N=2$, MC simulations have 
been carried out for the action Eq. (\ref{vortex_action}) on a three
dimensional lattice of size $L\times L\times L$ for two different cases. 
In the first case we simulate with unequal bare stiffnesses 
$|\psi^{(1)}|^2 = 1/2$ and $|\psi^{(2)}|^2 = 1$, $e^2 = 1/4$ and
$m_0^2 = 3/8$. The bare stiffnesses have been chosen to have well-separated bare energy scales
associated with the twist of the two types of phases. In the second
case we use equal phase stiffnesses
$|\psi^{(1)}|^2 = |\psi^{(2)}|^2 = 1$, $e^2 = 1/4$ and $m_0^2 = 1/2$.
The values for $m_0$ 
have been chosen such that they are of order the lattice spacing in the problem to
avoid difficult finite-size effects. One MC update consists of inserting a unitary 
vortex loop of random direction and species according to the Metropolis algorithm.

To calculate the critical exponents $\alpha$ and $\nu$ we performed
finite size scaling (FSS) analysis with bootstrap error estimates of the 
third moment of the action\cite{m3} $M_3=\langle(S_{\rm V}-\langle S_{\rm
  V}\rangle)^3\rangle/L^3$ where $S_{\rm V}$ is given in \eq{vortex_action}. The peak 
to peak value of this quantity scales with system size $L$ as $L^{(1+\alpha)/\nu}$, whereas 
the width between the peaks scales as $L^{-1/\nu}$. The advantage of this is that 
asymptotically correct behavior is reached for practical system sizes. 

To characterize the phase transitions further, we consider the correlation functions given in 
Eqs. (\ref{A_gaugeprop}), (\ref{Gpluss_corr}), and (\ref{hp_corr}). In the Higgs phase the gauge 
field mass $m_{\ve A}$ scales according to the Ansatz\cite{kajantie2004} given by 
Eqs. (\ref{def_gaugemass}) and (\ref{ansatz_gaugemass})
\begin{eqnarray}
\label{prop_ansatz}
\mathcal{G}_{\ve A}(\ve q)^{-1}\frac{2}{\beta} = m_{\ve A}^2 + C |\ve
q|^{2-\eta_{\ve A}} + \mathcal{O}(|\ve q|^\delta),
\end{eqnarray}
with a corresponding Ansatz for $\mathcal{G}^{(+)}_{\ve h}(\ve q)$. The masses of  
$\ve A$ and $\sum_{\alpha=1}^{N} \ve h^{(\alpha)}$ are therefore
defined through the $\ve q\to 0$ limit of the respective Ans\"{a}tze
\begin{eqnarray}
m_{\ve A}^2 & \equiv & \lim_{\ve q\to0} \frac{2}{\beta\mathcal{G}_{\ve A}(\ve q)} \nonumber \\
m_{\Sigma\ve h}^2 & \equiv & \lim_{\ve q\to0}\frac{2\beta\psi^2}{\mathcal{G}^{(+)}_{\ve h}(\ve q)}.
\label{gauge_field_masses}
\end{eqnarray}
The gauge field masses are found by measuring vortex correlators
followed by a fit for small $\ve q$ to their respective Ans\"{a}tze.

We briefly review the $N=1$ GL-model. The dual field theory of the neutral 
fixed point is a charged theory describing an incompressible vortex tangle 
\cite{hove2000}. The leading behavior of the vortex correlator
$G^{(+)}(\ve q)\sim\langle \ve m_{\ve q}\ve \cdot \ve m_{-\ve q}\rangle$ is\cite{hove2000}
\begin{equation}
\lim_{\ve q\to 0}G^{(+)}(\ve q) \sim \begin{cases}
  [1-C_2(T)]\ve q^2 & ; T<T_{c} \\
  \ve q^2-C_3(T)|\ve q|^{2+\eta_{\ve h}} & ; T=T_{\rm c} \\
  \ve q^2+C_4(T)\ve q^4 & ;T>T_{\rm c},
\end{cases}
\end{equation}
where $\eta_{\ve h}$ is the anomalous scaling dimension of the dual
gauge field $\ve h$. For $T < T_{\rm c}$ the mass of the dual gauge field given by
Eqs. (\ref{hp_corr}) and (\ref{gauge_field_masses}) (with $N=1$ and $e=0$)
is zero, however for $T>T_{\rm c}$ the $\ve q^{-2}$ terms in
\eq{hp_corr} 
cancel out exactly and the mass $m_{\ve h}$ attains an expectation value. 
At the charged fixed point of the GL model, the effective field theory
of the vortices is a neutral theory. The vortex tangle is compressible
with a scaling Ansatz for the vortex correlator 
\begin{equation}
\label{n1_corr}
\lim_{\ve q\to 0}G^{(+)}(\ve q)\sim \begin{cases} \ve q^2 & ; T<T_{\rm c} \\
|\ve q|^{2-\eta_{\ve A}} & ; T=T_{\rm  c} \\
c(T) & ; T> T_{\rm c},
\end{cases}
\end{equation}
where $c(T)$ is a nonzero constant. Consequently, from Eqs. (\ref{A_gaugeprop}), (\ref{hp_corr}), and
(\ref{gauge_field_masses}) (with $N=1$ and $e\neq 0$), the mass $m_{\ve A}$ drops to zero at $T_{\rm c}$, 
and the mass of the dual vector field $m_{\ve h}$ is finite for all temperatures and has 
a kink at $T_{\rm c}$ \cite{hove2000}. Renormalization group arguments yield 
$\eta_{\ve  A}=4-d$ where $d$ is the dimensionality \cite{HLM,Bergerhoff,Herbut_Tesanovic1996}, 
which has recently been verified numerically \cite{hove2000,kajantie2004}.

\subsection{Critical exponents $\alpha$ and $\nu$, $|\psi^{(1)}| < |\psi^{(2)}|$}
\begin{figure}[htb]
\centerline{\scalebox{0.34}{\rotatebox{-90.0}{\includegraphics{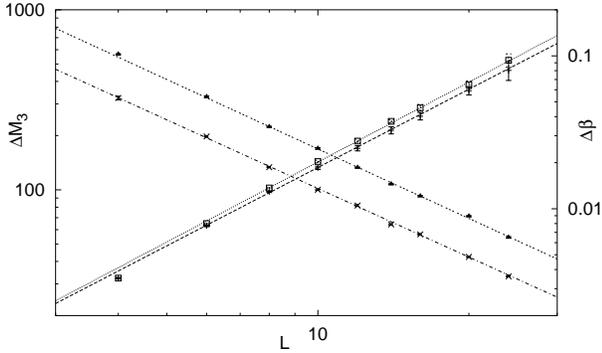}}}} 
\caption{\label{m3} The FSS of the peak to peak value of
  the third moment $\Delta M_3$ labeled ($\Box$) and (+) for
  $T_{\rm c1}$ and $T_{\rm c2}$, respectively. The scaling of
  the width between the peaks $\Delta\beta$ is labeled ($\blacktriangle$) 
  and ($\times$) for $T_{\rm c1}$ and $T_{\rm c2}$ respectively. The 
  lines are power law fits to the data for $L>6$ used to extract 
  $\alpha$ and $\nu$.}
\end{figure}

We observe two anomalies in the specific heat at $T_{\rm c1}$ and $T_{\rm c2}$ where 
$T_{\rm c1}< T_{\rm c2}$. We find $T_{\rm c1}$ and $T_{\rm c2}$ from scaling of the second moment of 
the action $\langle (S_{\rm V} - \langle S_{\rm V} \rangle )^2\rangle/L^3$
to be $T_{\rm c1} = 1.4(6)$ and $T_{\rm c2} = 2.7(8)$. The $M_3$ FSS
plots for system sizes $L=4, 6, 8, 10, 12, 14, 16, 20, 24$ are shown
in \fig{m3}. 
From the scaling we conclude that both anomalies are in fact critical points, and 
we obtain $\alpha = -0.02 \pm 0.02$ and $\nu=0.67\pm
0.01$ for $T_{\rm c1}$ and $\alpha = -0.03 \pm 0.02$ and $\nu = 0.67\pm 0.01$ for
	 $T_{\rm c2}$. These values are consistent with those of the
\xy and the {\it inverted} \xy universality classes found with high
precision in Refs. \onlinecite{Kleinert1999,Hasenbusch,Lipa}.
 
\subsection{Vortex correlator, Higgs mass, and anomalous scaling dimension, $|\psi^{(1)}| < |\psi^{(2)}|$}

\begin{figure}[htb]
\centerline{\scalebox{0.34}{\rotatebox{-90.0}{\includegraphics{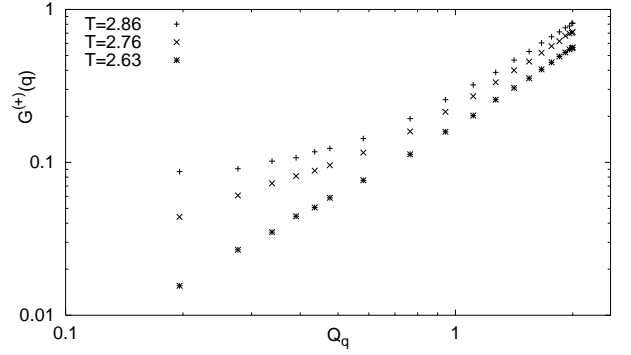}}}} 
\caption{\label{G_q} $G^{(+)}(\ve q)$ for $N=2$ and $L=32$, plotted
  for temperatures 
$T=2.86 > T_{\rm{c2}}$, $T=2.76\simeq T_{\rm c2}$, and $T=2.63 < T_{\rm{c2}}$,  
$\lim_{\ve q \to 0} G^{(+)}(\ve q) \sim c(T)$,  $\sim |\ve q|$, and
  $\sim \ve q^2$, 
respectively. The $q\to 0$ behavior of the correlator matches
  precisely the signature of a changed fixed point given in \eq{n1_corr}.}
\end{figure}

The vortex correlators for the $N=2$ case are
sampled in real space and $G^{(+)}(\ve q)$ given in
\eq{Gpluss_corr} is found by a discrete Fourier
transformation. At the lower transition $T_{\rm c1}$ the leading
behavior is $G^{(+)}(\ve q)\sim \ve q^2$ on both sides of the
transition. Consequently, due to Eqs. (\ref{A_gaugeprop}),
(\ref{hp_corr}), and (\ref{gauge_field_masses}), $m_{\ve A}$ and
$m_{\Sigma \ve h}$ are finite in this regime. This shows that the 
vortex tangle is incompressible and that the anomalous scaling
dimension $\eta_{\ve A}=0$, which corresponds to a neutral fixed
point. Fig. \ref{G_q} shows the correlator $G^{(+)}(\ve q)$ around $T_{\rm
  c2}$.
 Below $T_{\rm c2}$ the dominant behavior is $G^{(+)}(\ve q)\sim
\ve q^2$ whereas $G^{(+)}(\ve q)\sim c(T)$
above the transition. At the critical point $G^{(+)}(\ve q)\sim |\ve q|$, indicating
$\eta_{\ve A}=1$. Accordingly $m_{\ve A}$ is finite below the
transition and zero for $T \ge T_{\rm c2}$.

\begin{figure}[htb]
\centerline{\scalebox{0.34}{\rotatebox{-90.0}{\includegraphics{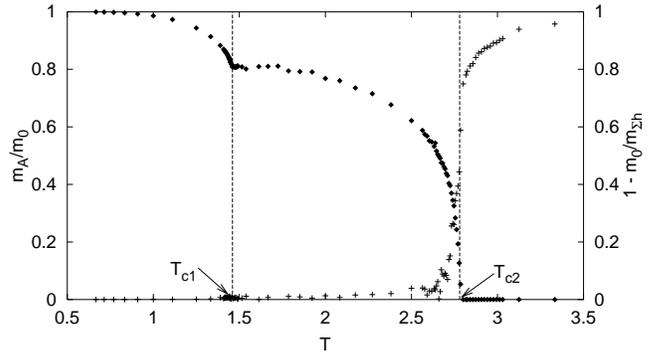}}}} 
\caption{\label{gauge_mass}
 The mass $m_{\ve A}$ ($\blacklozenge$) and
  $1-m_0/m_{\Sigma\ve h}$ (+) found from Eqs. (\ref{A_gaugeprop}) and (\ref{hp_corr}). 
Two  non-analyticities can be seen in $m_{\ve A}$ at $T_{\rm c1}$ and
  $T_{\rm c2}$, corresponding a neutral fixed point and a charged
  Higgs fixed point, respectively. An abrupt increase in $m_{\Sigma\ve h}$ due 
  to vortex
  condensation is located at $T_{\rm c2}$. } 
\end{figure}

For each coupling we fit $\mathcal{G}_{\ve A}(\ve q)^{-1}$ for $|\ve
Q_{\ve q}| < 0.9$ using system sizes
$L=8,12,20,32$ to Eq. (\ref{prop_ansatz}). The results for $m_{\ve A}$,
and $m_{\Sigma \ve h}$ which is found in a similar fashion, are given in
\fig{gauge_mass}. The system exhibits Higgs mechanism
when $m_{\ve A}$ drops to zero at $T_{\rm c2}$ with an anomaly in
$m_{\Sigma\ve h}$ due to vortex condensation. Furthermore $m_{\ve A}$
has a kink at $T_{\rm c1}$ due to ordering of the phase difference
$\theta^{(1)}-\theta^{(2)}$ with the phase stiffness
$|\psi^{(1)}|^2|\psi^{(2)}|^2/(2|\psi^{(1)}|^2 + 2|\psi^{(2)}|^2)$,
confirm \eq{charge_neutral}\cite{frac}.
The anomalies in $m_{\ve A}$ and 
$m_{\Sigma \ve h}$ coincide precisely with $T_{\rm c2}$ and $T_{\rm c1}$. 
Note also how $m_{\Sigma\ve h}$ changes abruptly at $T_{\rm c2}$.
This is due to a sudden change in screening by 
$\sum_{\alpha=1}^N\ve h^{(\alpha)}$, giving an abrupt increase in $m_{\Sigma\ve h}$. This is 
consistent with the flow equation Eq. (\ref{rg_flow}). Note that the mass of the
algebraic sum of the dual fields appears in Eq. (\ref{dual2}) after 
integrating out the gauge field $\ve A$.

We may understand the  transitions  as follows. Above  $T_{\rm{c2}}$,  $\bf{A}$ is 
massless, giving a compressible vortex tangle which accesses configurational entropy 
better than an incompressible one. Below $T_{\rm{c2}}$, ${\bf{A}}$ is massive and 
merely renormalizes $|\Psi|^4$ terms in Eq. (\ref{gl_action}). The theory is 
effectively a $|\Psi |^4$ theory in this regime. Thus, the remaining proliferated 
vortex species originating in the matter fields with lower bare stiffnesses form  vortex 
tangles  as if they originated in a neutral superfluid. For the general $N$ case, a Higgs 
mass is generated at the highest critical temperature, after which ${\bf{A}}$
renormalizes the $|\Psi|^4$ term, such that the Higgs  fixed point is followed by $N-1$ 
neutral fixed points as the temperature is lowered.

The picture that emerges from the above discussion of the gauge field and the dual gauge field 
correlators is the following. Below $T_{\rm c1}$ there is one massless "photon", namely 
$|\psi^{(2)}|^2\ve h^{(1)} - |\psi^{(1)}|^2\ve h^{(2)}$, while $\ve A$ is massive.  Above $T_{\rm
  c1}$ and  below $T_{\rm c2}$, 
both $|\psi^{(2)}|^2\ve h^{(1)} - |\psi^{(1)}|^2\ve h^{(2)}$ and $\ve A$ are massive, while above
$T_{\rm c2}$,  
$|\psi^{(2)}|^2\ve h^{(1)} - |\psi^{(1)}|^2\ve h^{(2)}$ is massive and $\ve A$ is massless. 

\subsection{Critical exponents $\alpha$ and $\nu$, $|\psi^{(1)}| = |\psi^{(2)}|$}

A special case is obviously presented by the case $|\psi^{(1)}| = |\psi^{(2)}|$ since then
$T_{\rm c1} = T_{\rm c2} \equiv T_{\rm c}$, and we have a transition directly from a low-temperature phase with 
one massless dual gauge  field $|\psi^{(2)}|^2\ve h^{(1)}- |\psi^{(1)}|^2\ve h^{(2)} = 
|\psi^{(1)}|^2(\ve h^{(1)}- \ve h^{(2)})$ to a high-temperature phase with one massless direct gauge 
field $\ve A$. This is the remarkable self-duality observed in Refs. \onlinecite{motrunich2004,sachdev2004,smiseth2004}.

\begin{figure}[htb]
\centerline{\scalebox{0.34}{\rotatebox{-90.0}{\includegraphics{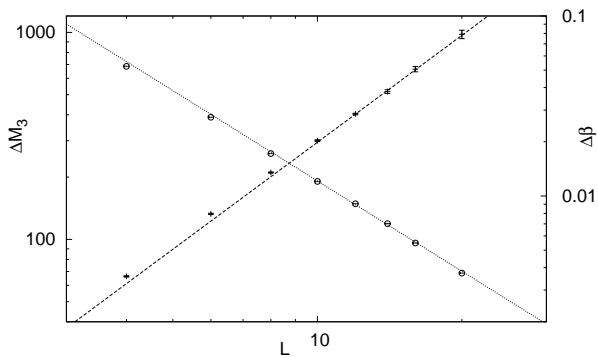}}}} 
\caption{\label{scaling_eqpsi} The FSS of the peak to peak value of
  the third moment $\Delta M_3$ labeled (+) for
  $T_{\rm c}$ and for $|\psi^{(1)}| = |\psi^{(2)}|$.  The scaling
  of the width between the peaks $\Delta\beta$ is labeled ($\circ$). The lines
are power law fits to the data for $L>8$ used to extract $\alpha$ and $\nu$.}
\end{figure}

The second moment of the action with $|\psi^{(1)}|^2 =
|\psi^{(2)}|^2 = 1$, $e^2=1/4$ and $m_0^2 = 1/2$ exhibits one anomaly at
$T_{\rm c}=2.7(8)$. Scaling plots of the third moment of the action 
are shown in Fig. \ref{scaling_eqpsi}. FSS yields $\alpha=0.03 \pm 0.04$ and
$\nu=0.60\pm0.02$.  The numerical value for $\nu$ is in agreement with the value found in
Ref. \onlinecite{motrunich2004}, $\nu = 0.60\pm0.05$. Note that our
result for $\alpha$ and $\nu$ is not in agreement with hyper scaling.

\subsection{Vortex correlator and Higgs mass, $|\psi^{(1)}| = |\psi^{(2)}|$}

\begin{figure}[htb]
\centerline{\scalebox{0.34}{\rotatebox{-90.0}{\includegraphics{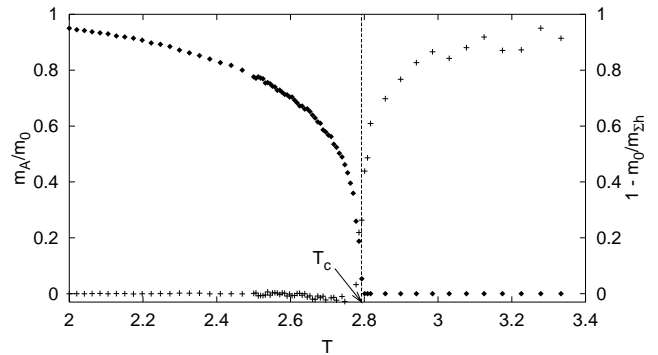}}}} 
\caption{\label{mass_A_eqpsi}
 The mass $m_{\ve A}$ ($\blacklozenge$) and
  $1-m_0/m_{\Sigma\ve h}$ (+) found from Eqs. (\ref{A_gaugeprop}) and (\ref{hp_corr}),
for $|\psi^{(1)}| = |\psi^{(2)}|$. One  non-analyticity can be seen in $m_{\ve A}$ 
at $T_{\rm c}$, corresponding to a fixed point which is
not in the \xy or inverted \xy universality class. 
An abrupt increase in $m_{\Sigma\ve h}$ due to vortex
  condensation is located at $T_{\rm c}=2.7(8)$. } 
\end{figure}

The mass of the gauge field $m_{\ve A}$ was found by fitting $\mathcal{G}_{\ve A}(\ve q)^{-1}$ data from system sizes
$L=8,12,20,32$ to Eq. (\ref{prop_ansatz}). The mass $m_{\Sigma \ve h}$
was found similarly. The results are presented in \fig{mass_A_eqpsi}.

\subsection{Discussion}

The result for the exponents $\alpha$ and $\nu$ at $T_{\rm c}$ for  $|\psi^{(1)}| = |\psi^{(2)}|$ 
 shows that when the \xy and inverted 
\xy critical points collapse onto each other, then instead of  a simple superposition,
one gets a new fixed point which is in a different universality class. This result is 
far from obvious. Naively one would perhaps have guessed from Eq. (\ref{Sdual_AH}) that 
for $N=2$ one has two 
decoupled vortex modes, one neutral mode exhibiting a phase transition in the \xy universality 
class and one charged mode exhibiting a phase transition in the inverted \xy universality class. 
At $|\psi^{(1)}| = |\psi^{(2)}|$ a naive guess would be that one would have two such phase 
transitions superimposed on each other, giving $\alpha$ and $\nu$ values in the \xy universality 
class. However, there is a principal distinction from the case when $|\psi^{(1)}| \ne |\psi^{(2)}|$. 
In the latter case the upper phase transition is always a charged critical point
because the neutral mode is not developed. Thus at the upper
transition the interaction of vortices is of short range, while
at the lower transition there is a proliferation of vortices
with long range interaction. However, in the case $|\psi^{(1)}| = |\psi^{(2)}|$, 
then  below the single phase transition {\it both} types of vortices have neutral vorticity 
along with charged vorticity and  thus this phase transition
can not be mapped onto a superposition of a neutral and a charged fixed
points. 

Also, it is the fact that the system is self-dual at this point that invalidates
the naive superposition conjecture, since the \xy and inverted \xy 
phase transitions do not describe  phase transitions of a self-dual system. 
Even though the value of $\nu$ appears to be in good agreement with
the 3D Ising value, we observe that the 3D Ising model is not self-dual either, and 
the new type of critical point for $|\psi^{(1)}| = |\psi^{(2)}|$ can 
therefore not be in the 3D Ising universality class.
The origin of the novel exponents is therefore essentially topological,
showing that when the vortex loop blowouts of the neutral and 
charged modes are not well separated, they interact in a 
non-trivial fashion. There will therefore exist a crossover regime
parametrized by the field $|\psi^{(1)}|^2 - |\psi^{(2)}|^2$
where the exponents $\alpha$ and $\nu$ change from \xy
values to the new values we find here (see Fig. 7 of Ref. \onlinecite{motrunich2004}).  In principle, 
it is possible to compute the relevant crossover 
exponents in order to shed further light on this new self-dual 
universality class.

\section{\label{mc_simsN=3}Monte Carlo simulations, $N=3$}

In the model Eq. (\ref{vortex_action}) with $N=3$ vortex flavors we expect
in general one charged critical point associated with the condensation of the
charged vortex mode and two neutral critical points where neutral
vortex modes proliferate. To study the phases of this model we have 
performed MC simulations with the action given in Eq. (\ref{vortex_action}) 
with bare phase stiffnesses given in \tab{tab:n3_stiffnesses}. We have 
applied the same methods for calculating the critical exponents $\alpha$ 
and $\nu$ as well as gauge masses as we did for the $N=2$ case.

It is useful to  give the superfluid modes specifically  for the $N=3$ case (see 
Appendix \ref{App:charged_neutral} for details of the derivation for the general-$N$ 
case). Using Eq. (\ref{charge_neutral}), we have for this case
 \begin{widetext}
 \begin{eqnarray}
\label{gl4}
S & = &
\int d^3\ve r \left[\frac{1}{\Psi^2} \Biggl( \frac{|\psi^{(1)}|^2}{2} \nabla\theta^{(1)} +
\frac{|\psi^{(2)}|^2}{2} \nabla \theta^{(2)} + \frac{|\psi^{(3)}|^2}{2} \nabla \theta^{(3)}- 
e \Psi^2 {\bf A}\Biggr)^2 + V(\{\psi^{(\alpha)}\}) + \frac{1}{2}(\nabla\times\ve{A})^2 \right.\nonumber \\
&+&\left.\frac{|\psi^{(1)}|^2 |\psi^{(2)}|^2}{2\Psi^2}(\nabla(\theta^{(1)}-\theta^{(2)}))^2 
 + \frac{|\psi^{(1)}|^2 |\psi^{(3)}|^2}{2\Psi^2}(\nabla(\theta^{(1)}-\theta^{(3)}))^2 
 + \frac{|\psi^{(2)}|^2 |\psi^{(3)}|^2}{2\Psi^2}(\nabla(\theta^{(2)}-\theta^{(3)}))^2\right]. 
\label{varsep}
\end{eqnarray}
Here, we have defined $\Psi^2 = |\psi^{(1)}|^2 + |\psi^{(2)}|^2+|\psi^{(3)}|^2 $.
In the regime of short penetration length, the combination of phase gradients which
is  coupled to the gauge field $\ve A$ can be gauged away at length scales of the
order of the penetration length $\lambda = 1/e \Psi$. 
The remaining gradient terms for the neutral  modes are given by
\begin{eqnarray}
S_{\rm n}&=&  
\int d^3\ve r\left[\frac{|\psi^{(1)}|^2  |\psi^{(2)}|^2}
{2 \Psi^2}(\nabla (\theta^{(1)}-\theta^{(2)}))^2\right.  
+\frac{|\psi^{(1)}|^2 |\psi^{(3)}|^2}
{2 \Psi^2}(\nabla (\theta^{(1)}-\theta^{(3)}))^2 
+\left.\frac{|\psi^{(2)}|^2 |\psi^{(3)}|^2 }
{2 \Psi^2}(\nabla (\theta^{(2)}-\theta^{(3)}))^2\right].
\label{neutral}
\end{eqnarray}
This action could be inferred also directly from Eq. (\ref{vortex_action}).
For the case $N=3$, we write the action in the vortex representation as 

\begin{eqnarray}
S_V & = & \sum_{\ve q} \sum_{\eta=1}^3 \sum_{\alpha=1}^3  2 \pi^2 \beta |\psi^{(\alpha)}|^2  \ve m^{(\alpha)}_{\ve q}
 \left( \frac{\lambda^{(\eta)} }{|\ve Q_{\ve q}|^2 + m_0^2}
+ \frac{\delta_{\alpha,\eta}-\lambda^{(\eta)}}{|\ve Q_{\ve q}|^2}\right)  \ve m^{(\eta)}_{\ve q} 
\end{eqnarray}
which when written out takes the form
\begin{equation}
\begin{split}
\frac{S_V}{2 \pi^2 \beta/\Psi^2} = &  \sum_{\ve q} \left\{ 
\frac{
(\sum_{\alpha} |\psi^{(\alpha)}|^2  \ve m^{(\alpha)}_{\ve q} ) \cdot
(\sum_{\eta}  |\psi^{(\eta)} |^2  \ve m^{(\eta)}_{-\ve q} )
}
{|\ve Q_{\ve q}|^2 + m_0^2} 
 + 
\frac{
|\psi^{(1)}|^2  |\psi^{(2)} |^2 (\ve m^{(1)}_{\ve q} - \ve m^{(2)}_{\ve q})
\cdot (\ve m^{(1)}_{-\ve q} - \ve m^{(2)}_{-\ve q})
}
{|\ve Q_{\ve q}|^2}  \right. \\
 + &    \left.
\frac{
 |\psi^{(1)}|^2  |\psi^{(3)} |^2 (\ve m^{(1)}_{\ve q} - \ve m^{(3)}_{\ve q})
\cdot (\ve m^{(1)}_{-\ve q} - \ve m^{(3)}_{-\ve q})
}
{|\ve Q_{\ve q}|^2} 
 + 
\frac{
|\psi^{(2)}|^2  |\psi^{(3)} |^2 (\ve m^{(2)}_{\ve q} - \ve m^{(3)}_{\ve q})
\cdot (\ve m^{(2)}_{-\ve q} - \ve m^{(3)}_{-\ve q})
}
{|\ve Q_{\ve q}|^2} \right\} .
\end{split}
\label{neutral_vortex}
\end{equation}  
\end{widetext}
The three last terms in Eq. (\ref{neutral_vortex}) are nothing but the vortex representation 
of Eq. (\ref{neutral}). Notice also how all cross-terms between different vortex species
cancel out for arbitrary bare phase stiffnesses when $m_0^2 =0$. 

Thus, for  the case $N=3$, we have three phase variables yielding three neutral gauge 
invariant combinations  of phase differences. This amounts to two true neutral modes, 
the remaining degree of freedom is associated with the composite charged mode, which 
absorbs $\ve A$ and yields a massive vector field  via the Higgs mechanism. If all three 
bare phase stiffnesses $|\psi^{(1)}|$, $|\psi^{(2)}|$, and $|\psi^{(3)}|$ are different, 
this yields one charged inverse \xy critical point where the Meissner effect sets in, and 
two neutral \xy critical points at lower temperatures,  all separate. Consider now 
$|\psi^{(1)}| = |\psi^{(2)}| < |\psi^{(3)}|$.  
Then the charged mode proliferates at the highest critical temperature where the Meissner-effect 
sets in, and the two neutral modes proliferate simultaneously at a lower temperature.
The highest transition is still an inverted \xy transition, the lower one is a neutral 
\xy critical point. {\it Note how this is dramatically different from the case $N=2$, when 
the original neutral \xy critical point was collapsed on top of the inverted \xy critical 
point, resulting in a new universality class of the phase transition, essentially due 
to the self-duality of the $N=2$ system. It is also evident that collapsing a neutral 
and a charged fixed point is quite different from collapsing two neutral fixed points.} 

For the case $|\psi^{(1)}| = |\psi^{(2)}| < |\psi^{(3)}|$, in terms of the masses 
of $\ve A$ and the two  dual gauge fields associated with the neutral modes, $m_{\ve A}$ 
is non-zero below the {\it upper} critical temperature, while the two dual gauge fields become 
massive above the {\it lower} critical temperature.
In this case, the degenerate lower critical point is therefore
a \xy critical point, while the upper critical point is an inverted \xy critical point. 

A further interesting possibility is to set $|\psi^{(1)}| <
|\psi^{(2)}| = |\psi^{(3)}|$.
Consider the masses of $\ve A$ and the two dual gauge fields associated with the neutral mode in this 
case. At the lower critical temperature, one neutral vortex mode proliferates
in a \xy transition, generating a mass to the dual gauge field (thus breaking
one dual gauge symmetry). This mode is therefore dual-higgsed out of the problem at
higher temperatures. The gauge field $\ve A$ becomes massive {\it below} the upper critical 
temperature, while the dual gauge field associated with the remaining neutral mode
becomes massive {\it above} the same upper critical temperature. Hence, the
situation at the upper critical point corresponds precisely to the case 
$N=2$, $|\psi^{(1)}| = |\psi^{(2)}|$, for which we have already seen that a 
non-\xy critical point emerges.  When all bare stiffnesses are equal, 
$|\psi^{(1)}| = |\psi^{(2)}| = |\psi^{(3)}|$, all three fixed point collapse. 
We present MC simulations for the three cases given in
\tab{tab:n3_stiffnesses}, of which the case  $|\psi^{(1)}| <
|\psi^{(2)}| < |\psi^{(3)}|$ is the most pertinent 
to mixtures of superconducting condensates of for instance hydrogen and deuterium, 
or hydrogen and tritium.

\begin{table}[h]
\label{Sweeps}
\caption{\label{tab:n3_stiffnesses}Phase stiffnesses
  $|\psi^{(\alpha)}|$ and bare masses $m_0^2$ for the $N=3$ MC
  simulations. In all cases the charge $e=1/2$.}
\begin{tabular}{ccccc}
\hline
Case & $|\psi^{(1)}|^2$ & $|\psi^{(2)}|^2$ & $|\psi^{(3)}|^2$ & $m_0^2$  \\
\hline
1    &   1/3              &       2/3        &     4/3          &  7/12  \\ 
2    &   1/2              &       1/2        &     4/3          &  7/12  \\ 
3    &   7/9              &       7/9        &     7/9          &  7/12  \\ 
\hline
\end{tabular} 
\end{table}

\subsection{Critical exponents $\alpha$ and $\nu$, $|\psi^{(1)}| <
  |\psi^{(2)}| < |\psi^{(3)}|$}

MC simulations are performed for a $N=3$ system with bare phase
stiffnesses $|\psi^{(1)}|^2 =1/3$, $|\psi^{(2)}|^2=2/3$,
$|\psi^{(3)}|^2=4/3$ and system sizes $L=4,6,8,10,12,14,16$. 
We sample the second moment of the action Eq. (\ref{vortex_action}) and
find three anomalies for temperatures $T_{\rm c1}$, $T_{\rm c2}$, and
$T_{\rm c3}$, which from FSS are found to be  $T_{\rm c1}=0.98$,
$T_{\rm c2}=1.92$, and $T_{\rm c3}=3.63$.  

\begin{figure}[htb]
\centerline{\scalebox{0.6}{\rotatebox{-90.0}{\includegraphics{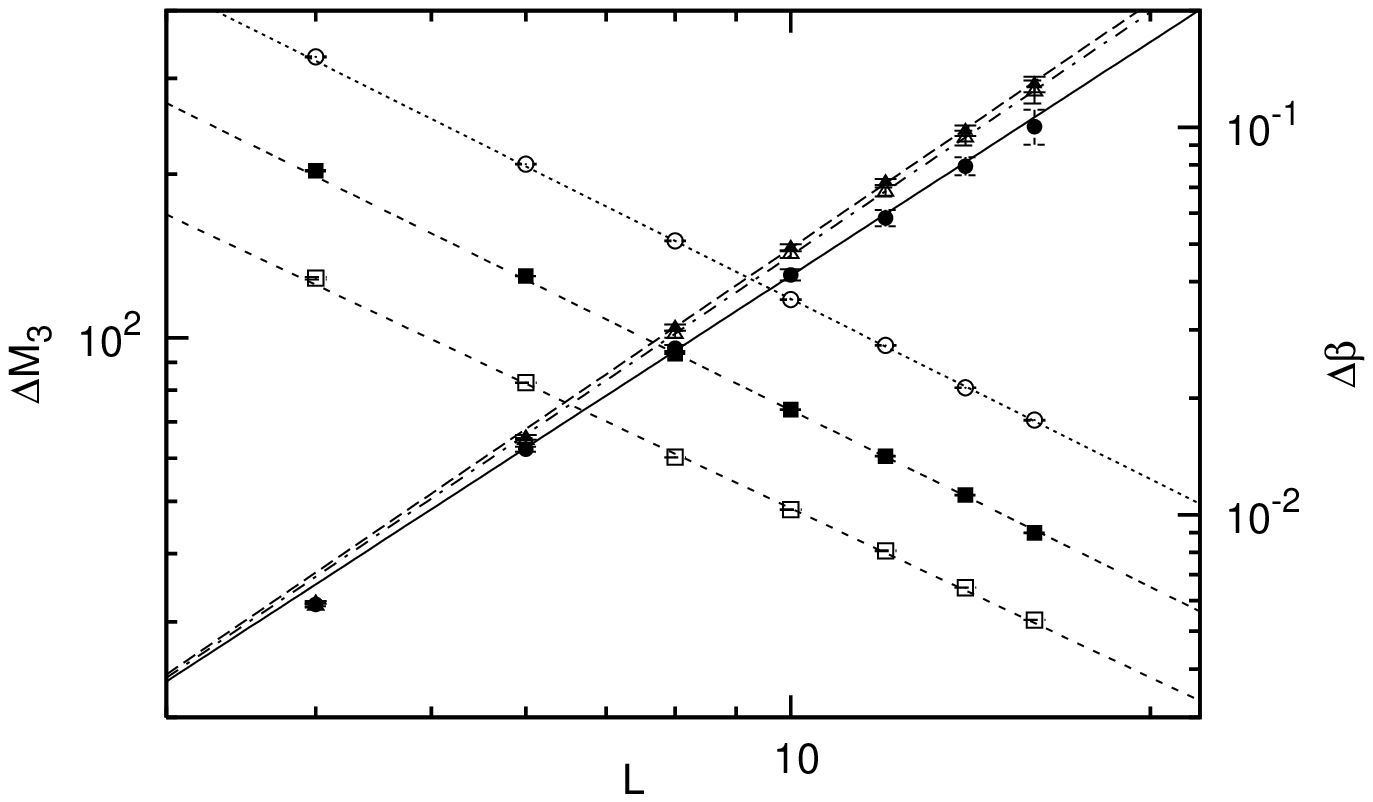}}}} 
\caption{\label{m3_fss_n3_1} FSS of the peak to peak value of the third moment of action 
 $\Delta M_3$ for $N=3$ with $|\psi^{(1)}|^2 =1/3$, $|\psi^{(2)}|^2=2/3$, 
$|\psi^{(3)}|^2=4/3$ labeled ($\blacktriangle$), ($\vartriangle$), and ($\bullet$), 
 for $T_{\rm c1}$, $T_{\rm c2}$ and $T_{\rm c3}$, respectively. The scaling of the 
 width between the peaks $\Delta\beta$ ($\circ$), ($\blacksquare$), and labeled 
 ($\Box$), for $T_{\rm c1}$, $T_{\rm c2}$, and $T_{\rm c3}$ respectively. The lines 
 are power law fits to the data for $L>6$ used to extract $\alpha$ and $\nu$.
}
\end{figure}

From a FSS analysis of the third moment of the action, 
we have measured the critical exponents $\alpha$ 
and $\nu$. The FSS plots are given in \fig{m3_fss_n3_1}. We find $\alpha =
-0.03 \pm 0.02$ and $\nu = 0.65 \pm 0.02$ for $T_{\rm c1}$, $\alpha =
-0.02 \pm 0.02$ and $\nu = 0.66 \pm 0.01$ for $T_{\rm c2}$, and $\alpha =
-0.01 \pm 0.03$ and $\nu = 0.69 \pm 0.02$ for $T_{\rm c3}$. These values
are consistent with the values for the \xy and the inverted \xy universality classes.

\subsection{Vortex correlator, Higgs mass, and anomalous scaling
 dimension, $|\psi^{(1)}| < |\psi^{(2)}| < |\psi^{(3)}|$}
In the Higgs phase, we expect the gauge field correlator
$\mathcal{G}_{\ve A}(\ve q)$ in Eq. (\ref{A_gaugeprop}) 
to scale according to the Ansatz Eq. (\ref{prop_ansatz}). For each coupling 
we fit $\mathcal{G}_{\ve  A}(\ve q)^{-1}$ from the MC simulations 
for system sizes $L=8,12,20$ and estimate the gauge field mass $m_{\ve A}$. 

\begin{figure}[htb]
\centerline{\scalebox{0.60}{\rotatebox{-90.0}{\includegraphics{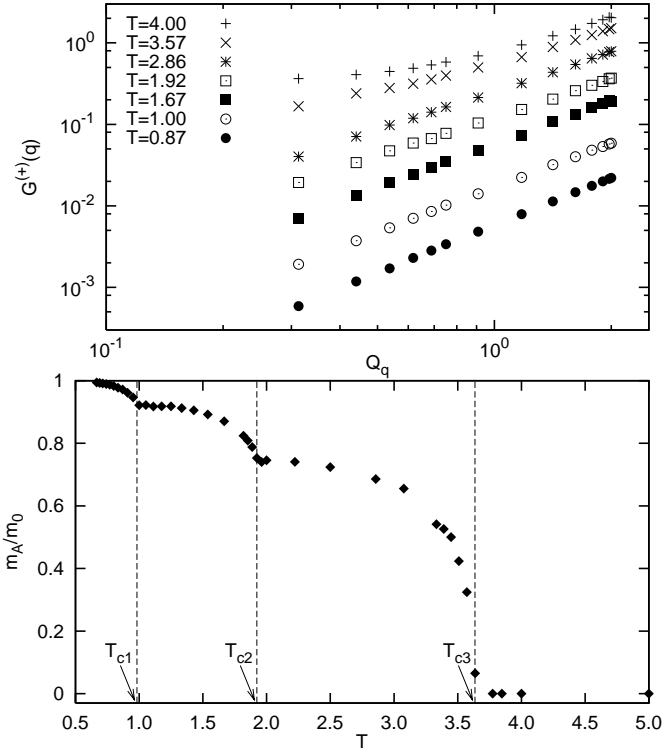}}}} 
\caption{\label{Gq_m3_1} Results for the vortex correlator Eq. (\ref{Gpluss_corr}), 
and the Higgs mass Eq. (\ref{gauge_field_masses}) for the case $|\psi^{(1)}|^2 =1/3 ,|\psi^{(2)}|^2=2/3, 
|\psi^{(3)}|^2=4/3$. The upper panel shows 
$G^{(+)}({\ve q})$ as a function of $|\ve Q_{\ve q}|$ for seven temperatures starting from 
above: Above and close to $T_{\rm {c3}}$,  above and close to $T_{\rm {c2}}$, 
above and close to $T_{\rm {c1}}$, and below $T_{\rm c1}$. Above $T_{\rm{c3}}$, the vortices are seen to 
have condensed, $G^{(+)}({\ve q}) \sim \rm{const}$ while for all temperatures 
below $T_{\rm{c3}}$, including above and below $T_{\rm {c1}}$ and 
$T_{\rm {c2}}$, $G^{(+)}({\ve q}) \sim q^2$ for small $\ve q$. The lower panel 
shows the Higgs mass as a function of temperature, showing the onset of Meissner 
effect at $T_{\rm{c3}}$, and the additional anomalies at $T_{\rm {c1}}$ and 
$T_{\rm {c2}}$ due to the appearance of additional neutral
modes at these temperatures.
}
\end{figure}

The results for the vortex correlator $G^{(+)}({\ve q})$ in Eq. (\ref{Gpluss_corr}) and 
the Higgs mass Eq. (\ref{gauge_field_masses}) are given in \fig{Gq_m3_1}. Note how the 
$\ve q$-dependence of $G^{(+)}({\ve q})$ changes when the temperature is 
varied from above to below $T_{\rm{c3}}$ from $G^{(+)}({\ve q}) \sim \rm{const}$ 
to $G^{(+)}({\ve q}) \sim q^2$, respectively. Note also how the $\ve q$-behavior
of the vortex correlator remains unchanged when the temperature is varied through 
$T_{\rm{c2}}$ and $T_{\rm{c1}}$, i.e. it remains $G^{(+)}({\ve q}) \sim q^2$.
This reflects the fact that the field $\ve A$ has been higgsed out of the problem
at $T_{\rm{c3}}$ such that the vortex tangle is incompressible below this 
temperature. From Eq. (\ref{gauge_field_masses}) it is therefore clear that a Higgs 
mass is generated at $T_{\rm{c3}}$ by the establishing of a charged superconducting
mode. Moreover, when the two additional neutral superfluid modes are established 
at $T_{\rm{c2}}$ and $T_{\rm{c1}}$, this adds to the total superfluid density and 
hence leads to kinks in the London penetration length and thereby $m_{\ve A}$.

Precisely at $T_{\rm{c3}}$, $m_{\ve A}$ vanishes, and the scaling
Ansatz given by \eq{n1_corr}
may be used to extract $\eta_{\ve A}$. From \fig{Gq_m3_1} and
$G^{(+)}(\ve q)$ at $T_{\rm{c3}}$, we extract $\eta_{\ve A}=1$, from
which we  
conclude that the critical point at $T_{\rm{c3}}$ is an inverted \xy critical point. 
Likewise, from the $G^{(+)}(\ve q) \sim q^2$ behavior at $T_{\rm{c1}}$ and $T_{\rm{c2}}$ 
we conclude that these two critical points feature $\eta_{\ve A}=0$ and hence represent 
\xy critical points.

\subsection{Critical exponents $\alpha$ and $\nu$, $|\psi^{(1)}| =
  |\psi^{(2)}| < |\psi^{(3)}|$}

\begin{figure}[htb]
\centerline{\scalebox{0.6}{\rotatebox{-90.0}{\includegraphics{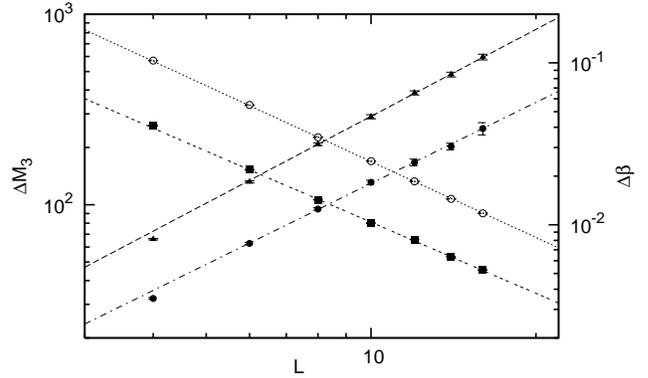}}}} 
\caption{\label{m3_fss_n3_2} FSS of the peak to peak value of the third moment of action 
 $\Delta M_3$ for $N=3$ for $|\psi^{(1)}|^2 = |\psi^{(2)}|^2 =
 1/2$ and $|\psi^{(3)}|^2 = 4/3$, labeled ($\blacktriangle$), and ($\bullet$), 
 for $T_{\rm c1}$ and $T_{\rm c2}$, respectively. The scaling of the 
 width between the peaks $\Delta\beta$ labeled ($\circ$),
 ($\blacksquare$), for $T_{\rm c1}$ and $T_{\rm c2}$ respectively. The lines 
 are power law fits to the data for $L>6$ used to extract $\alpha$ and $\nu$.}
\end{figure}

MC simulations have been performed for a $N=3$ system with bare phase
stiffnesses $|\psi^{(1)}|^2 = |\psi^{(2)}|^2 = 1/2$ and
$|\psi^{(3)}|^2=4/3$ and system sizes $L=4,6,8,10,12,14,16$.
By measuring the second moment of the action Eq. (\ref{vortex_action})
we find two anomalies for the temperatures $T_{\rm c1}$ and $T_{\rm
  c2}$, which from FSS are found to be  $T_{\rm c1}=1.46$ and $T_{\rm c2}=3.63$. 
From a FSS analysis of the third moment of the action 
we have measured the critical exponents $\alpha$ 
and $\nu$. The FSS plots are given in \fig{m3_fss_n3_2}. We find $\alpha =
-0.03 \pm 0.02$ and $\nu = 0.65 \pm 0.02$ for $T_{\rm c1}$, and $\alpha =
-0.03 \pm 0.03$ and $\nu = 0.68 \pm 0.02$ for $T_{\rm c2}$. These values
are consistent with the values for the \xy and the inverted \xy universality classes.

\subsection{Vortex correlator, Higgs mass, and anomalous scaling
 dimension, $|\psi^{(1)}| = |\psi^{(2)}| < |\psi^{(3)}|$}

Like the previous case, we extract the gauge field mass by fitting the
gauge field correlators for small $\ve q$ to the Ansatz Eq. (\ref{prop_ansatz}) for
system sizes $L=8,12,20$. 

\begin{figure}[htb]
\centerline{\scalebox{0.60}{\rotatebox{-90.0}{\includegraphics{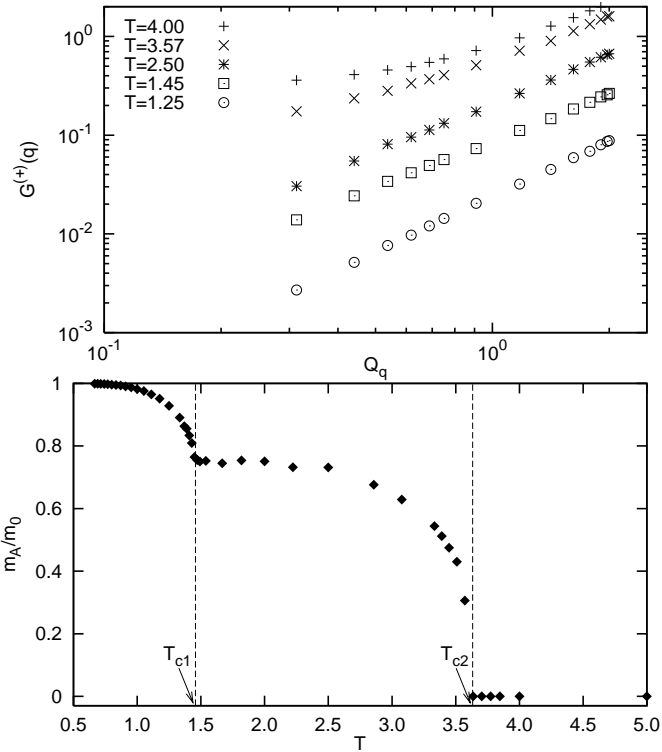}}}} 
\caption{\label{Gq_m3_2} Results for the vortex correlator Eq. (\ref{Gpluss_corr}), 
and the Higgs mass Eq. (\ref{gauge_field_masses}) for the case
$|\psi^{(1)}|^2 = |\psi^{(2)}|^2 = 1/2$ and $|\psi^{(3)}|^2 = 4/3$. The upper panel shows 
$G^{(+)}({\ve q})$ as a function of $|\ve Q_{\ve q}|$ for five temperatures starting from 
above: Above and close to $T_{\rm {c2}}$, 
above and close to $T_{\rm {c1}}$, and below $T_{\rm c1}$. Above $T_{\rm{c2}}$, the vortices are seen to 
have condensed, $G^{(+)}({\ve q}) \sim \rm{const}$ while close to
$T_{\rm{c2}}$, $G^{(+)}({\ve q}) \sim |\ve q|$. For all temperatures 
below $T_{\rm{c2}}$, including above and below $T_{\rm {c1}}$, 
$G^{(+)}({\ve q}) \sim q^2$ for small $\ve q$. The lower panel 
shows the Higgs mass as a function of temperature, showing the onset of Meissner 
effect at $T_{\rm{c2}}$, and an additional anomaly at $T_{\rm {c1}}$ due to the appearance of additional neutral
modes at this temperature. }
\end{figure}

The results for the vortex correlator $G^{(+)}({\ve q})$ in Eq. (\ref{Gpluss_corr}) and the Higgs mass 
defined in \eq{gauge_field_masses}  are
given in \fig{Gq_m3_2}. Note how the $\ve q$-dependence of
$G^{(+)}({\ve q})$ changes when the temperature is 
varied from above to below $T_{\rm{c2}}=3.63$ from $G^{(+)}({\ve q}) \sim \rm{const}$ 
to $G^{(+)}({\ve q}) \sim q^2$, respectively. Note also how the $\ve q$-behavior
of the vortex correlator remains unchanged when the temperature is varied through 
$T_{\rm{c1}}=1.46$, i.e. it remains $G^{(+)}({\ve q}) \sim q^2$.
This reflects the fact that the field $\ve A$ has been higgsed out of the problem
at $T_{\rm{c2}}=3.63$ such that the vortex tangle is incompressible below this 
temperature. From Eq. (\ref{gauge_field_masses}) it is therefore clear that a Higgs 
mass is generated at $T_{\rm{c2}}$ by the establishing of a charged superconducting
mode. Moreover, when the two additional neutral superfluid modes are established 
at $T_{\rm{c2}}$ this adds to the total superfluid density and 
hence leads to a kink in the London penetration length and thereby $m_{\ve A}$.

Precisely at the charged transition $T_{\rm{c2}}$, $m_{\ve A}$ vanishes
and we find the gauge field correlator has the form $G^{(+)}(\ve q)
\sim |\ve q|^{2-\eta_{\ve A}}$. From the $G^{(+)}(\ve
q)$ data in \fig{Gq_m3_2} we extract $\eta_{\ve A}=1$, from which we 
conclude that the critical point at $T_{\rm{c2}}$ is an inverted \xy critical point. 
Likewise, from the behavior of $G^{(+)}(\ve q) \sim q^2$ at $T_{\rm{c1}}$
conclude that this critical point features $\eta_{\ve A}=0$ and hence represents 
a \xy critical point.

\subsection{Critical exponents $\alpha$ and $\nu$, $|\psi^{(1)}| =
  |\psi^{(2)}| = |\psi^{(3)}|$}

\begin{figure}[htb]
\centerline{\scalebox{0.6}{\rotatebox{-90.0}{\includegraphics{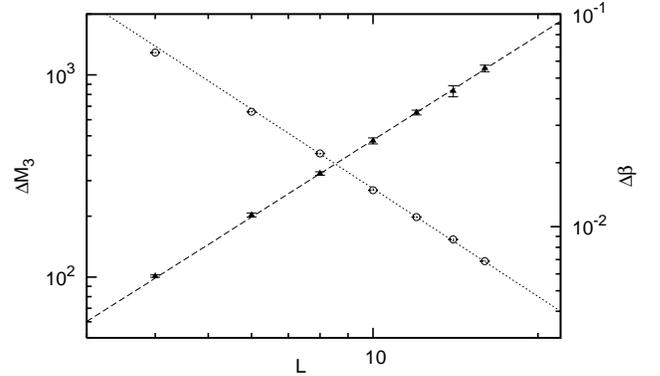}}}} 
\caption{\label{m3_fss_n3_3} FSS of the peak to peak value of the third moment of action 
 $\Delta M_3$ for $N=3$ for $|\psi^{(1)}|^2 = |\psi^{(2)}|^2 =
 |\psi^{(3)}|^2 = 7/9$ labeled ($\blacktriangle$). The scaling of the  
 width between the peaks $\Delta\beta$ labeled ($\circ$). The lines 
 are power law fits to the data for $L>6$ used to extract $\alpha$ and $\nu$.}
\end{figure}

MC simulations are performed for a $N=3$ system with equal bare phase
stiffnesses $|\psi^{(1)}|^2 = |\psi^{(2)}|^2 = |\psi^{(3)}|^2 = 7/9$
and system sizes $L=4,6,8,10,12,14,16$.
From measurements of the second moment of the action Eq. (\ref{vortex_action})
we find one anomaly for temperature the $T_{\rm c}$, which from FSS is  
found to be  $T_{\rm c}=2.19$. 
From a FSS analysis of the third moment of the action we have measured
the critical exponents $\alpha$  
and $\nu$. The FSS plots are given in \fig{m3_fss_n3_3}. We find $\alpha =
0.02 \pm 0.03$ and $\nu = 0.59 \pm 0.02$. The values appear not to
agree with hyper scaling. They are \textit{not} consistent with the 
\xy universality class. 

The above values for $\alpha$ and $\nu$  are however in agreement with 
those found for the case $N=2$, $|\psi^{(1)}| = |\psi^{(2)}|$. We 
observe, based on the numerical results for the two cases
$N=2$, $|\psi^{(1)}| = |\psi^{(2)}|$ and
$N=3$, $|\psi^{(1)}| = |\psi^{(2)}|=|\psi^{(3)}|$ compared to
the other cases that we have considered, that collapsing two neutral
critical points in the \xy universality class leads to a single
critical point also in the \xy universality class. On the other
hand, it appears that collapsing
$N-1$ neutral critical points in the \xy universality class
{\it and one charged fixed point in the inverted} \xy universality class
leads to an $N$-fold degenerate single critical point in a universality class 
(which in principle depends on $N$) which is not that of the \xy or inverted 
\xy type. For $N=2$, we may define the universality class as that of a 3D 
self-dual $U(1)\times U(1)$ gauge theory, while it is less clear what it is for other 
$N \geq 3$.

\subsection{Vortex correlator, Higgs mass, and anomalous scaling
 dimension, $|\psi^{(1)}| = |\psi^{(2)}| = |\psi^{(3)}|$}

We extract the gauge field mass $m_{\ve A}$ by fitting the 
gauge field correlators for small $\ve q$ to the Ansatz Eq. (\ref{prop_ansatz}) for
system sizes $L=8,12,20,32$. 

\begin{figure}[htb]
\centerline{\scalebox{0.60}{\rotatebox{-90.0}{\includegraphics{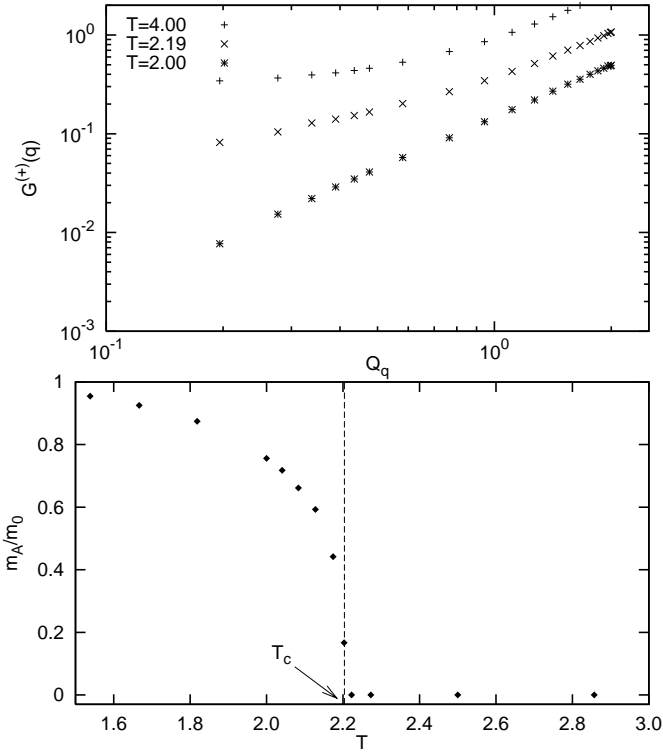}}}} 
\caption{\label{Gq_m3_3} Results for the vortex correlator Eq. (\ref{Gpluss_corr}), 
and the Higgs mass Eq. (\ref{gauge_field_masses}) for the case
$|\psi^{(1)}|^2 = |\psi^{(2)}|^2 = |\psi^{(3)}|^2 = 7/9$. The upper panel shows 
$G^{(+)}({\ve q})$ as a function of $|\ve Q_{\ve q}|$ for temperatures
above and close to $T_{\rm {c}}$, and below $T_{\rm c}$. Above
$T_{\rm{c}}$, the vortices have condensed, $G^{(+)}({\ve q}) \sim
\rm{const}$. 
Below $T_{\rm{c}}$, $G^{(+)}({\ve q}) \sim q^2$ for small $\ve q$. The lower panel 
shows the Higgs mass as a function of temperature, showing the onset of Meissner 
effect at $T_{\rm{c}}$.}
\end{figure}

The results for the vortex correlator $G^{(+)}({\ve q})$ in Eq. (\ref{Gpluss_corr}) and the Higgs mass 
defined in Eq. (\ref{gauge_field_masses}) are given in \fig{Gq_m3_3}. 
Note how the $\ve q$-dependence of $G^{(+)}({\ve q})$ changes when the temperature is 
varied from above to below $T_{\rm{c}}=2.20$ from $G^{(+)}({\ve q}) \sim \rm{const}$ 
to $G^{(+)}({\ve q}) \sim q^2$, respectively.  From
Eq. (\ref{gauge_field_masses}) it is therefore clear that a Higgs  
mass is generated at $T_{\rm{c}}=2.19$ by the establishing of a charged superconducting
mode. From $G^{(+)}({\ve q})$ measurements at $T_c$ we find the
anomalous scaling dimension to be $\eta_{\ve A}=1$.

\subsection{General $N$}

The critical properties of the $N$-component system are governed solely by excitations of 
vortex loops with fractional flux. That is, in the $N=2$ case, $T_{\rm c1}$ is governed by 
proliferation of the vortex loops with phase windings ($\Delta \theta^{(1)}=2\pi, \Delta
\theta^{(2)}=0$), while  $T_{\rm c2}$ marks the onset of proliferation of the loops of  
vortices with windings ($\Delta \theta^{(1)}=0, \Delta \theta^{(2)}=2\pi$). Remarkably, 
for general $N$, below the temperature $T_{{\rm c}N-1}$, where $T_{\rm
  c1} < \cdots < T_{{\rm c}N-1}< T_{{\rm c}N}$, topological excitations 
with nontrivial windings only in one phase has a logarithmically divergent energy
\cite{frac,smiseth2004}. Moreover,  the composite vortex loops 
($\Delta \theta^{(1)}=2\pi, \Delta \theta^{(2)}=2 \pi$) which in contrast have finite 
energy per unit length, do not play a role as far as  critical properties  are concerned. 

For the
case $N=2$, the 
critical point at $T_{\rm c2}>T_{\rm c1}$ is a charged fixed point. Proliferation of the vortex loops 
($\Delta \theta^{(1)}=2\pi, \Delta \theta^{(2)}=0$) at $T_{\rm c1}$  eliminates the neutral mode. 
On the other hand, the composite vortices ($\Delta \theta^{(1)}=2\pi, \Delta \theta^{(2)}=2\pi$) 
do not feature neutral vorticity at any temperature and thus can be mapped onto vortices in a 
$N=1$ superconductor with bare phase stiffness $|\psi^{(1)}|^2+|\psi^{(2)}|^2$. 
A characteristic temperature of proliferation of such vortex loops is higher than $T_{\rm c2}$, 
which excludes the composite vortices from  the sector of  critical fluctuations in the system.
The same argument applies to the $N>2$ case.

\begin{figure}[htb]
\centerline{\scalebox{0.8}{\rotatebox{0.0}{\includegraphics{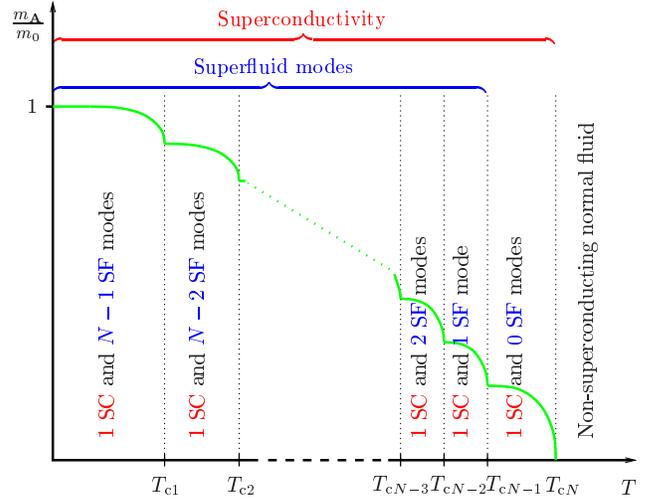}}}} 
\caption{  \label{transitions}  (Color online). Phase transitions in the $N$-flavor London superconductor 
with different bare stiffnesses of the $N$ order parameter components. The green line is 
the gauge field mass $m_{\ve A}$. At the highest temperature the system becomes superconducting 
via a phase transition in the inverted \xy universality class.  At the lower transitions 
the system develops composite neutral superfluid modes in the superconducting state via a 
series of $N-1$ phase transitions, all in the \xy universality class.} 
\end{figure}

Summarizing the previous two sections, the resulting schematic phase diagram of the 
$N$-flavor London superconductor in the absence of external field is presented in 
\fig{transitions}. Assuming the bare stiffnesses have been chosen to have 
well separated bare energy scales associated with the twist of phases of every flavor, 
we find $N$ distinct critical points. At the highest critical temperature, the charged 
vortex mode condenses and the gauge field acquires a mass, driving the system into a 
superconducting phase. For lower critical temperatures, neutral vortex loops condense 
and the system develops superfluid modes. Hence, in zero magnetic field there are 
$N-1$ superfluid modes arising in a superconducting state.


\section{\label{N=2_extfield} $N=2$ system in an external magnetic field, lattice and 
sub-lattice melting, and metallic superfluidity}

We next discuss the situation  when the system is subjected to an external magnetic field.
{ Two important aspects  of the physics to be described below, are
{\it{ i)}} three dimensionality and {\it{ii)}}  a significant difference in the bare 
stiffnesses of the condensates.}
As discussed recently  \cite{BSA,SSBS}, when an external magnetic field is applied to a 
three dimensional type-II 
$N$-component superconductor,  it  changes its properties much more dramatically than in the
ordinary  $N=1$ case.
The composite charged vortices have finite energy per unit length and
couple to the magnetic field, 
and hence are relevant for magnetic properties.
If the bare stiffnesses of the fields are different,
 the existence of composite purely charged vortices
results in a particularly rich phase diagram 
with several novel phases and phase transitions. 
Note that in the following two chapters we denote a constituent vortex originating in a 
$2\pi$ phase-winding in $ \theta^{(\alpha)}$ a \textit{type-$\alpha$
  vortex}, where  $\alpha\in [1,\dots,N]$.

\subsection{$N=2$ system in external field at $T=0$}
\begin{figure}[htb]
\centerline{\scalebox{0.09}{\rotatebox{0.0}{\includegraphics{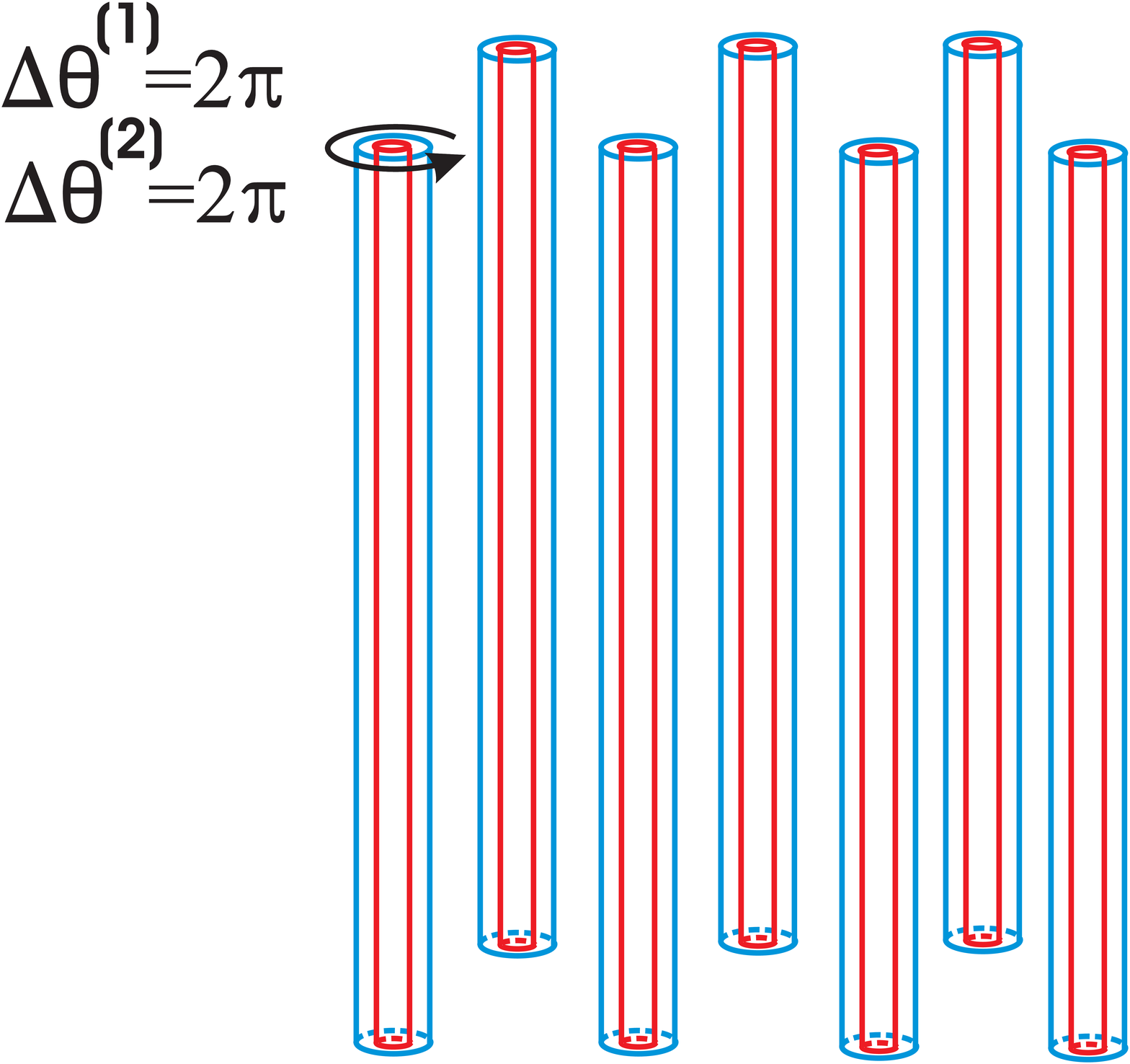}}}} 
\caption{ \label{lattice} (Color online). A type-II, $N=2$ system at zero temperature in external magnetic 
field forms a lattice of composite Abrikosov vortices. A composite
vortex may be viewed as co-centered type-1 (red) and type-2 (blue) vortices $(\Delta \theta^{(1)}=2\pi, \Delta \theta^{(2)}=2\pi)$.}
\end{figure}

In the presence of an external magnetic field, but in the absence of thermal 
fluctuations, the formation of an Abrikosov lattice of non-composite
vortices is forbidden because these defects have a logarithmically
divergent energy\cite{frac,smiseth2004}, cf. discussion following Eq. (\ref{potential_0}). 
In a type-II $N$-component system, the system forms a
lattice of  
{\it composite vortices} for which $\Delta \theta^{(\alpha)}=2\pi$ for every 
$\alpha\in [1,\dots,N]$. A schematic picture of the resulting 
lattice of composite vortices in an $N=2$ superconductor is  shown in Fig. \ref{lattice}. 
In the discussion below we consider the type-II limit, but not extreme type-II since 
the interaction between vortices of different species is depleted at the length scales 
smaller than the penetration length, cf. Eq. (\ref{potential}) and the discussion following
Eq. (\ref{potential_0}). We do not discuss effects of this depletion assuming 
a moderately short penetration length scale.

\subsection{Effects of low-temperature fluctuations on field-induced composite vortices}
In this subsection, we will consider the effects of thermal
fluctuations, and how it affects the Abrikosov vortex  
lattice of composite vortices defined above.

\subsubsection{Thermal generation of loop-like splitting of line vortices}
\begin{figure}[htb]
\centerline{\scalebox{0.09}{\rotatebox{0.0}{\includegraphics{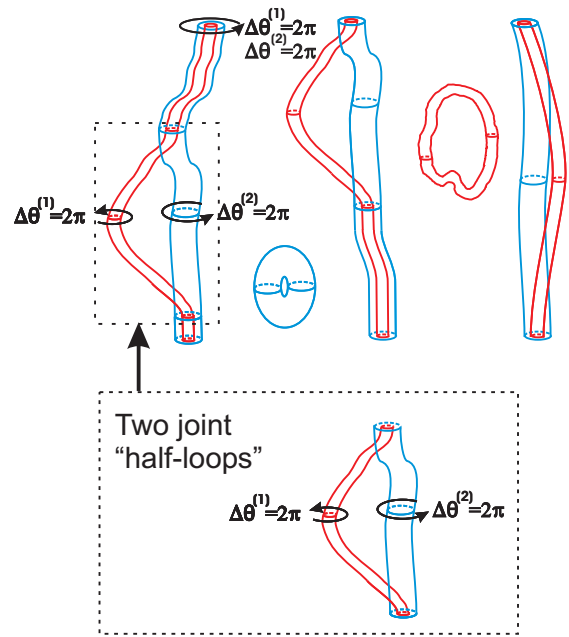}}}} 
\caption{\label{fluct1}  (Color online). Low-temperature fluctuations
in the $N=2$ system subjected to a magnetic field. Thermal fluctuations generate closed 
loops of composite fractional flux vortices and  {\it local} splitting of field-induced 
composite vortex lines. The type-1 vortices (red) are the vortices of the component with the 
lowest  bare phase stiffness.  When these vortices are viewed as world-lines of bosons, 
they constitute the ``lighter" of the vortex species. These ``light" vortices fluctuate 
more strongly than the ``heavier"  type-2 vortices (blue).} 
\end{figure}

At finite temperature, the $N$-component system subjected to a magnetic field $\ve B$ will 
exhibit thermal excitations in the form of vortex loops with fractional flux similar to the 
${\ve B}=0$ discussion in the first part of this paper. 
We observe that since the field-induced composite vortices are logarithmically bound
\cite{frac,smiseth2004}, thermal fluctuations will induce a {\it local splitting} of composite 
vortices {\it in a configuration of two half-loops connected to  a straight line} 
\cite{BSA,SSBS} as shown in Fig. \ref{fluct1}.
We observe that every branch of  a ``split loop" formed on a field-induced  vortex 
line features neutral as well as charged vorticity. The interaction between these two 
branches is mediated by a neutral vortex mode exclusively associated
with the phase 
difference $\gamma=\theta^{(1)}-\theta^{(2)}$. The \textit{screened charged mode} does not contribute  
to the interaction between the two branches.
This is implicit in
Eq. (\ref{potential}) as follows. The
vortex segments of different flavors do not interact at short distances
much smaller than $\lambda = m_0^{-1}$, where the charge, or $m_0^2$ appearing 
in the interaction matrix \eq{potential}, can be ignored. On such length scales, the 
screened part of the interaction matrix is essentially unscreened,
and is canceled by  the inter-flavor interaction, which is 
unscreened on all length scales. Hence, as also discussed in Section \ref{model}, 
the intra-vortex interaction is strongly reduced at length scales smaller than 
$\lambda$.
%
%

Moreover, in terms of the field $\gamma=\theta^{(1)}-\theta^{(2)}$, two split branches of a composite 
field-induced  vortex have opposite vorticities ($\Delta\gamma=2\pi$ on one branch and 
$\Delta\gamma=-2\pi$ on another branch). On the other hand, such a loop emits two integer flux 
vortices at its top and bottom, which do {\it not} feature neutral vorticity. So the process of 
such a thermal local splitting of a field-induced line may be mapped onto a thermally generated 
proliferation of  {\it closed} vortex loops in the artificial phase field $\gamma$ as those in 
the neutral $N=1$ model in absence of magnetic field \cite{tesanovic1999,nguyen1999}. Hence, somewhat 
counterintuitively such a splitting transition
should be in the  \xy universality class \cite{BSA,SSBS} .  This transition, being topological in its origin, should not be 
confused with the topological Kosterlitz-Thouless transition known to occur in planar systems.

\subsubsection{Melting}
Apart from  the splitting of composite vortices and generation of closed vortex loops, the thermal 
fluctuations will produce one more competing process. That is, the lattice of composite vortices
can be mapped onto an ordinary vortex lattice in a one-component superconductor. 
Sufficiently strong thermal fluctuations drive {\it a first-order melting transition} of the 
field-induced Abrikosov lattice \cite{Hetzel1992,Fossheim_Sudbo_book,nguyen1999}. A counterpart to 
this effect for the case $N=2$ when $|\psi^{(1)}| \ne |\psi^{(2)}|$ is much more complicated. 
We next consider this process in the regimes of low and high magnetic fields, separately.

\subsection{Sublattice melting in low magnetic fields}
\begin{figure}[htb]
\centerline{\scalebox{0.09}{\rotatebox{0.0}{\includegraphics{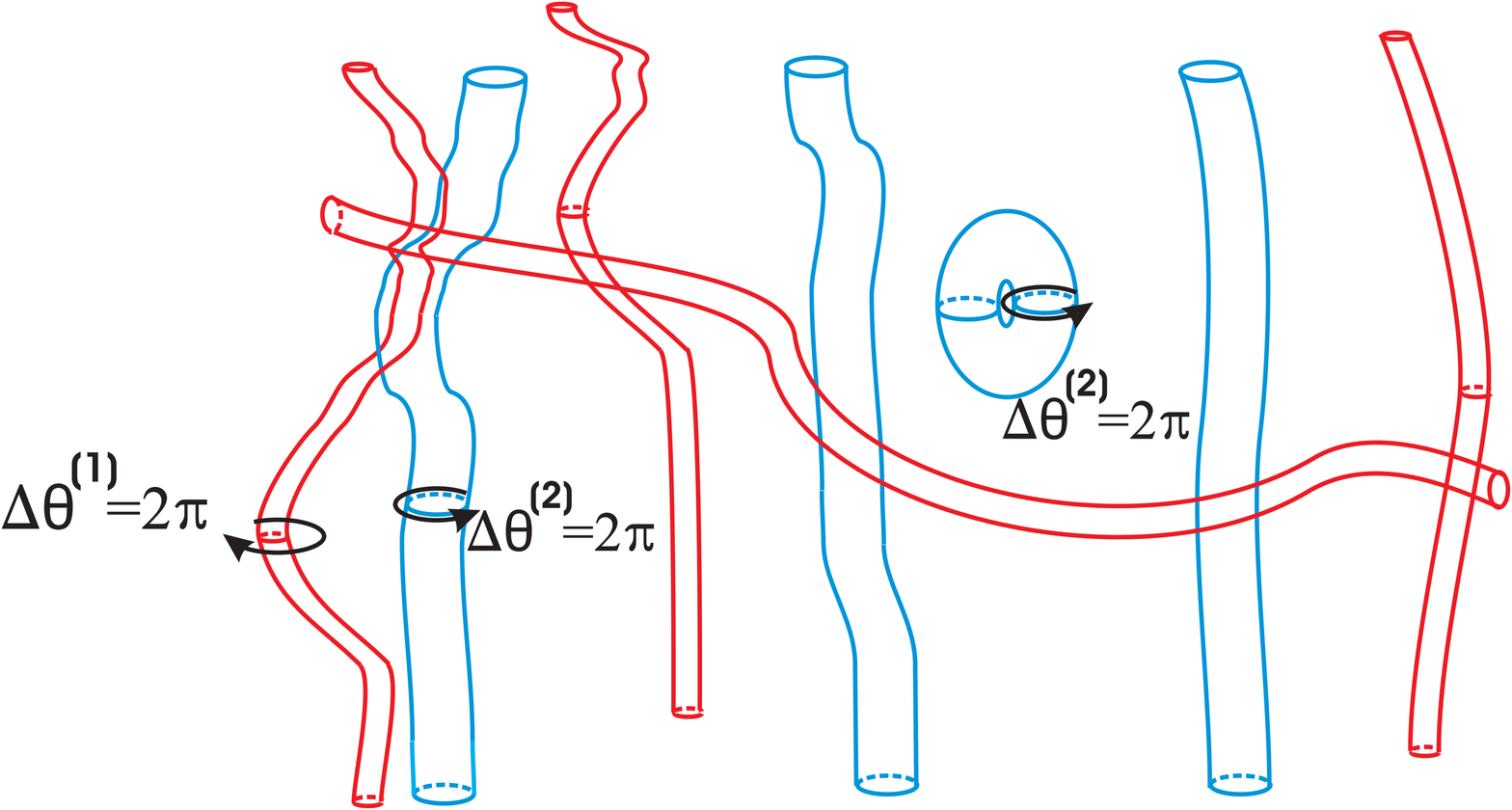}}}} 
\caption{ \label{sl} (Color online). A vortex liquid of type-1 vortices (red) immersed
  in a background of a type-2 vortex lattice (blue) in the $N=2$ system in the regime 
$|\psi^{(1)}| \ll |\psi^{(2)}|$. This is the type-1 vortex sublattice melting.
There is a temperature region in low magnetic 
field when ``light" vortices are decoupled and form a liquid. ``Light" vortex loops are 
proliferated, while ``heavy" vortices form a lattice immersed a liquid of ``heavy" vortex 
loops. Both heavy and light vortices carry only a fraction of magnetic flux quantum in 
this state.} 
\end{figure}

Consider the case of weak magnetic field (much smaller than the upper critical magnetic 
field for which superconductivity is essentially destroyed) for the situation where 
$|\psi^{(1)}| \ll |\psi^{(2)}|$. Introducing a characteristic temperature 
associated with a melting of the type-2 vortex
lattice in the absence of the condensate 
$\Psi_0^{(1)}$, then at sufficiently low magnetic field this melting temperature will be much 
higher than the characteristic temperature of thermal decomposition of a composite vortex line 
into two individual vortex lines. Thus, the first transition that would be encountered upon heating 
the system, is the thermal splitting of field-induced composite
vortices into separate type-1 and type-2 vortices.
This would be accompanied by a  proliferation of closed loops of type-1 vortices, {\it while the 
vortices of type-2 will remain arranged in a lattice}. We will denote this phase transition as 
{\it sublattice melting} \cite{BSA,SSBS}. The critical temperature of
this phase transition is denoted $T_{\rm SLM}$ (see \fig{fd}). A schematic picture of the sublattice vortex liquid 
is given in \fig{sl}. As discussed above, upon thermal decomposition of the composite  vortices, 
the emerging individual vortices can be mapped onto positively and negatively electrically 
charged strings which logarithmically interact with each other. 

{\it Quite remarkably, the Abrikosov lattice order for 
the component with the highest phase stiffness survives the decomposition 
transition}, for the following reason.  The dominant interaction between individual vortices 
is the long-ranged interaction mediated by neutral vorticity, cf. Eq. (\ref{potential}).
This permits a  mapping of such vortices onto positively and negatively charged strings. Upon thermal 
decomposition, the effective long-range Coulomb interaction
mediated by the neutral mode  is screened without affecting 
the charged modes.
Consider the case when $|\psi^{(1)}|\ll|\psi^{(2)}|$. 
Then the stiffness $|\psi^{(2)}|$ is large enough to keep
the type-2 vortices arranged in a lattice while the stiffness $|\psi^{(1)}|$  
is too weak to constrain type-1 vortices to the lattice. Thus, the
``light" type-1 vortex lines are in their molten phase. This is the physical origin of the sublattice melting process. 
The situation is illustrated in \fig{sl}. We emphasize that the existence of the regime 
of sublattice melting follows from the fact that the stiffness of the
neutral mode, which keeps composite 
vortices bound at low temperatures, is always smaller than the smallest stiffness of the individual 
condensates, namely
\begin{eqnarray}
J_{\rm neutral}=\frac{|\psi^{(1)}|^2|\psi^{(2)}|^2}{|\psi^{(1)}|^2+|\psi^{(2)}|^2} < |\psi^{(1)}|^2.
\end{eqnarray}

\subsection{Composite vortex lattice melting in strong magnetic fields}

\begin{figure}[htb]
\centerline{\scalebox{0.09}{\rotatebox{0.0}{\includegraphics{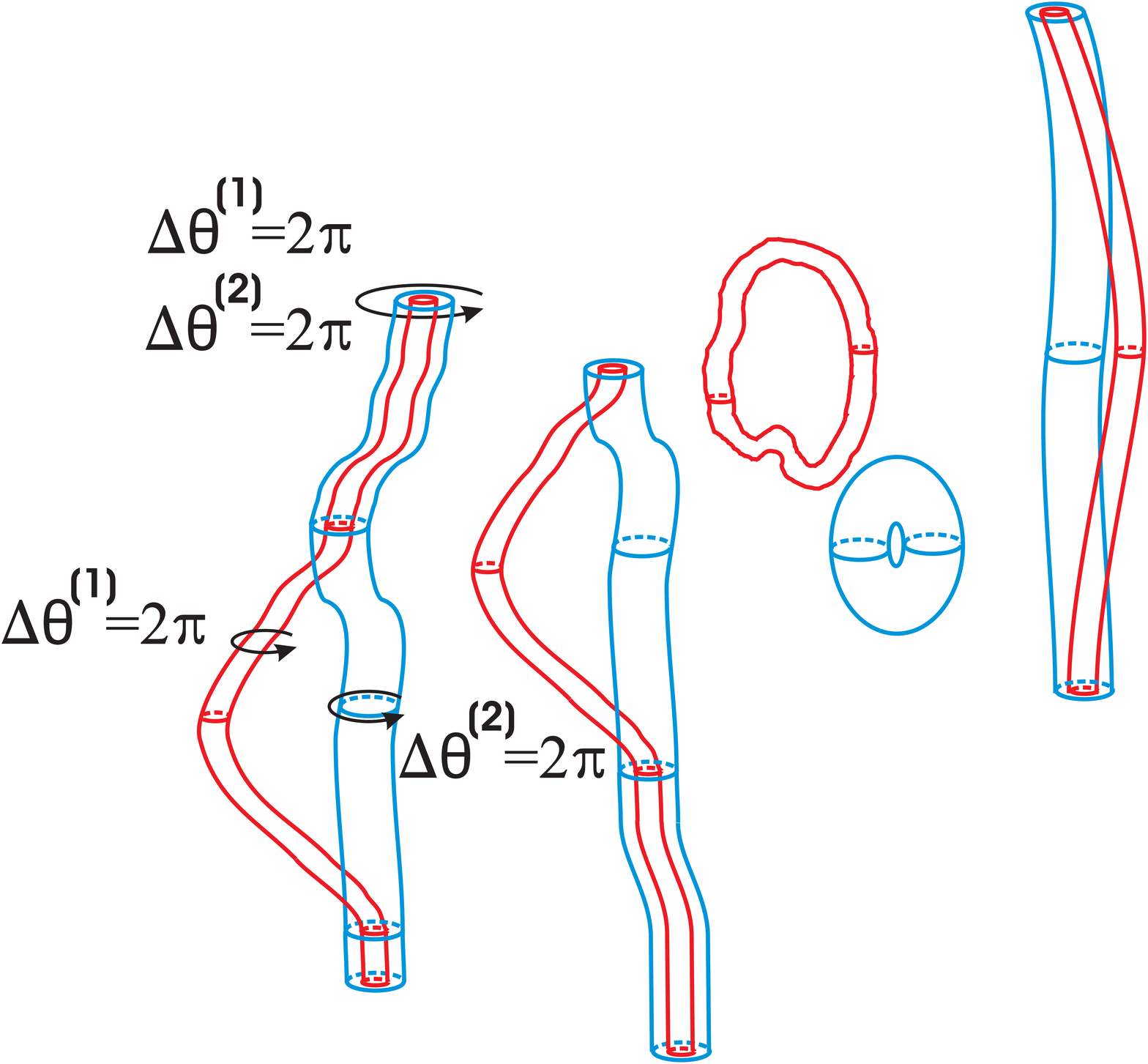}}}} 
\caption{\label{liquid} (Color online). Liquid of composite vortices in the $N=2$ model
immersed in a liquid of non-proliferated vortex loops. 
It is realized for $|\psi^{(1)}| \ll |\psi^{(2)}|$ in strong magnetic fields.} 
\end{figure}

It is known  from the $N=1$ system that an increase in magnetic field suppresses the melting
temperature of the vortex lattice\cite{Fossheim_Sudbo_book}. Thus, an important 
and characteristic  feature of the phase diagram of the $N=2$ system
is that the composite vortex lattice melting curve should at some point cross the 
decomposition curve. Thus, the phase diagram should feature a composite vortex liquid 
phase in the low-temperature, high-magnetic field corner. A schematic picture of this 
phase is given in Fig. \ref{liquid}

\subsection{Vortex line plasma in the $N=2$ model}

\begin{figure}[htb]
\centerline{\scalebox{0.09}{\rotatebox{0.0}{\includegraphics{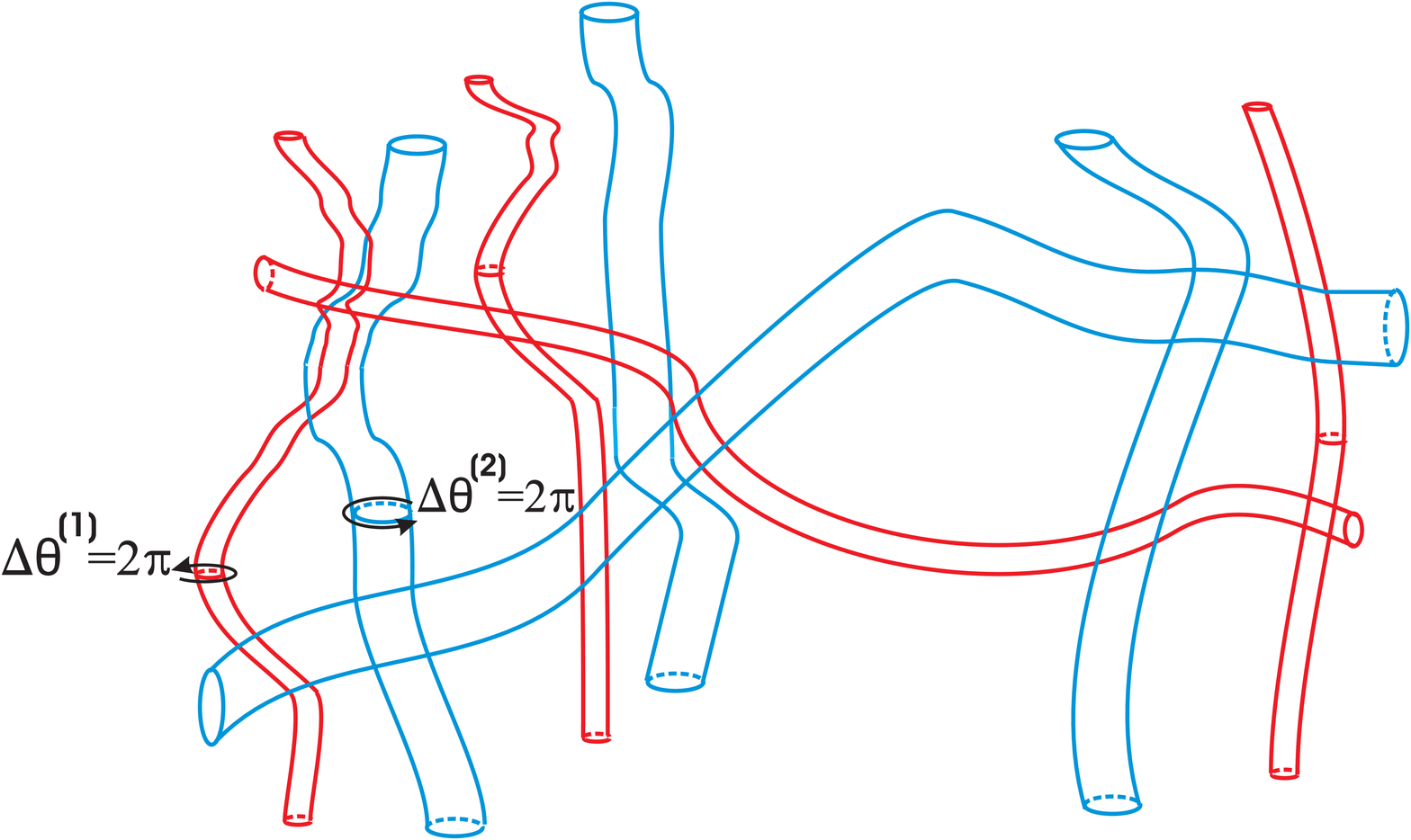}}}} 
\caption{\label{normal} (Color online). Plasma  of fractional vortices in the $N=2$ model in the regime 
$|\psi^{(1)}| \ll |\psi^{(2)}|$ at high temperatures.} 
\end{figure}

\begin{figure}[htb]
\centerline{\scalebox{0.049}{\rotatebox{0.0}{\includegraphics{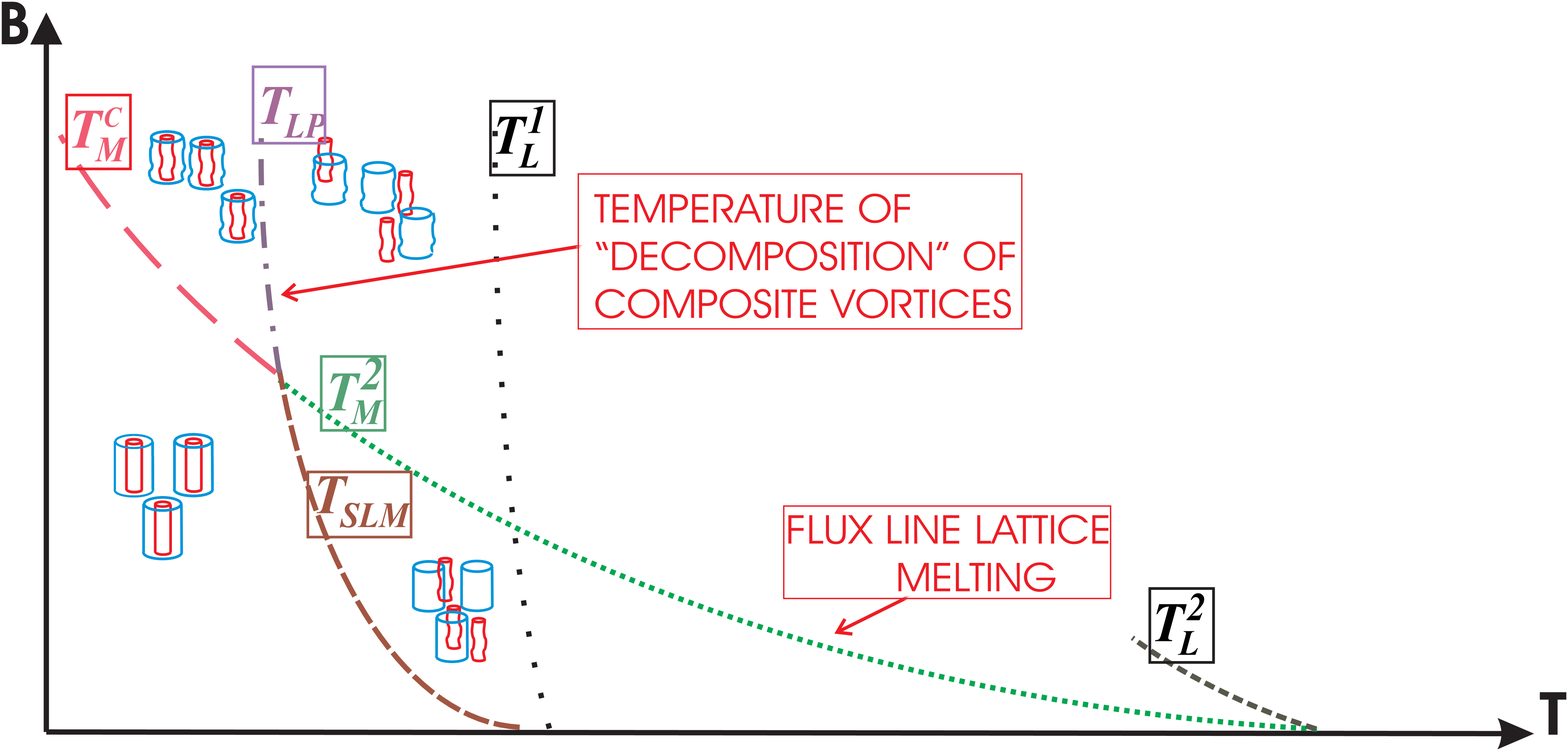}}}} 
\caption{\label{fd} (Color online). A schematic phase diagram of different phases of vortex matter 
and phase transition lines in the $N=2$ model in the regime 
$|\psi^{(1)}| \ne |\psi^{(2)}|$. At temperatures $T_{\rm
  M}^{\rm c}$, $T_{\rm M}^{\rm 2}$, and $T_{\rm SLM}$ the melting of
the composite vortex lattice, the sublattice of heavy vortices and the
sublattice of the light vortices occurs, respectively. At $T_{\rm LP}$
the composite vortices decompose. The temperatures $T_{\rm L}^{\rm 1}$ 
and $T_{\rm L}^{\rm 2}$ denote temperatures where a phase transition via a
proliferation of vortex loops would take place in the absence of a magnetic 
field in models with bare phase stiffnesses $|\tilde{\psi}^{(\alpha)}|(T)$ 
equal to $|\psi^{(\alpha)}|(B,T)$ (where $\alpha=1,2$), if the effect of a magnetic were to be 
taken into account only via the depletion of the modulus of the order 
parameter.} 
\end{figure}

If the temperature is raised either at strong or weak magnetic fields,
a  situation arises 
where all field-induced composite vortices are decomposed and disordered. In addition, closed loops 
have proliferated\cite{tesanovic1999,nguyen1999,Fossheim_Sudbo_book}.
A schematic picture of this state is shown in Fig. \ref{normal}.
The resulting phase diagram of the $N=2$ GL model featuring the various transitions described 
above, is shown in Fig. \ref{fd}.

\subsection{Physical interpretation of the external field-induced
  phases of the $N=2$ model}

We next discuss the  physical interpretation of the various phases that appear as a result 
of the  above described vortex matter transitions. The resulting phases, which exhibit
some quite unusual properties, come about as a result of the interplay between the topology 
of the system and thermal fluctuations. This is rather remarkable, given the 
three-dimensionality of the systems we consider.

\subsubsection{Vortex lattice melting and the disappearance of superconductivity.}
Consider first the melting transition of an interacting ensemble of composite Abrikosov 
vortices. This phase transition, which is of first order \cite{Hetzel1992}, corresponds
to the lines $T_{\rm M}^{\rm c}(B)$ and $T_{\rm M}^2(B)$
shown in Fig. \ref{fd}. It is only the  gauge-charged mode that couples to the external field, 
while the neutral mode does not. The charged mode at low temperature forms an Abrikosov vortex 
lattice with a melting temperature that is suppressed with increasing magnetic field
\cite{Nelson1988,Houghton1989,Blatter1994,Fossheim_Sudbo_book}. The melting temperature of the 
Abrikosov vortex lattice can be suppressed below the temperature where the neutral mode proliferates 
and where the composite vortex lines decompose. For $N=1$, it is 
known that when the Abrikosov  lattice melts, superconductivity is lost also along the 
direction of the magnetic field \cite{Nonomura1998,nguyen1999}. The situation in the 
$N=2$ model is much more complex, since then there still exists a superfluid mode (the gauge neutral 
mode) which is decoupled from external magnetic field. {\it Thus, upon  melting of the Abrikosov 
lattice we arrive at emergent effective neutral superfluidity  existing in a system of charged 
particles  \cite{BSA}}. This  is a genuinely novel state of condensed matter, and moreover one 
which should be realizable in liquid metallic  states of light atoms at in principle experimentally 
accessible pressures in the range of $400 {\rm GPa}$ \cite{BSA}.

So in the absence of an external magnetic 
field, the system thermally excites only fractional flux vortices in the forms of loops, with phase 
windings only in individual condensates, and these fluctuations are responsible for critical 
properties. In contrast, the purely charged vortices (i.e. the composite one-flux-quantum vortices 
with no neutral super-flow) are not relevant in the absence of  external field and  the system is 
either a superconductor (below $T_{\rm c2}$) or a superconductor with neutral mode (below $T_{\rm c1}$). 
{ \it Thus, the effect of a sufficiently strong magnetic field essentially inverts the temperatures 
of the transitions by  melting the lattice of charged modes at  $T_{\rm M}^{\rm c}(B)$ while 
leaving neutral modes intact.} 

The phase transition from a {\it superconducting superfluid} phase where the neutral mode is superfluid 
{\it and} the Abrikosov vortex lattice is intact such that longitudinal superconductivity (parallel to 
the magnetic field) exists\cite{nguyen1999,Nonomura1998}, to a {\it metallic superfluid} phase where 
the system is superfluid, but  longitudinal superconductivity is lost due to the melting of the vortex 
lattice, can be mapped onto a lattice melting transition in the $N=1$ model
, because it is governed only by composite vortices and neutral modes are not involved. Thus it is a 
first order phase transition \cite{Hetzel1992,Fossheim_Sudbo_book}. 

\subsubsection{Decomposition. The disappearance of superfluidity.}
Analogously, the physical meaning of the  sublattice 
melting transition $T_{\rm SLM}$  (see \fig{fd}) is a transition from a {\it superconducting superfluid} 
to an {\it ordinary 
one-gap superconductor}, because a disordering of the phase $\theta^{(1)}$
destroys the massless neutral boson associated with the gauge invariant phase
difference $\theta^{(1)}-\theta^{(2)}$.

If we heat the system further, the ordinary superconductivity will disappear via disordering 
of the phase $\theta^{(2)}$ when we reach the melting transition of the remaining sublattice 
of ``heavy" vortices at $T_{\rm M}^2$.

\begin{figure}[htb]
\centerline{\scalebox{0.04}{\rotatebox{0.0}{\includegraphics{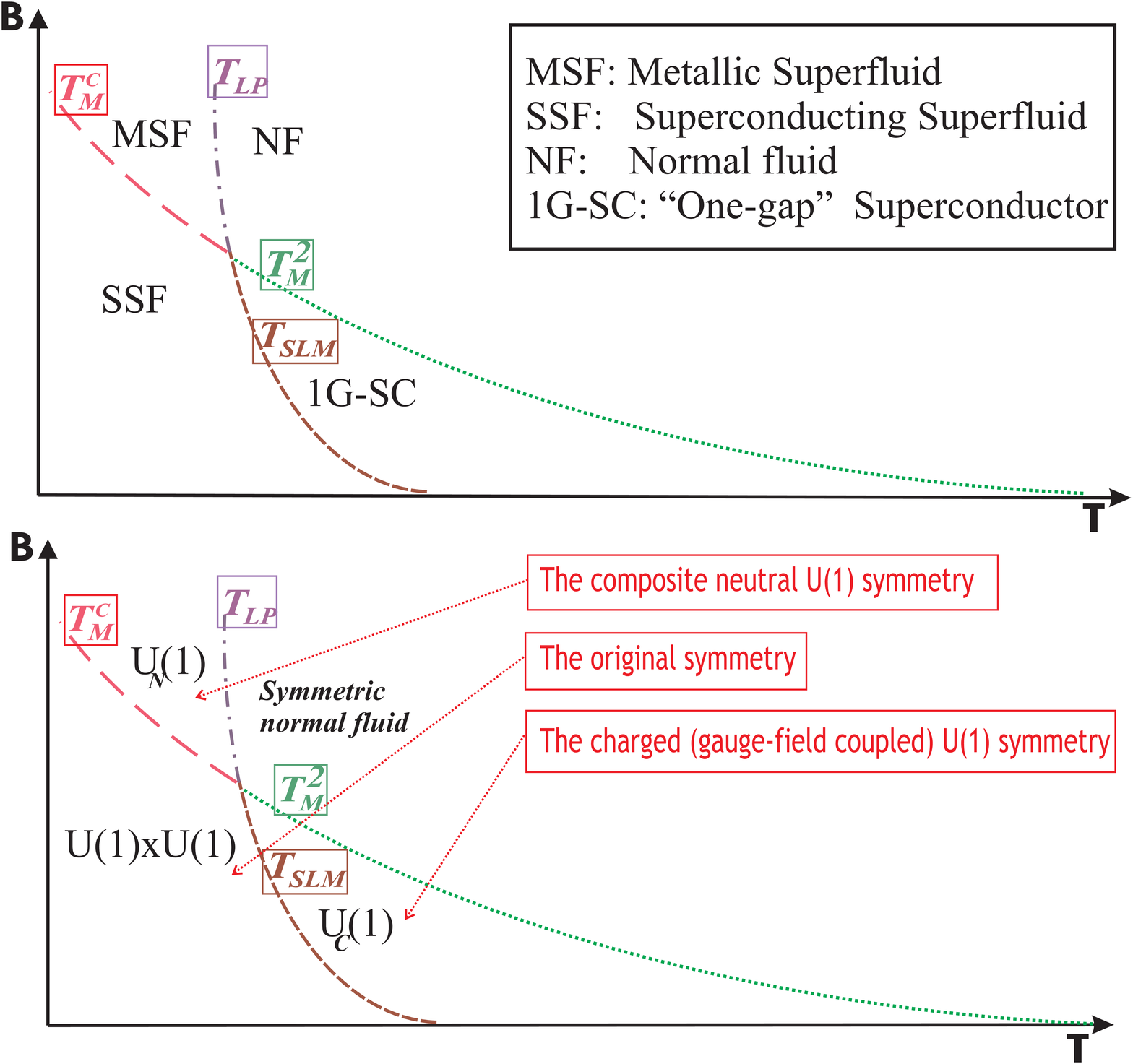}}}} 
\caption{\label{fd2} (Color online). A schematic phase diagram of physical states appearing in the 
$N=2$ model in the regime $|\psi^{(1)}| \ne |\psi^{(2)}|$ as a consequence 
of vortex matter phase transitions. Increasing the magnetic field suppresses the 
melting transition of the composite vortex lattice formed by the charged mode
below the proliferation line for the neutral mode, which does not couple to magnetic 
field. In the absence of disorder (pinning of vortices), superconductivity only remains 
{\it along} the direction of the magnetic field, provided that the vortex system remains 
in a lattice phase. When the composite vortex lattice melts, the system looses ability 
to carry dissipation-less charge currents, but at large enough magnetic field, the neutral 
mode should still be superfluid above the melting temperature \cite{BSA}. Thus, we have a first 
order phase transition from a {\it superconducting superfluid to a  metallic superfluid}. 
The neutral mode proliferates through a second order phase transition in the \xy 
universality class. Therefore, at large enough magnetic fields, a \xy anomaly 
in the specific heat should appear inside the vortex liquid phase. The separation
between the first order specific heat anomaly due to vortex lattice melting and the 
\xy anomaly due to loop-proliferation should increase with increasing magnetic field. 
At low magnetic fields one has another   phase transition inside the Abrikosov 
vortex lattice phase, from a {\it superconducting superfluid to a one-gap (``ordinary")
superconducting state}. 
} 
\end{figure}

The system features one more phase transition. That is a transition from the metallic superfluid 
to a normal fluid, which has a purely topological origin. That is, from the vortex matter point of 
view, this manifests itself as a decomposition of a liquid of composite vortices to a ``plasma" of 
individual vortices at the characteristic temperature $T_{\rm LP}$, and such a transition has no 
counterpart in an $N=1$ superconductor. A schematic diagram of the resulting physical phases is 
shown in Fig. \ref{fd2}.

\subsubsection{A direct SSF $\to$ NSF transition}

\begin{figure}[htb]
\centerline{\scalebox{0.05}{\rotatebox{0.0}{\includegraphics{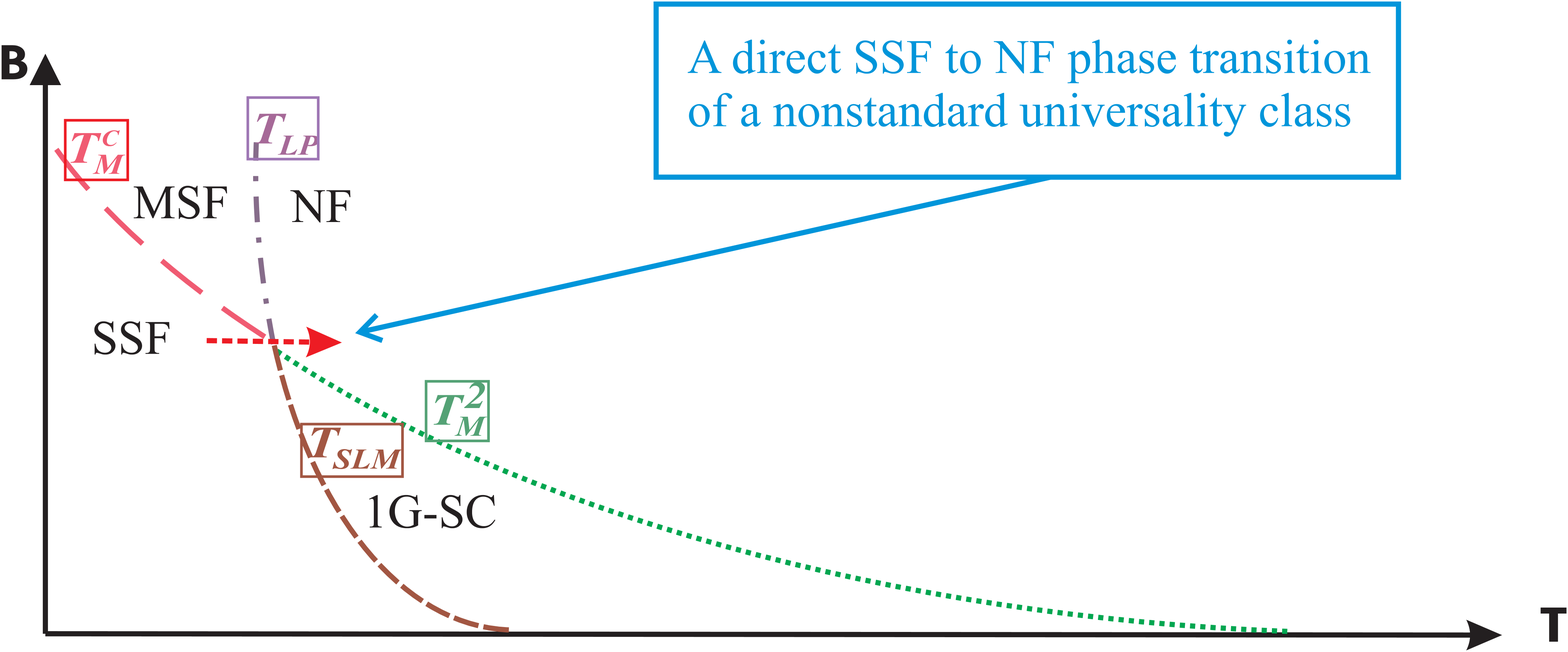}}}} 
\caption{\label{direct_transition} (Color online). A direct phase transition from SSF to NF phase 
(shown with red arrow).} 
\end{figure}

We note also the possibility of an existence of a phase transition directly from a 
superconducting superfluid (SSF) to a metallic normal fluid (NF),  shown in 
Fig. \ref{direct_transition}.
What is remarkable is that this  resembles (while indeed being a different type of 
transition) the type of direct phase transition from a low temperature phase with 
Higgs mass and superfluid density of the neutral mode, to a phase with zero Higgs 
mass and zero helicity modulus of the neutral mode that we would find in zero magnetic
field when  the bare phase stiffnesses of Eq. (\ref{gl_action}) are equal, i.e.
$|\psi^{(1)}| = |\psi^{(2)}|$ \ \ \cite{sachdev2004,motrunich2004,smiseth2004}.
The SSF phase features one massless dual Higgs photon, and one massive Higgs photon 
while the NF phase features one massless photon and one massive dual Higgs photon. 
These phases are therefore {\it self-dual}, in analogy to the situation encountered 
when Eq. (\ref{gl_action}) is viewed as a quantum anti-ferromagnet with easy-plane 
anisotropy \cite{sachdev2004,motrunich2004}. However, there is a significant difference. 
The novel critical point encountered in the case of the quantum antiferromagnet was a 
result of a superposition of a \xy and an inverted \xy critical point. In the $2$-flavor 
London model in finite magnetic field, the crossing point is a superposition between a \xy 
critical line and first order phase transition line, so it  should have a different character 
than the self-dual critical point discussed in Ref. \onlinecite{motrunich2004}.
It can, however, not be a \xy or inverted \xy critical point, since neither of 
these are critical points between self-dual phases. 

Thus, both for high and low magnetic fields we have genuinely novel physics in that for a three 
dimensional system, {\it i)} a critical phenomenon takes place inside either the vortex liquid 
phase (high magnetic fields) or the vortex lattice phase (low magnetic fields), and {\it ii)} we 
have three new equilibrium states which have no counterpart in the $N=1$ case. Moreover,
by the discussion given above, it is clear that  the crossing point between the vortex
lattice melting line and the neutral mode proliferation line warrants further study. 
This is best left for a separate computational analysis.

\subsection{Monte-Carlo results, finite magnetic field, $N=2$}

We now present large-scale Monte Carlo results for the case $N=2$ in finite magnetic field
at low magnetic fields when the temperature is varied \cite{SSBS}. We consider the model 
based on \eq{gl_action} for $N=2$ on an $L^3$ lattice (with $L$ up $96$) with periodic 
boundary conditions for coupling constants $|\psi^{(1)}|^2=0.2$,
$|\psi^{(2)}|^2=2$, and $e^2=1/10$. The ratio  $|\psi^{(2)}|^2/|\psi^{(1)}|^2 = 10$ brings out 
one second order phase transition at $T_{\rm SLM}(B)$ in the \xy universality class well below the 
melting temperature $T_{\rm M}^2$ of the vortex lattice. In LMH $|\psi^{(2)}|^2/|\psi^{(1)}|^2 \approx 2000$, 
but the physical picture remains. For real estimates of $T_{\rm{SLM}}$ and $T_{\rm M}^2$ in LMH, 
see Ref. \onlinecite{Ashcroft1999}. The Metropolis algorithm with local updating is used in 
combination with Ferrenberg-Swendsen reweighting. The external magnetic field 
$\ve B$ studied is $B^x=B^y=0$, $B^z = 2\pi/32$, thus there are $32$ plaquettes in the  
$(x,y)$-plane per flux-quantum. This is imposed by splitting the gauge field into a static 
part $\ve A_0$ and a fluctuating part $\ve A_{\rm fluct}$. The former is kept fixed to 
$(A_0^x,A_0^y(\ve r),A_0^z) = (0, 2\pi xf,0)$ where $f=1/32$ is the magnetic filling fraction, 
on top of which the latter field is free to fluctuate. Together with periodic boundary conditions 
on $\ve A_{\rm fluct}$,  the constraint $\oint_C(\ve A_{0}+\ve A_{\rm fluct}) d\ve l=2\pi fL^2$, 
where $C$ is a contour enclosing the system in the (x,y)-plane, is ensured. It is imperative to 
fluctuate $\ve A$, otherwise type-$1$ and type-$2$ vortices do not interact \cite{frac,smiseth2004}. 
To investigate the transition at $T_{\rm{SLM}}$ we have performed finite size scaling (FSS) of the third 
moment of the action. The simulations are done by using vortices directly \cite{smiseth2004}, but 
with a finite magnetic induction $B^z=2\pi/32$. 

We compute the specific heat $C_V$ and the third moment of the action. To probe the structural order 
of the vortex system we compute the planar structure function $S^{(\alpha)}(\ve k_\perp)$ of the 
{\it local vorticity} $\ve n^{(\alpha)}(\ve r) 
= \left( \ve \nabla \times  \left[ \ve \nabla \theta^{(\alpha)} -e ~  \ve  A  \right] \right)/2 \pi$, 
given by
\begin{equation}
S^{(\alpha)}(\ve k_\perp) = 
\frac{1}{(f L^3)^2}~ \langle | \sum_{\ve r} ~ n_z \f{\alpha}(\ve r) ~ e^{i \ve k_\perp \cdot \ve r_\perp} |^2 \rangle,
\end{equation}
where $\ve r$ runs over dual lattice sites and $\ve k_\perp$ is perpendicular to $\ve B$. This function 
will exhibit sharp peaks for the characteristic Bragg vectors $\ve K$ of the type-$\alpha$ vortex lattice and will 
feature a ring-structure in its corresponding liquid of type-$\alpha$ vortices. The signature of vortex 
sub-lattice melting will be a transition from a six-fold symmetric Bragg-peak structure to a ring 
structure in $S^{(1)}(\ve K)$ while the peak structure remains intact in $S^{(2)}(\ve K)$. Furthermore, 
we compute the {\it vortex co-centricity} $N_{\rm{co}}$ of type-$1$ and type-$2$  vortices, defined as 
$N_{\rm co} \equiv  N^{+}_{\rm{co}} - N^{-}_{\rm{co}}$, where
\begin{equation}
N^{\pm}_{\rm{co}}
 \equiv \frac{\sum_{\ve r}|n_z\f 2(\ve r)|\delta_{n_z\f 1(\ve r),\pm
 n_z\f 2(\ve r)}}{\sum_{\ve r}|n_z\f 2(\ve r)|},
\end{equation}
where $\delta_{i,j}$ is the Kronecker-delta. The reason for considering $N_{\rm co}$ 
is that we then eliminate the effect of random overlap of vortices in the high-temperature 
phase $T > T_{\rm SLM}$ due to vortex-loop proliferation, and focus on the {\it compositeness} 
of field-induced vortices. 

\begin{figure}[htb]
\centerline{\scalebox{0.55}{\rotatebox{0.0}{\includegraphics{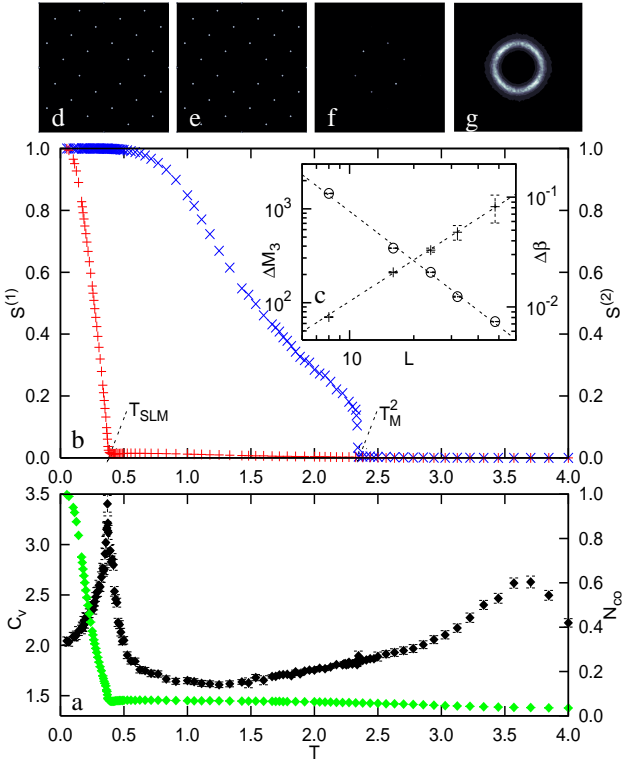}}}} 
\caption{ \label{MC_results} (Color online). 
  MC results for $N=2$ $|\psi^{(1)}|^2=0.2$, $|\psi^{(2)}|^2=2$, and $e=1/\sqrt{10}$. 
  Panel a: $C_V$ (black) and  $N_{\rm co}$  (green). The $C_V$-anomaly at $T_{\rm{SLM}}=0.37$, 
  where type-$1$ vortices proliferate, matches the point at which $N_{\rm co}$ drops to
  zero. Thus, type-$1$ vortices are torn off type-$2$
  vortices. The remnant of the zero-field anomaly in $C_V$ is 
  seen as a hump at $T \sim 3.6$. Panel b:   
  $S^{(1)}(\ve K)$ (red) and $S^{(2)}(\ve K)$ (blue) for
  the particular Bragg vector $\ve K =(\pi/4,-\pi/4)$. 
  $S^{(1)}(\ve K)$ vanishes continuously at $T_{\rm{SLM}}$,
  while $S^{(2)}(\ve K)$ vanishes discontinuously at $T_{\rm M}^2=2.34$. 
  Panel c: FSS plots of the $M_3$ from which the exponents 
  $\alpha=-0.02 \pm 0.05$ and $\nu=0.67 \pm 0.03$
  is extracted, showing that the sub-lattice melting is a  
  \xy phase transition. Panels d, e, f, and g: plots of 
  $S^{(2)}(\ve k_\perp)$ for the temperatures $T_{\rm d}=0.35$,
  $T_{\rm e}=0.4$, $T_{\rm f}=1.66$, and $T_{\rm g}=2.85$,
  respectively. At $T_{\rm d}$, $T_{\rm e}$, and $T_{\rm f}$, 
  the vortex lattice remains intact. The vortex lattice melts at 
  $T_{\rm M}^2$ to give a vortex liquid ring pattern at $T_{\rm g}$.}  
\end{figure}

The quantity $N_{\rm{co}}$ is the fraction of type-$2$ vortex segments that are co-centered 
with type-$1$ vortices, providing a measure of the extent to which vortices of type-$1$ and 
type-$2$ form a {\it composite} vortex system. Hence, it probes the splitting processes 
visualized  in Fig. \ref{fluct1}. The results are shown in Fig. \ref{MC_results}. 

At $T_{\rm{SLM}}$, $C_V$ has a pronounced peak associated with the \xy transition, 
and a broader less pronounced peak which is the finite field remnant of the zero-field 
inverted \xy transition \cite{nguyen1999}. Scaling of $M_3$ at $T_{\rm{SLM}}$ shown in the inset 
c in Fig. \ref{MC_results} yields the critical exponents $\alpha =-0.02\pm0.05$ and 
$\nu=0.67\pm0.03$ in agreement with the \xy universality class. A novel result is that 
$S^{(1)}(\ve K)$ vanishes {\it continuously} as the temperature approaches $T_{\rm{SLM}}$ 
from below, precisely the hallmark of the decomposition transition that separates 
the two types of vortex states depicted in Figs. \ref{fluct1} and \ref{sl}.
A related feature is the {\it vanishing} of $N_{\rm co}$ at 
$T_{\rm{SLM}}$ as a function of temperature, discussed in detail below. The 
first-order melting transition takes place at  $T_{\rm M}^2$, where $S^{(2)}(\ve K)$ 
vanishes discontinuously. This is the temperature at which the translational invariance 
is restored through melting of the type-$2$ vortex lattice. In the temperature interval 
$T < T_{\rm{SLM}}$ the system {features superconductivity and superfluidity simultaneously} 
\cite{BSA}, since there is long-range order both in the charged and the neutral vortex modes. 
In the temperature interval $T_{\rm{SLM}} < T < T_{\rm M}^2$ long-range order in the neutral mode 
is destroyed by loop-proliferation of type-$1$ vortices, hence superfluidity is lost 
\cite{BSA}. However, {longitudinal one-component superconductivity} is retained along the 
direction of the external magnetic field. For $T > T_{\rm M}^2$ superconductivity is also lost, 
hence this is the normal metallic state, which is a two-component vortex liquid. 

\begin{figure}[htb]
\centerline{\scalebox{0.13}{\rotatebox{0.0}{\includegraphics{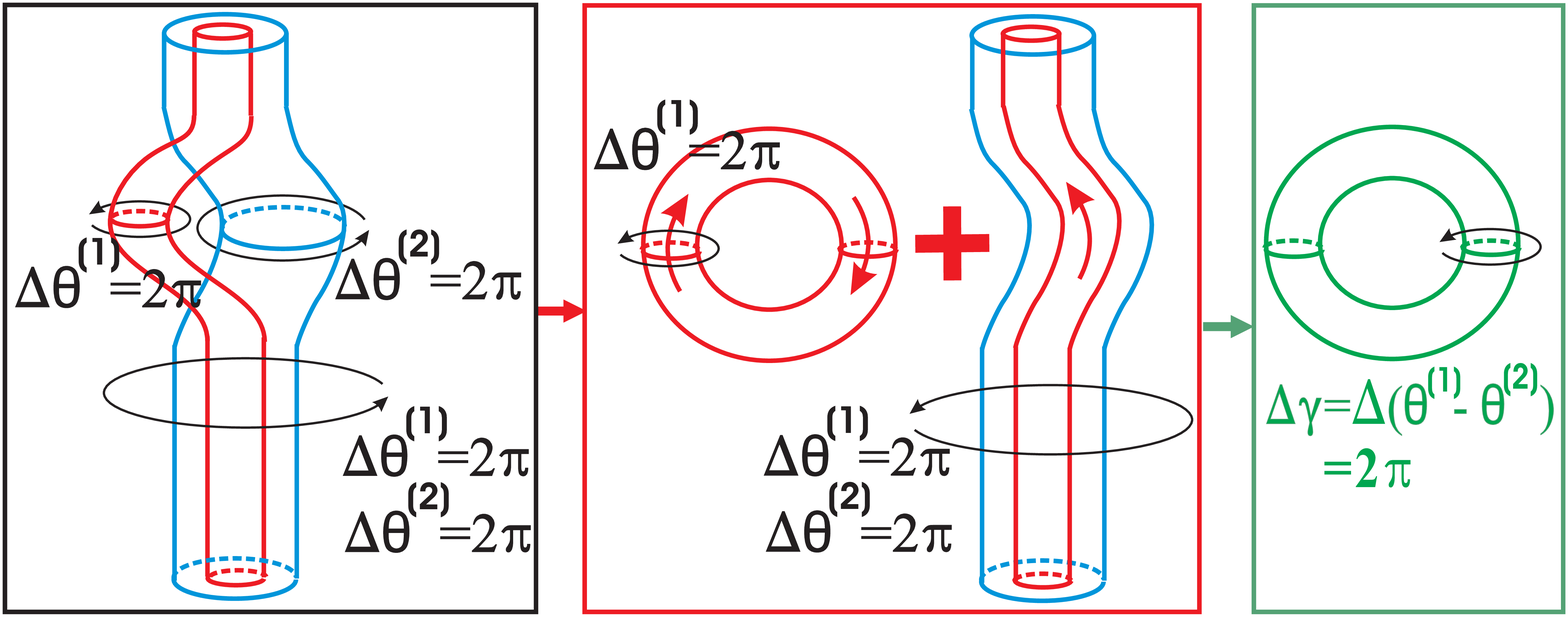}}}} 
\caption{ \label{splitting} (Color online). 
Detailed illustration of the low-temperature thermal fluctuations in a vortex lattice of composite 
vortices. A local excursion of  a type-$1$ vortex  away from the composite vortex lattice may be 
viewed as a type-$1$ bound vortex loop superposed on the composite vortex lattice. 
The composite vortex line does not interact with a vortex with nontrivial winding in
$\Delta \gamma = \Delta(\theta^{(1)}-\theta^{(2)})$. A splitting 
of the composite vortex lattice 
may be thus viewed as a  {\it zero-field} vortex-loop proliferation of type-$1$ vortices; 
a \xy phase transition universality \cite{tesanovic1999,nguyen1999,Fossheim_Sudbo_book}.  }
\end{figure}

The most unusual and surprising feature is the continuous variation of $S^{(1)}(\ve K)$ with 
temperature, even at $T_{\rm{SLM}}$ where it vanishes.
The explanation for this is the 
proliferation of type-$1$ vortices (which destroys the neutral superfluid mode) in the 
background of a composite vortex lattice, which the type-$1$ vortices essentially do not see, cf. Fig. 
\ref{splitting}. As far as the composite neutral Bose field $\theta^{(1)}-\theta^{(2)}$ 
is concerned, {\it it is precisely as if the composite vortex lattice were not present at all}. 
Hence, $S^{(1)}(\ve K)$ vanishes for a completely different reason than  $S^{(2)}(\ve K)$, 
namely due to {\it critical fluctuations, i.e. vortex-loop proliferation} in the condensate 
component with lowest bare stiffness. Such a phase transition does not completely restore 
broken translational invariance associated with a vortex lattice, since for the type-$2$ vortices 
{\it quite remarkably, the vortex lattice order survives the decomposition transition}, due to 
interaction between heavy vortices mediated by charged modes. The vanishing of $N_{\rm{co}}$ 
is particularly 
interesting, and finds a natural explanation within the framework of the above discussion. 
That is, for $T \ll T_{\rm{SLM}}$, we have $N_{\rm{co}} \approx 1$, so the vortex system 
consists practically exclusively of composite vortices. As the temperature increases, thermal 
fluctuations induce excursions such as those illustrated in  
in Fig. \ref{splitting}, which reduces $N_{\rm co}^{\rm{+}}$ from its low-temperature value, 
reaching a {\it minimum} at $T_{\rm{SLM}}$ and then {\it increase} for $T > T_{\rm{SLM}}$. 

We may view the splitting process as a type-$1$ closed vortex loop superposed on a vortex lattice of 
(slightly) fluctuating composite vortices. An important point to notice is that a 
type-$\alpha$ vortex does not interact with a composite vortex by means of a neutral mode. 
This follows from a topological argument that two split branches will feature 
nontrivial winding in the composite neutral field $\theta^{(1)}-\theta^{(2)}$, 
while a composite vortex line does not. Hence, the splitting transition may be 
viewed as {\it a type-$1$ vortex loop-proliferation in a neutral superfluid}. 
This is illustrated in Fig. \ref{splitting}.
Thus, we may utilize the well-known results for the critical properties of the \xy model 
for neutral superfluids described as a vortex-loop proliferation 
\cite{tesanovic1999,nguyen1999,Fossheim_Sudbo_book}. 
This ``vortex sublattice melting" phase transition is therefore in the  \xy universality 
class \cite{tesanovic1999,nguyen1999,Fossheim_Sudbo_book}, not a first order melting 
transition. The resulting phase is one where superfluidity is lost and longitudinal 
superconductivity retained in the component $\Psi_0^{(2)}$.

Conversely, $N_{\rm co}^{\rm{-}}$ remains essentially zero until  $T_{\rm{SLM}}$, thereafter 
increasing monotonically. For temperatures above, but close to $T_{\rm{SLM}}$, fluctuations 
in vortices originating in $\Delta \theta^{(2)}$ are still small, so the variations in 
$N_{\rm{co}}=N_{\rm co}^{\rm{+}}-N_{\rm co}^{\rm{-}}$ reflect thermal fluctuations 
in vortices originating in $\Delta \theta^{(1)}$. The increase of $N_{\rm co}^{\rm{\pm}}$ 
means that type-$1$ vortex loops  are thermally generated, and thus tend to {\it randomly} 
overlap more with the moderately fluctuating type-$2$ vortices. At their first order melting 
transition, type-$2$ vortices fluctuate only slightly. {\it Thus, the vanishing of $N_{\rm{co}}$ 
above $T_{\rm{SLM}}$ reflects the increase in the density of thermally generated type-$1$ vortex 
loops in the background of a slightly fluctuating type-$2$ vortex lattice}.

\subsection{Graphical representation of phase disordering transitions
  in the $N=2$ model.}
\begin{figure}[htb]
\centerline{\scalebox{0.25}{\rotatebox{0.0}{\includegraphics{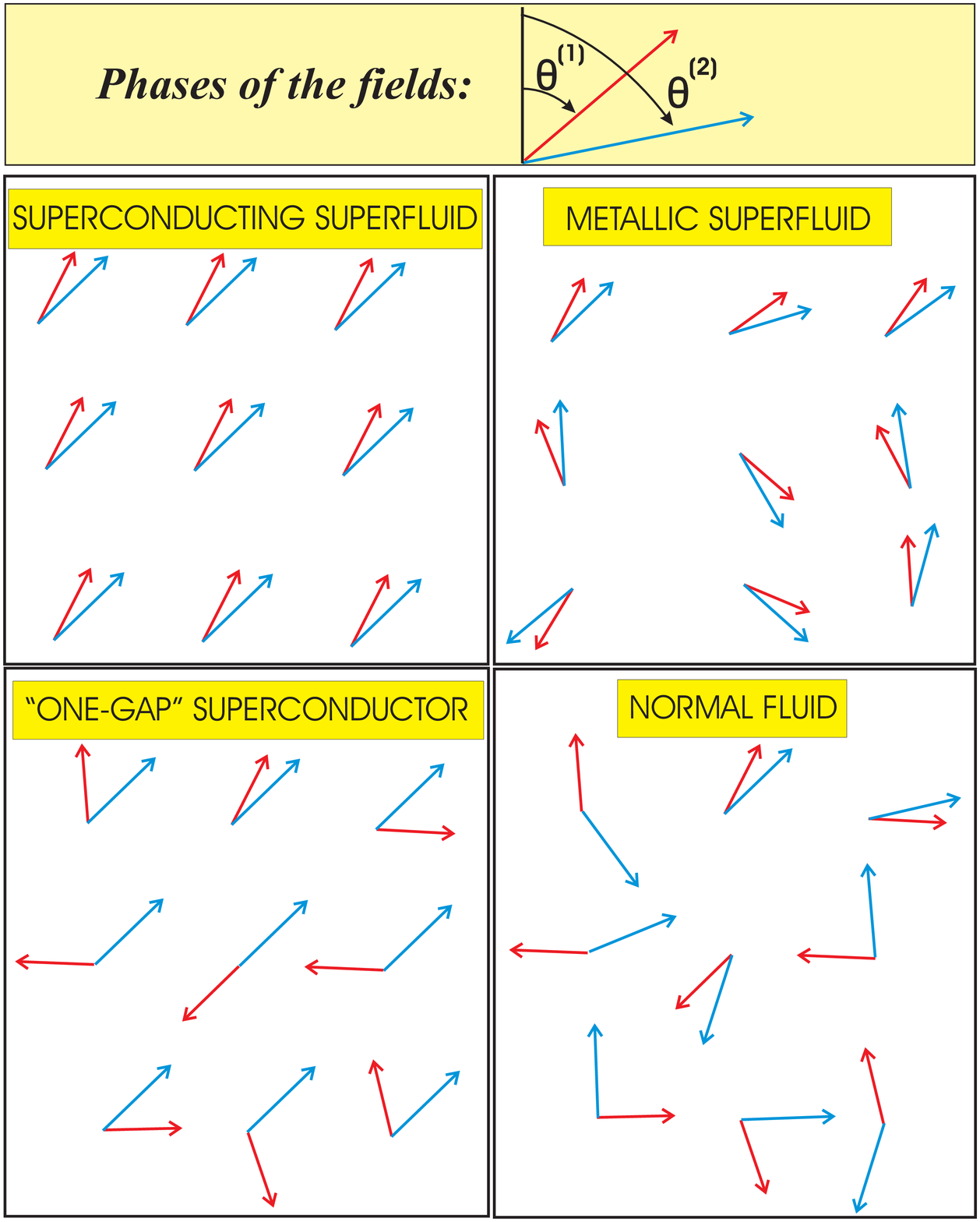}}}} 
\caption{ \label{2ph1} (Color online). Phases of the order parameters in the  various 
states for $N=2$. In the upper left  panel, both $\theta^{(1)}$ and $\theta^{(2)}$ 
are ordered, this is the superconducting superfluid state. In the upper right panel,
neither of the phases $\theta^{(1)}$ and $\theta^{(2)}$ are ordered, however the 
combination $\theta^{(1)} - \theta^{(2)}$ exhibits long-range order, this is the 
metallic superfluid state. In the lower left panel, $\theta^{(1)}$ is disordered
and $\theta^{(2)}$ is ordered. In this case, the neutral superfluid mode is 
destroyed and we are left with one charged superconducting mode, this is the 
analog of the one-gap superconducting state. In the lower right panel,
neither of the phases $\theta^{(1)}$ and $\theta^{(2)}$ are ordered and
the combination $\theta^{(1)} - \theta^{(2)}$ does not exhibit long-range order, this is the 
metallic normal fluid state. The states illustrated in the upper left and lower right and left
panels exist at zero as well as finite magnetic fields. The state illustrated in the upper right
panel only exists at finite magnetic fields.} 
\end{figure}

In Fig.  \ref{2ph1} we present a schematic picture 
of configuration of the order parameters phases $\theta^{(1)}$ and $\theta^{(2)}$ in various points in 
physical space, when vortex matter drives the system into one of the above discussed 
superconducting and superfluid states. 

\section{\label{N>2_extfield} The $N >2$ model in external magnetic field}
We next consider the novel features that are encountered, compared to the $N=1$-
and $N=2$ cases, when an $N>2$ system is subjected to an external magnetic field. 
The new features are  due to the fact that we have more than one neutral vortex 
mode, and that a vortex with phase winding in any single phase field $\theta^{(\alpha)}$ 
will excite {\it $N-1$ neutral modes} (see Appendix \ref{App:charged_neutral}).
We consider first the case $N=3$, followed by  the case $N=4$. En route we 
introduce the useful concept of ``color charge'' which  facilitates a discussion
of the universality class of the phase transitions that occur in multi-component
superconductors in external magnetic field when composite vortices decompose
due to thermal fluctuations.

\subsection{Decomposition transitions, $N=3$} 
 
We stress that the system is not 
mapped onto a $U(1)\times U(1)$ neutral system because the neutral modes remain topologically 
coupled, as a consequence of multiple connectedness of space introduced by the vortex core. 
It  manifests itself  in the fact that any  {\it single} phase variable 
$\theta^{(\alpha)}; \alpha \in [1,2,3]$ excites {\it two} neutral  modes, as illustrated in 
detail below.

We introduce bare phase stiffnesses for the neutral modes
in \eq{neutral} as follows
\begin{eqnarray}
J^{12}&=&\frac{|\psi^{(1)}|^2 |\psi^{(2)}|^2}{\Psi^2}, \nonumber \\
J^{23}&=&\frac{|\psi^{(1)}|^2 |\psi^{(3)}|^2}{\Psi^2}, \nonumber \\
J^{13}&=&\frac{|\psi^{(2)}|^2 |\psi^{(3)}|^2}{\Psi^2}.
\label{stiffnesses}
\end{eqnarray}
Hence, a vortex with phase windings $(\Delta\theta^{(1)}=2\pi,\Delta\theta^{(2)}=0,\Delta\theta^{(3)}=0)$,
can be mapped onto two co-centered vortices in a two-component neutral superfluid with bare stiffnesses 
$J^{12}$ and $J^{13}$. Thus, at a distance larger than the penetration length, such a vortex interacts with 
a vortex  $(\Delta\theta^{(1)}=-2\pi,\Delta\theta^{(2)}=0,\Delta\theta^{(3)}=0)$ like two vortices in 
a neutral superfluid with bare phase stiffness $\tilde{J}=J^{12}+J^{13}$.

Intra-vortex interaction, e.g. of the vortex
$(\Delta\theta^{(1)}=2\pi,\Delta\theta^{(2)}=0,\Delta\theta^{(3)}=0)$
with a vortex $(\Delta\theta^{(1)}=0,\Delta\theta^{(2)}=2\pi,\Delta\theta^{(3)}=0)$
or with a vortex  $(\Delta\theta^{(1)}=0,\Delta\theta^{(2)}=0,\Delta\theta^{(3)}=2\pi)$
is more complicated.
It can most conveniently be described by introduction of  the ``color charge" concept,
which we explain in the next subsection, \ref{subsection_color}. 

First, however, we observe that only a composite vortex  $(\Delta\theta^{(1)}=2\pi,\Delta\theta^{(2)}=2\pi,\Delta\theta^{(3)}=2\pi)$
has finite energy.  The key feature of a system with 
$|\psi^{(1)}|\neq|\psi^{(2)}|\neq|\psi^{(3)}|$, is
that  the three elementary constituent vortices are bound with
different strength to such 
a composite vortex. For example, when
$|\psi^{(1)}|\ll|\psi^{(2)}|\ll|\psi^{(3)}|$, the  neutral modes excited by a vortex $(\Delta\theta^{(1)}=2\pi,\Delta\theta^{(2)}=0,\Delta\theta^{(3)}=0)$
have bare phase stiffnesses $(J^{12},J^{13}) \ll J^{23}$.  This in turn implies that in the composite 
vortex $(\Delta\theta^{(1)}=2\pi,\Delta\theta^{(2)}=2\pi,\Delta\theta^{(3)}=2\pi)$, the 
constituent elementary vortex $(\Delta\theta^{(1)}=2\pi,\Delta\theta^{(2)}=0,\Delta\theta^{(3)}=0)$ is 
most loosely bound. Thus, in contrast to the $N=2$ case, the effect of
thermal fluctuations for $N=3$ is a two step transition. In
the general $N$ case the process of stripping a composite vortex of 
its $N$ constituent vortices is an $N-1$-step process occurring successively, starting at $T_{\rm c1}$ 
and progressing up through $T_{\rm c2}$ up to $T_{{\rm c}N-1}$, at which point the vortex system is 
fully decomposed.

\begin{figure}[htb]
\centerline{\scalebox{0.13}{\rotatebox{0.0}{\includegraphics{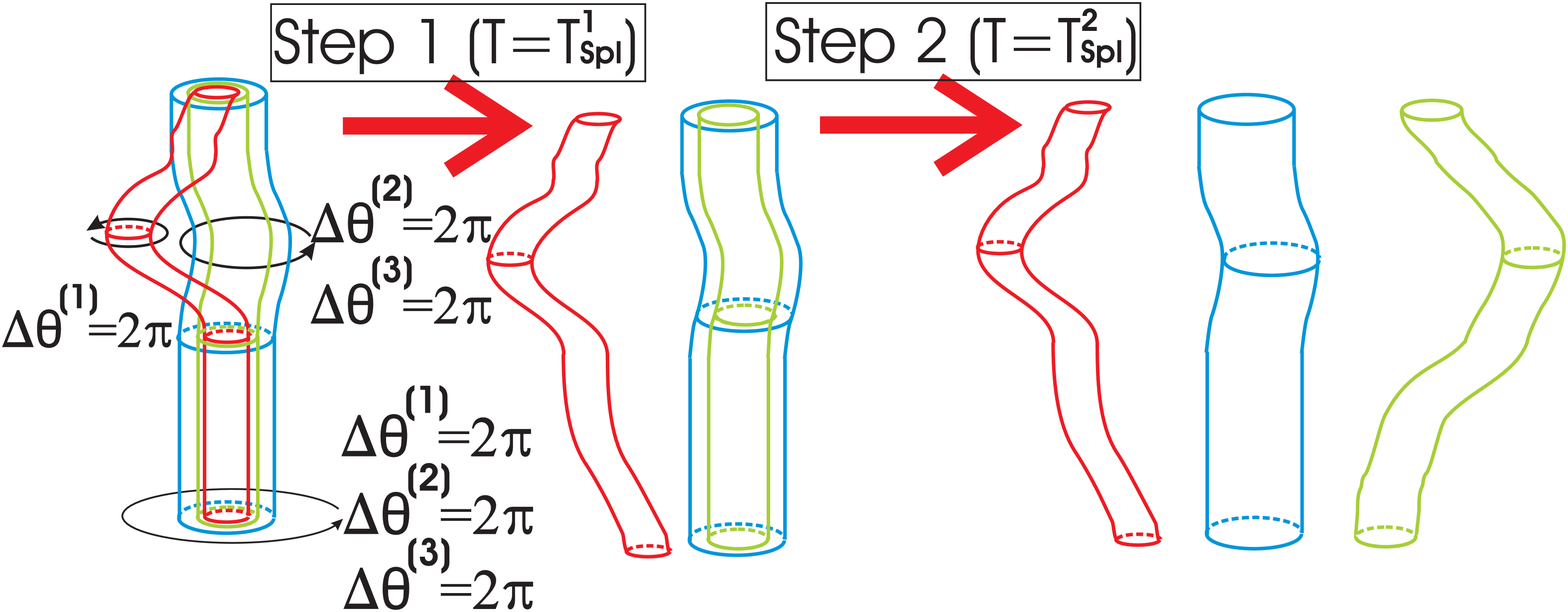}}}} 
\caption{\label{3spl_2} (Color online). A two-step decomposition transition in the
$N=3$ model in external magnetic field. 
The ``lightest" vortex component, originating 
in the order parameter component with the lowest bare phase stiffness, tears itself 
loose from the composite Abrikosov vortex  of the two stiffer order parameter 
components at $T_{\rm Spl}^{1}$. At a higher temperature $T_{\rm Spl}^{2}$, 
the ``next-to-lightest" vortex component 
(green vortex), originating in the order parameter component with the next-to-lowest bare 
phase stiffness, tears itself loose from the vortex  of the 
stiffest order parameter component in the background of proliferated vortices
originating in non-trivial phase-windings of the phase with lowest phase stiffness
(red vortex).} 
\end{figure}

For $N=3$, at a low temperature determined by the smallest bare phase stiffness $|\psi^{(1)}|$
and by $J^{12}$ and $J^{13}$, 
there should therefore take place a partial decomposition of the vortex
$(\Delta\theta^{(1)}=2\pi,\Delta\theta^{(2)}=2\pi,\Delta\theta^{(3)}=2\pi)$
into two vortices $(\Delta\theta^{(1)}=2\pi,\Delta\theta^{(2)}=0,\Delta\theta^{(3)}=0) +
(\Delta\theta^{(1)}=0,\Delta\theta^{(2)}=2\pi,\Delta\theta^{(3)}=2\pi)$,
illustrated in \fig{3spl_2}. Then, upon increasing 
the temperature there should take place a phase transition, also illustrated in \fig{3spl_2},
into a fully decomposed state defined by the phase windings
$(\Delta\theta^{(1)}=2\pi,\Delta\theta^{(2)}=0,\Delta\theta^{(3)}=0) +
(\Delta\theta^{(1)}=0,\Delta\theta^{(2)}=2\pi,\Delta\theta^{(3)}=2\pi)
\to(\Delta\theta^{(1)}=2\pi,\Delta\theta^{(2)}=0,\Delta\theta^{(3)}=0) +
(\Delta\theta^{(1)}=0,\Delta\theta^{(2)}=2\pi,\Delta\theta^{(3)}=0)+
(\Delta\theta^{(1)}=0,\Delta\theta^{(2)}=0,\Delta\theta^{(3)}=2\pi)$.

We also stress that apart from the neutral mode
 $(\theta^{(2)}-\theta^{(3)})$, which provides an attractive
 interaction for the vortex pair
$(\Delta\theta^{(1)}=0,\Delta\theta^{(2)}=2\pi,\Delta\theta^{(3)}=0)+
(\Delta\theta^{(1)}=0,\Delta\theta^{(2)}=0,\Delta\theta^{(3)}=2\pi)$,
the first of these vortices excites the neutral mode  $(\theta^{(1)}-\theta^{(2)})$
which consists of oppositely directed currents in the condensates
$\Psi_0^{(1)}$ and $\Psi_0^{(2)}$. The second vortex
excites the mode $(\theta^{(1)}-\theta^{(3)})$
which is associated with oppositely directed currents in condensates
$\Psi_0^{(1)}$ and $\Psi_0^{(3)}$. These modes, apart from giving the pair 
$(\Delta\theta^{(1)}=0,\Delta\theta^{(2)}=2\pi,\Delta\theta^{(3)}=0)+
(\Delta\theta^{(1)}=0,\Delta\theta^{(2)}=0,\Delta\theta^{(3)}=2\pi)$
logarithmically divergent energy per unit length\cite{frac,smiseth2004}, also
yield some repulsive interaction in this pair, because both the modes
$(\theta^{(1)}-\theta^{(2)})$ and $(\theta^{(1)}-\theta^{(3)})$
feature unscreened currents of condensate $\Psi_0^{(1)}$. However,
such a repulsive interaction in this pair is negligibly small in the 
considered regime $|\psi^{(1)}|\ll|\psi^{(2)}|\ll|\psi^{(3)}|$, 
compared to the interactions mediated by the mode $(\theta^{(2)}-\theta^{(3)})$.

\subsection{\label{subsection_color}Color electric charge}
Formally, the partial decomposition process can be described by introducing 
the concept of ``color electric charges''. That is, we may introduce e.g. ``+red", 
``+green" and ``+blue"  charges associated with $2\pi$ windings in $(\theta^{(1)}-\theta^{(2)})$, 
$(\theta^{(1)}-\theta^{(3)})$ and $(\theta^{(2)}-\theta^{(3)})$,
respectively (see Tab. \ref{tab:colors_n3}). If we
have a $-2\pi$ winding in $(\theta^{(1)}-\theta^{(2)})$, $(\theta^{(1)}-\theta^{(3)})$
or $(\theta^{(2)}-\theta^{(3)})$,  that would correspond to `` $-$red", ``$-$green" and ``$-$blue"  
color electric charges, respectively. We stress once more that in order to preserve 
single-valuedness of the order parameters, the $\pm 2\pi$ gains in phase differences may only 
come as $\pm 2\pi$ gains in individual phases. For example if we would have 
$(\Delta\theta^{(1)}=3\pi/4 ,\Delta\theta^{(2)}=-5\pi/4 )$ then $(\theta^{(1)}-\theta^{(2)})$  
would change by $2\pi$. However, such a configuration would be unphysical because individual 
order parameters $\Psi_0^{(1)}$ and $\Psi_0^{(2)}$ would loose their single-valuedness. Then, 
a vortex $(\Delta\theta^{(1)}=2\pi,\Delta\theta^{(2)}=0,\Delta\theta^{(3)}=0)$ which excites  
two neutral modes associated with  $(\theta^{(1)}-\theta^{(2)})$ and $(\theta^{(1)}-\theta^{(3)})$ 
with stiffnesses  $J^{12}$ and $J^{13}$, respectively, may be viewed as a color charged string 
with color charge ``$+$red" and ``$+$green". 

The regime $|\psi^{(1)}|\ll|\psi^{(2)}|\ll|\psi^{(3)}|$, i.e. when
$J^{12}\ll J^{13}\ll J^{23}$,  corresponds to the situation where red
electric charges are much weaker than green charges, which in turn are much
weaker than the blue charges. The blue charge then dominates the  
binding of the vortices $(\Delta\theta^{(1)}=0,\Delta\theta^{(2)}=2\pi,\Delta\theta^{(3)}=0)$
and $(\Delta\theta^{(1)}=0,\Delta\theta^{(2)}=0,\Delta\theta^{(3)}=2\pi)$. The tightly bound 
composite vortex $(\Delta\theta^{(1)}=0,\Delta\theta^{(2)}=2\pi,\Delta\theta^{(3)}=2\pi)$ then 
has electric charge $(-\rm{red},-\rm{green})$ which loosely binds it with 
$(+\rm{red},+\rm{green})$ color charged vortex 
$(\Delta\theta^{(1)}=2\pi,\Delta\theta^{(2)}=0,\Delta\theta^{(3)}=0)$
into a color charge neutral 
finite energy  one-flux-quantum vortex 
$(\Delta\theta^{(1)}=2\pi,\Delta\theta^{(2)}=2\pi,\Delta\theta^{(3)}=2\pi)$.

\begin{figure}[htb]
\centerline{\scalebox{0.14}{\rotatebox{0.0}{\includegraphics{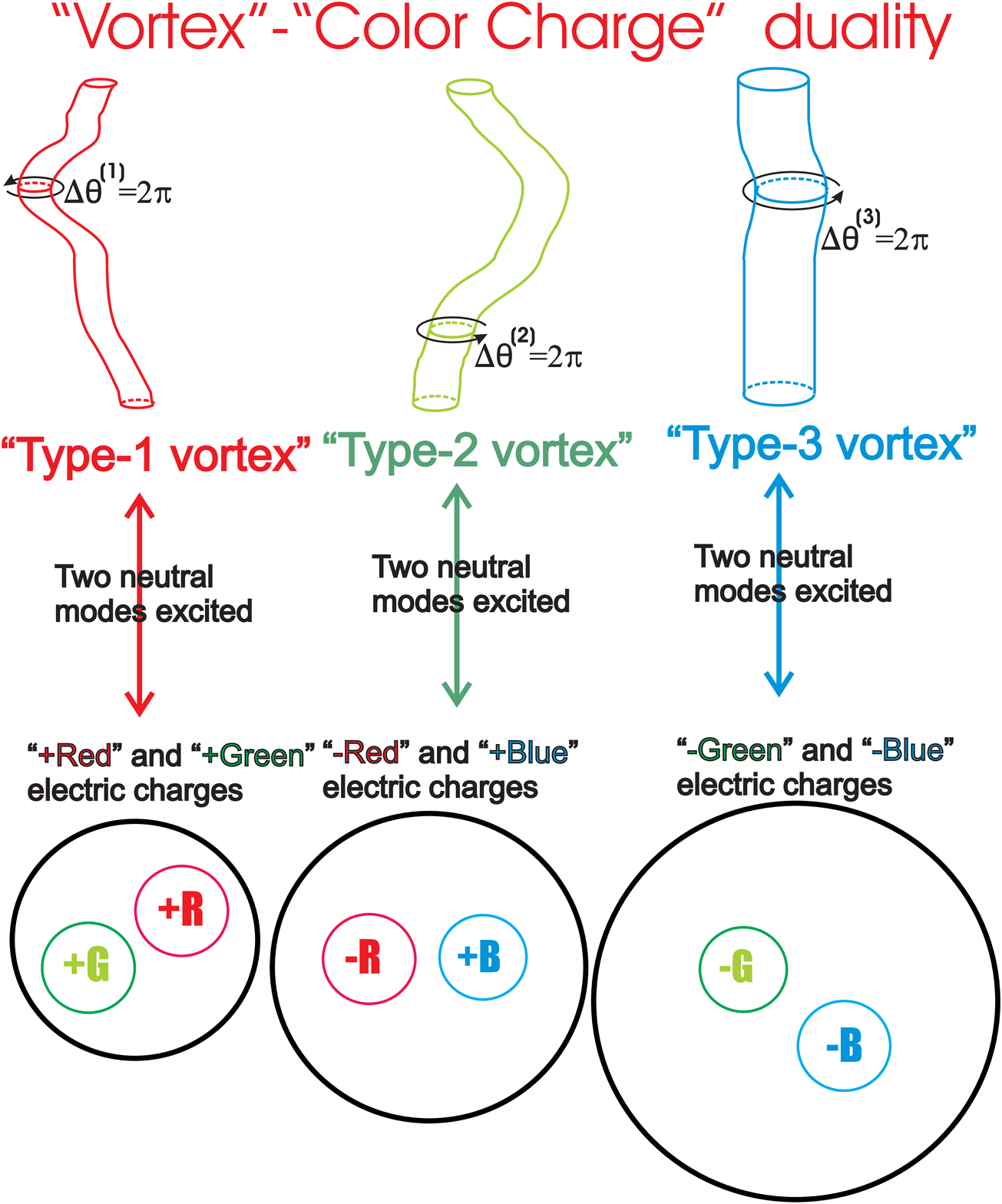}}}} 
\caption{ \label{bridge_vortex_charge}  (Color online). 
The connection between vortex illustrations and bound states of color charges, for 
the case $N=3$. Each type of vortex is a bound state of 2 color
charges. The radius of each black circle is for graphical
convenience is used to differentiate between ``heavy" and
``light" vortices. The order parameter component  with the lowest bare
phase stiffness is taken to be the vortex with the ``smallest 
diameter". A vortex originating in a non-trivial phase-winding in $\theta^{(\alpha)}$
is denoted a type-$\alpha$ vortex. {\it The color of the vortex on the top
of the Figure should not be confused with the color of the electric charges
in the ``dual" charge representation}.} 
\end{figure}

In Fig. \ref{bridge_vortex_charge}, we illustrate how to connect the
vortex picture to 
the picture of color charges, for the case $N=3$.  For $N=3$, 
each type of vortex is a bound state of two color charges.
\begin{table}
\caption{\label{tab:colors_n3}A color charge is defined as a $ \pm 2\pi$
  winding in the mode $(\theta^{(\alpha)}-\theta^{(\eta)})$ with the
  following mapping for $N=3$. }
\begin{ruledtabular}
\begin{tabular}{rccc}
  Mode: & $(\theta^{(1)}-\theta^{(2)})$ &
  $(\theta^{(1)}-\theta^{(3)})$ & $(\theta^{(2)}-\theta^{(3)})$\\
  Color charge:        & red  & green  & blue\\
\end{tabular}
\end{ruledtabular}
\end{table}

\begin{figure}[htb]
\centerline{\scalebox{0.27}{\rotatebox{0.0}{\includegraphics{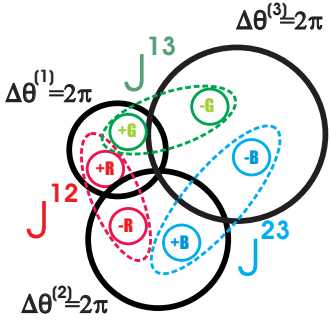}}}} 
\caption{ \label{colorcharges_31}  (Color online). A color charge representation for $N=3$ of the 
composite vortex lattice phase illustrated also in the left part of Fig. \ref{3spl_2}.
This low-temperature phase, where the fluctuations of the composite vortex lines only 
involve small excursions of each of the constituent vortex lines away from the
main composite object, may be viewed as a 3-color dielectric phase. 
The main composite vortex line is ``anchored" on the thickest vortex. We assume we 
are at temperatures and fields such that the thickest vortex line does not undergo a 
melting transition. Each constituent vortex may be viewed as a bound state of certain 
combinations of color charges. The composite vortex line may, on the other hand, be 
viewed as bound states of $\pm {\rm red}$, $\pm {\rm green}$, and $\pm {\rm blue}$ 
charges, as indicated by the dotted ellipses. This is a three-color dielectric 
``insulating" phase. We strongly emphasize that the above illustration is meant 
to illustrate what the situation is in a typical cross section  along the lines, 
{\it which are not rigid straight vortex lines}.} 
\end{figure}

A schematic picture of the low-temperature composite vortex lattice phase, the partial 
decomposition transition in the color electric charge representation, and the complete 
decomposition transition,  are given in Figs.  \ref{colorcharges_31}, \ref{colorcharges_32}, 
and \ref{colorcharges_33}.

\begin{figure}[htb]
\centerline{\scalebox{0.15}{\rotatebox{0.0}{\includegraphics{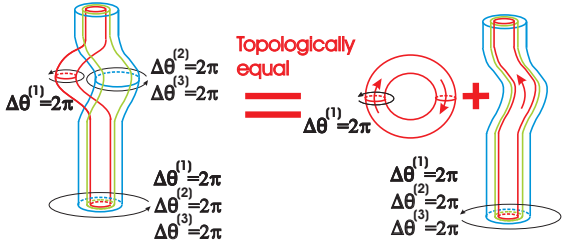}}}} 
\caption{ \label{superim} (Color online). A local thermally driven detachment of the
  type-$1$ vortex line (red color) 
from the composite vortex line can be viewed as a superposition of thermally created 
{\it closed type-$1$ vortex loop} on a completely composite vortex line. Both processes are 
topologically equivalent because  the vorticity of the type-$1$ constituent vortex in the 
composite vortex line is exactly canceled by a superimposed type-$1$ vortex ring
with the opposite vorticity. We stress that if left segment of a vortex loop has a 
counterclockwise vorticity then the opposite right segment has a clockwise vorticity.
} 
\end{figure}

\begin{figure}[htb]
\centerline{\scalebox{0.28}{\rotatebox{0.0}{\includegraphics{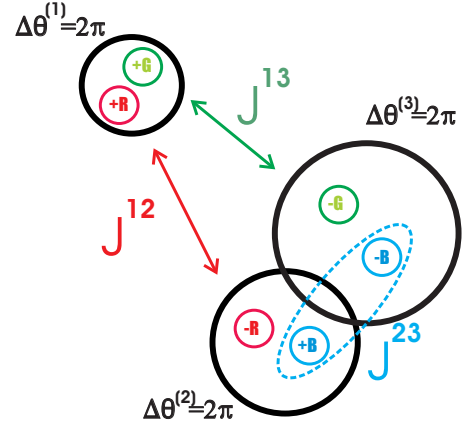}}}} 
\caption{ \label{colorcharges_32}  (Color online). A color charge representation for $N=3$ of the 
partial decomposition transition 
In the color charge representation, this 
may be viewed as a $\rm{red}$-$\rm{green}$ dielectric-metal transition, 
while the $\rm{blue}$ dielectric phase remains intact. As explained in the text, this 
partial 2-color metal-insulator transition involving fluctuating line charges, is in 
the \xy universality class. 
The dotted ellipse indicates which color is involved in forming the remaining 
dielectric phase.} 
\end{figure}

\begin{figure}[htb]
\centerline{\scalebox{0.28}{\rotatebox{0.0}{\includegraphics{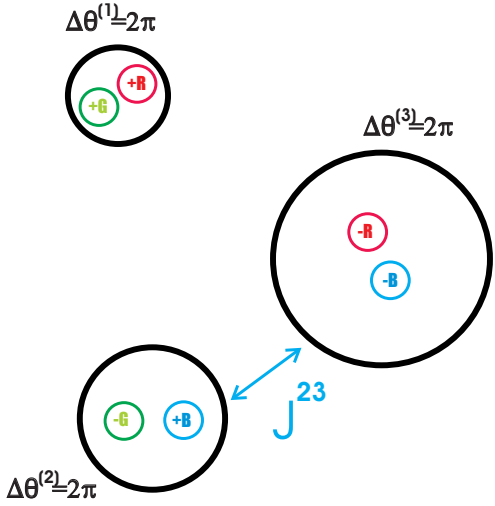}}}} 
\caption{ \label{colorcharges_33} (Color online). A color charge representation of the complete 
decomposition transition also illustrated in Fig. \ref{3spl_2}.  
In the color charge representation, this may be viewed as a $\rm{blue}$  
dielectric-metal transition, in the background of a $\rm{red}$-$\rm{green}$ metallic 
phase. As explained in the text, this complete ``1-color" metal-insulator transition 
involving fluctuating line charges is in the same universality class as the
partial decomposition transition, namely the \xy universality class. 
} 
\end{figure}

For the purposes of determining the universality class of this partial decomposition, 
{\it which is a genuine phase transition}, it may also be viewed as follows. The 
completely intact composite vortex is color charge neutral in the sense that it contains 
a positive and negative electric charge of each color. The fluctuation snapshot  in the 
left part of Fig. \ref{3spl_2} can be viewed as a completely intact composite vortex line 
with a small type-1 vortex loop superimposed on it as shown on Fig. \ref{superim}.
 The type-$1$ vortex 
loop which carries a red and green electric 
charges does not interact with the  composite vortex
which is color charge neutral. This can also
be seen by examining the vortex interaction matrix given in \eq{potential}. 
That is, when we add up the
neutral-mode-mediated interactions between the type-$1$ vortex
with all three  type-$1,2,3$ vortices in the composite vortex 
$(\Delta \theta^{(1)}=2\pi,\Delta \theta^{(2)}=2\pi, \Delta \theta^{(3)}=2\pi)$
{\it they add  to zero}. The situation in the right 
part of Fig. \ref{3spl_2} is topologically equivalent
to a completely intact composite vortex 
line superimposed with one segment of an {\it unbound} vortex loop.
{\it Therefore the transition may be viewed as an Onsager vortex loop proliferation 
transition \cite{Onsager1949,tesanovic1999,nguyen1999}
 of type-$1$ vortex loops in the background of a 
color charge neutral composite vortex lattice \cite{BSA,SSBS}.} 
That is because type-$1$ vortex loops, from the point of view of superfluid modes, 
cannot ``see'' color charge neutral vortex lines, so this is equivalent to a type-$1$ 
vortex-loop proliferation transition in  the complete absence of a composite
color charge neutral vortex lattice. 
Since these vortices excite neutral modes, this transition, which is the first stage in 
decomposing a color charge neutral composite  vortex line, is a vortex
loop proliferation 
transition in the \xy universality class \cite{BSA,SSBS}. 
In the color charge representation given in Fig. \ref{colorcharges_32},
this also means that the first-stage partial decomposition transition of three 
charged fluctuating {\it line objects} with different charges (but such that the 
algebraic sum of their charges add up to zero) is also a $2$-color metal-insulator
phase transition in the \xy universality class, {\it involving flexible color
line charges}.

The usefulness of the color charge representation becomes particularly clear when we 
go on to describe the second stage of the decomposition transition, illustrated in 
Fig. \ref{3spl_2}. This  transition may be viewed as a proliferation of type-$2$ vortex 
loops in the background of liberated type-$1$ vortex loops, all superimposed on a background
composite vortex lattice. The situation therefore is more complicated than in the first 
stage illustrated in Fig. \ref{3spl_2}, since that was a proliferation of type-$1$ loops 
in  vacuum.

However, let us view this transition in the color charge picture, illustrated in going from 
Fig. \ref{colorcharges_32} to Fig. \ref{colorcharges_33}. This is a metal-insulator transition 
for the $\rm{blue}$-charge sector in the background of coexistent $\rm{red}$ and $\rm{green}$ 
metallic phases. 
However, $\rm{red}$ and $\rm{green}$ charges cannot screen $\rm{blue}$ 
charges, while these charges eliminate the neutral modes associated with them 
(that is the ones with bare stiffnesses $J^{12}$ and $J^{13}$). Therefore, this is 
a metal-insulator transition for the sector of the $\rm{blue}$ charges, while $\rm{red}$ 
and $\rm{green}$ charges screen themselves and do not affect this transition.

In the color charge representation given in Fig. \ref{colorcharges_33}, this also means that 
the second-stage decomposition transition depicted in Fig. \ref{3spl_2} is a $1$-color 
metal-insulator transition in the \xy universality class, involving flexible 
color charged strings.

\subsection{Decomposition transitions, $N=4$} 

\begin{figure}[htb]
\centerline{\scalebox{0.28}{\rotatebox{0.0}{\includegraphics{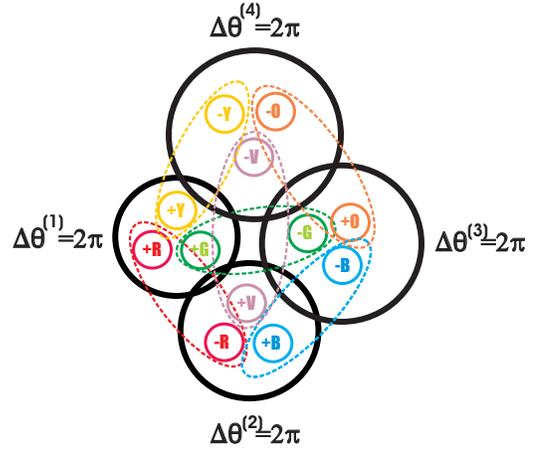}}}} 
\caption{ \label{colorcharges_41}  (Color online). 
A color charge representation for $N=4$ of the composite vortex lattice phase. 
This low-temperature phase, where the fluctuations of the composite vortex lines only 
involve small excursions of each of the constituent vortex lines away from the
main composite object, may be viewed as a ``6-color dielectric" phase. Moreover,
for $N=4$, each type-$\alpha$ vortex in the $N=4$ case is a bound state of $3$ color charges.
The dotted ellipses indicate which colors are involved in forming the dielectric phase.}
\end{figure}

\begin{figure}[htb]
\centerline{\scalebox{0.28}{\rotatebox{0.0}{\includegraphics{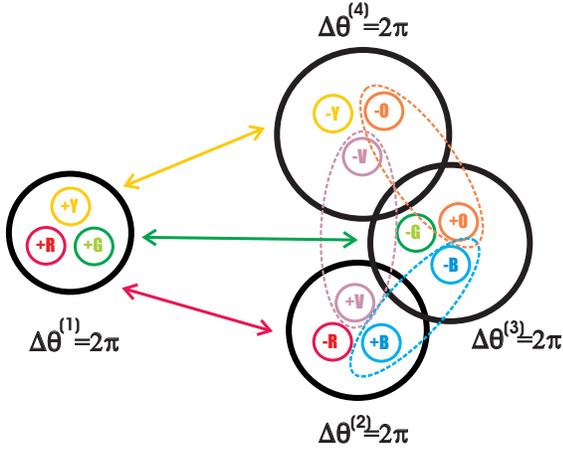}}}} 
\caption{ \label{colorcharges_42}  (Color online). A color charge representation for $N=4$ of the 
first partial decomposition transition. In the color charge representation, 
this may be viewed as a $\rm{red}$-$\rm{green}$-$\rm{yellow}$  dielectric-metal transition, 
while the $\rm{violet}$-$\rm{blue}$-$\rm{orange}$ dielectric phase remains intact. As explained 
in the text, this partial 3-color metal-insulator transition involving fluctuating line charges, 
is in the \xy universality class. The arrows indicate which three colors are involved in the
metal-insulator transition. The dotted ellipses indicate which colors are involved in forming 
the remaining dielectric phase.} 
\end{figure}

\begin{figure}[htb]
\centerline{\scalebox{0.28}{\rotatebox{0.0}{\includegraphics{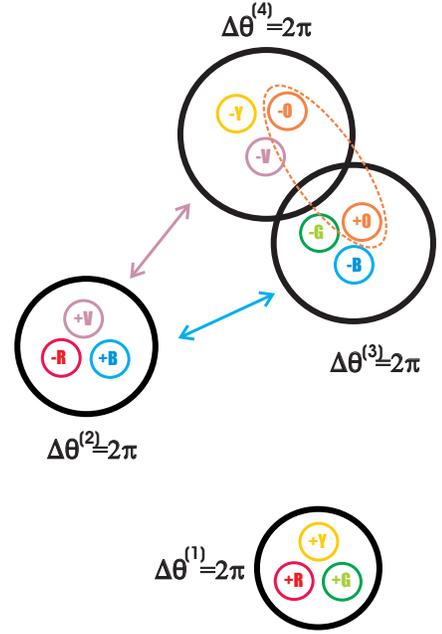}}}} 
\caption{ \label{colorcharges_43}  (Color online). A color charge representation for $N=4$ of the 
second partial decomposition transition. In the color charge representation, 
this may be viewed as a $\rm{violet}$-$\rm{blue}$  dielectric-metal transition, 
while the $\rm{orange}$ dielectric phase remains intact. As explained in the text, this 
partial 2-color metal-insulator transition involving fluctuating line charges, is in 
the \xy universality class. The arrows indicate which two colors are involved in the
metal-insulator transition. The dotted ellipse indicates which color is involved in 
forming the remaining dielectric phase.} 
\end{figure}

\begin{figure}[htb]
\centerline{\scalebox{0.28}{\rotatebox{0.0}{\includegraphics{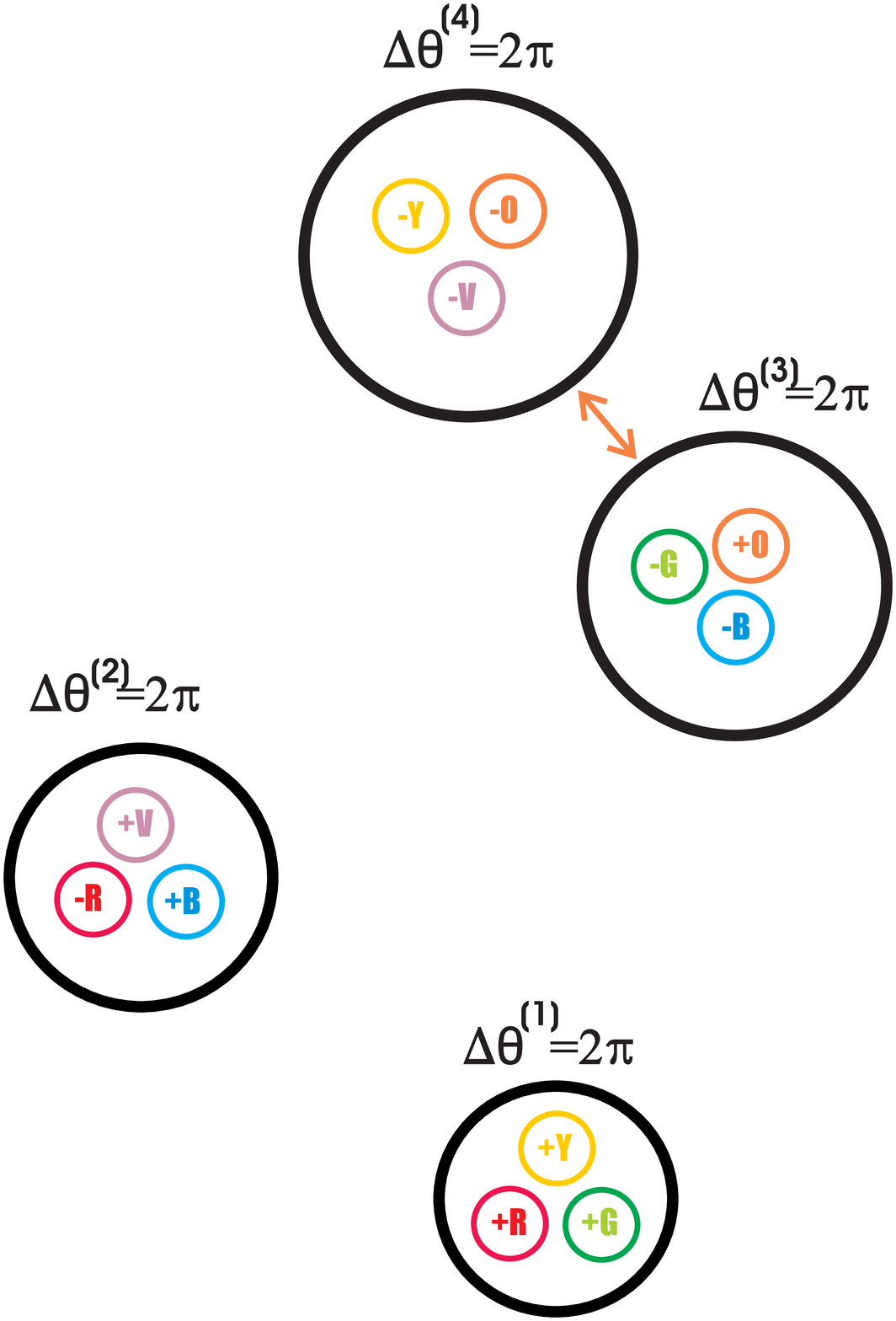}}}} 
\caption{ \label{colorcharges_44}  (Color online). A color charge representation for $N=4$ of the 
third and complete  decomposition transition. In the color charge 
representation, this may be viewed as an $\rm{orange}$  dielectric-metal transition, 
while the color-dielectric phase is completely destroyed. As explained in the text, this 
partial 1-color metal-insulator transition involving fluctuating line charges, is in 
the \xy universality class. The arrow indicates which  color is involved in the
metal-insulator transition.} 
\end{figure}

In Figs. \ref{colorcharges_41}, \ref{colorcharges_42}, \ref{colorcharges_43}, and 
\ref{colorcharges_44} we illustrate the partial decomposition
transitions for the case $N=4$ in the color charge picture. That is, we  introduce 
as in the case $N=3$ case  ``$\pm$ red", ``$\pm$ green" and ``$\pm$ blue"  charges 
associated with $\pm 2\pi$ windings in $(\theta^{(1)}-\theta^{(2)})$, 
$(\theta^{(1)}-\theta^{(3)})$ and $(\theta^{(2)}-\theta^{(3)})$, respectively. 
In addition, we introduce ``$\pm$ yellow", ``$\pm$ violet", and ``$\pm$ orange" 
charges associated with $\pm 2\pi$ windings in $(\theta^{(1)}-\theta^{(4)})$, 
$(\theta^{(2)}-\theta^{(4)})$ and $(\theta^{(3)}-\theta^{(4)})$,
respectively (see Tab. \ref{tab:colors_n4}).
Therefore, the case $N=4$ features one new aspect which was absent in the case $N=3$, 
namely that for $N=4$ we need more color charges than number of order parameter
components in order to completely cover all the possible ways that a 
neutral mode $(\theta^{(\alpha)}-\theta^{(\beta)})$ can be excited.   
Each type-$\alpha$ vortex is a bound state of three color charges,
as indicated in Fig. \ref{colorcharges_41}.

The low-temperature phase is a $4$-composite color charge neutral vortex system, which
may alternatively be viewed as a $6$-color dielectric phase, Fig. \ref{colorcharges_41}. 
The first-stage partial decomposition involves a type-$1$ vortex tearing itself off the
$2$-composite vortex, i.e. a  vortex loop proliferation of type-$1$ loops carrying 
$+\rm{red}, +\rm{green},+\rm{yellow}$ color charges in the background
of color charge neutral 
objects. So this is a phase transition in the \xy universality class. It may 
alternatively be viewed as a $3$-color metal insulator transition in the
$\rm{red},\rm{green},\rm{yellow}$ line-charge sectors, Fig. \ref{colorcharges_42},
leaving a $3$-color ($\rm{violet,blue,orange}$) dielectric phase. 
The second-stage decomposition process involves a type-$2$ vortex tearing itself off 
a $3$-composite vortex in the background of a system of proliferated  type-$1$ loops. 
Due to screening of $\rm{red},\rm{green}$ and $\rm{yellow}$ charges, this transition 
may be viewed in a simplified manner. It may be considered as the first-stage 
decomposition in a $N=3$ system consisting of type-$2$, type-$3$, and type-$4$ 
vortices, involving $\rm{violet}, \rm{blue}$, and $\rm{orange}$ charges. This 
we have already argued is a \xy (type-$2$) vortex loop proliferation transition, 
when we considered the $N=3$ case. Alternatively, it may be viewed as a $2$-color 
metal-insulator transition in the $\rm{violet}$ and $\rm{blue}$ line-charge sectors, 
Fig. \ref{colorcharges_43}, leaving a $1$-color ($\rm{orange}$) dielectric phase. 
The third-stage decomposition may be viewed as a type-$3$ vortex loop proliferation 
in the background of type-$1$ and type-$2$ 
proliferated vortex lines. Due to screening of $\rm{violet}$ and $\rm{blue}$ charges 
this may be viewed in a simplified manner. It may be considered as a vortex loop 
proliferation of loops carrying $\rm{orange}$ charges in the background  of an 
$\rm{orange}$-neutral vortex lattice, or vacuum. This is a vortex loop 
proliferation in the \xy universality class \cite{BSA,SSBS}. Alternatively, it may be viewed  
as a $1$-color ($\rm{orange}$) metal-insulator transition, (see Fig. \ref{colorcharges_44}),
leaving the $6$-color dielectric phase completely destroyed. 

\begin{table}
\caption{\label{tab:colors_n4} A color charge is defined as a $ \pm 2\pi$
  winding in the mode $(\theta^{(\alpha)}-\theta^{(\eta)})$ with the
  following mapping for $N=4$. }
\begin{ruledtabular}
\begin{tabular}{rccc}
  Mode: & $(\theta^{(1)}-\theta^{(2)})$ &
  $(\theta^{(1)}-\theta^{(3)})$ & $(\theta^{(2)}-\theta^{(3)})$ \\
  Color charge:        & red  & green  &  blue \\
\hline
  Mode: & $(\theta^{(1)}-\theta^{(4)})$ & $(\theta^{(2)}-\theta^{(4)})$ & $(\theta^{(3)}-\theta^{(4)})$\\
  Color charge: & yellow & violet & orange \\
\end{tabular}
\end{ruledtabular}
\end{table}

\subsection{General $N$ }
In the general $N$-component case, the number of color charges that needs to be introduced
to give an equivalent description as the above, may be counted as follows, starting from the 
third term in Eq. (\ref{charge_neutral}). Each combination $\theta^{(\alpha)}-\theta^{(\beta)}$ 
is given a color. We start with one phase, the one with the lowest bare stiffness say $\theta^{(1)}$, 
and 
introduce a non-trivial phase winding $\pm 2 \pi$ in this phase. This excites $N-1$ neutral 
modes since $N-1$ gauge-invariant phase
differences which involve $\theta^{(1)}$ can be formed. Introducing non-trivial 
phase-windings in the next phase $\theta^{(2)}$, the one with the next-to-lowest 
bare stiffness say,  will also excite $N-1$ neutral modes, but only $N-2$ {\it new} neutral 
modes. Non-trivial phase-windings in the third phase 
$\theta^{(3)}$ will excite $N-3$ new neutral modes, and 
so on.  The number of different colors $N_{\rm color}$ we will have to introduce 
for a theory with $N$ flavors of scalar fields is therefore given by 
$N_{\rm color}(N) = (N-1) +(N-2)+(N-3) + \dots + 2 + 1 = N(N-1)/2$,
i.e.$N_{\rm color}(2) = 1$, $N_{\rm color}(3) = 3$, 
$N_{\rm color}(4) = 6
$, and $N_{\rm color}(5)=10$. A completely composite vortex, which we denote as an $N$-composite 
vortex, consists of $N$ constituent vortices originating in nontrivial phase windings in each of the 
individual phases $\theta^{(\alpha)}, \alpha \in [1,\dots,N]$. Since a non-trivial phase winding in  any phase 
$\theta^{(\eta)}$, $\eta \in [1,\dots,N]$, excites $N-1$ neutral modes $\theta^{(\eta)} - \theta^{(\alpha)}$, 
it is clear that a  type-$\eta$ vortex may be viewed as a bound state of $N-1$ 
color charges. The particular combination of $N-1$ color charges, out of the 
total collection of $N(N-1)/2$ color charges, that will enter the $N-1$-body bound state
in each vortex, will depend on $\eta$. {\it The  $N$-composite vortex is a 
color charge neutral object.} 

Small fluctuations in the $N$-composite vortex may therefore be viewed
 as a dielectric insulating phase of an $N(N-1)/2$-component
dielectric. The first stage in the $N-1$ stage decomposition process of the $N$-composite 
vortex line, where a type-$1$ vortex tears itself off the $N$-composite vortex, 
is therefore a metal-insulator transition where $N-1$ color charges of the $N(N-1)/2$-colors 
dielectric system {\it simultaneously} undergo a metal insulator transition, in the 
\xy universality class. The next stage, where a type-$2$  vortex tears itself off the 
remaining $N-1$-composite vortex in the background of a system of proliferated type-$1$ 
loops, is phase where $N-2$ color charges {\it simultaneously} undergo a metal insulator 
transition in the \xy universality class by the same argument as used for the $N=3$ case, 
and so on. The complete decomposition of the $N$-composite vortex proceeds in $N-1$ steps of 
metal-insulator transitions for color charges, where step number  $1 \leq {\cal N} \leq N-1$ 
may be viewed as either a type-${\cal N}$ vortex tearing itself off an $N-{\cal N}-1$-composite 
vortex line, or equivalently a {\it simultaneous} metal-insulator transition  for $N-{\cal N}$ 
new color charges that have not been involved in the previous ${\cal N}-1$ metal-insulator 
transitions. All the $N-1$ partial decomposition transitions, or metal-insulator transitions 
for color charges, are in the \xy universality class. 

We should emphasize that, as follows from  Eq. (\ref{charge_neutral}) in the limit  
$N\to\infty$, the strength of each of the electric charges goes to zero. At the same time 
the number of colors of electric charges $N_c$ tends to infinity.
From Eq. (\ref{charge_neutral}), it follows that even in the limit $N\to\infty$,
the energy binding of a type-$\alpha$ vortex to a color charge neutral composite vortex 
is finite, even though the strength of each individual color charge tends to zero.

\subsection{Graphical representation of phase disordering transitions
  in the $N=3$ model.} 
\begin{figure}[htb]
\centerline{\scalebox{0.20}{\rotatebox{0.0}{\includegraphics{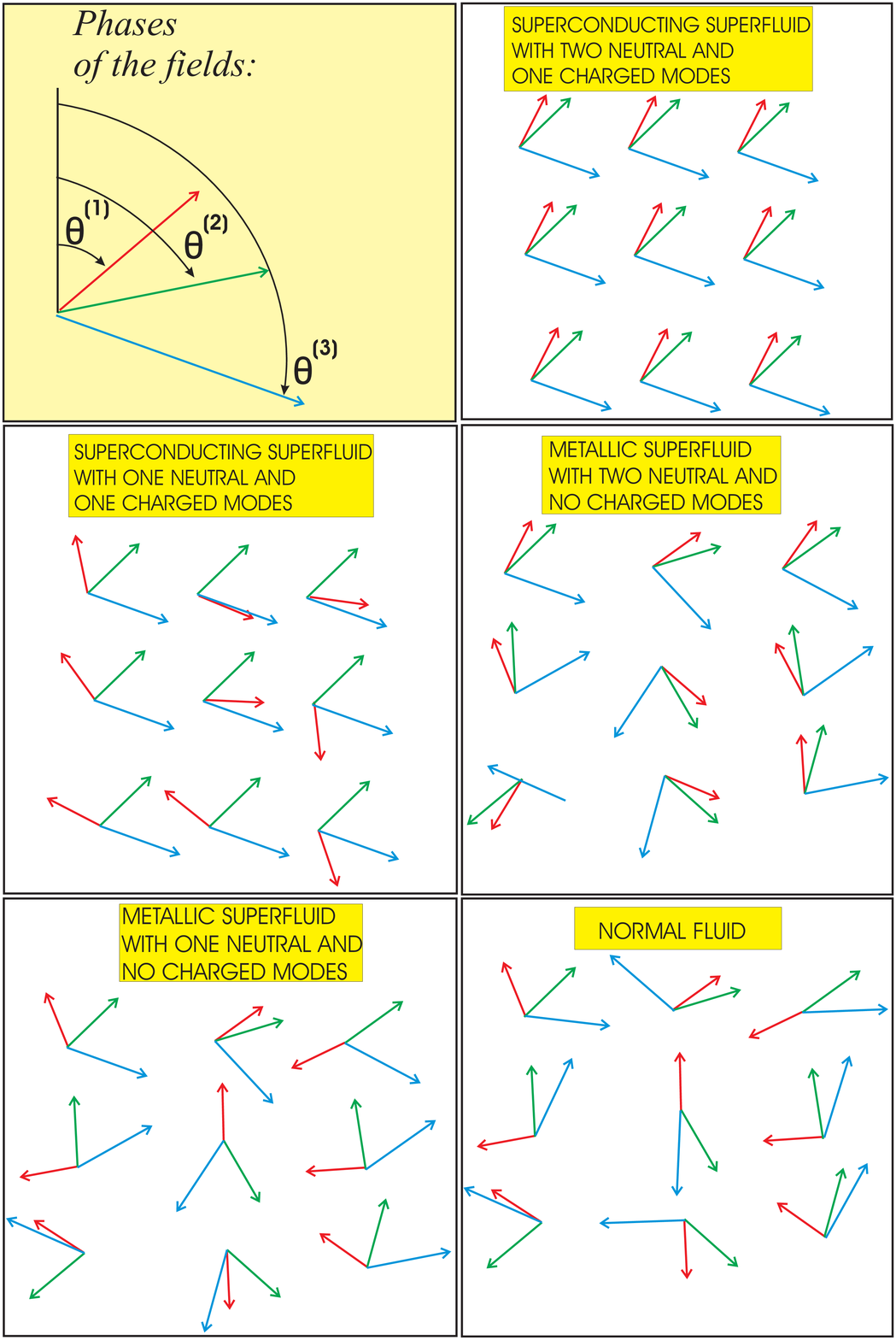}}}} 
\caption{ \label{ph1} (Color online). A schematic plot of states in the $N=3$ system. The upper left panel 
shows the phases 
of the condensate order parameters that are involved. The upper right panel shows the state where 
all three phases $\theta^{(1)}$,  $\theta^{(2)}$, and $\theta^{(3)}$ are ordered individually. This 
state is the low-temperature (ground) state and features one superconducting charged mode and two 
superfluid neutral modes. The middle left panel is a phase where $\theta^{(1)}$ is disordered, while 
$\theta^{(2)}$ and $\theta^{(3)}$ are ordered. Thus, this is a state which features one charged 
superconducting mode and one neutral superfluid mode. The middle right panel illustrates a state where 
all of the phases $\theta^{(1)}$,  $\theta^{(2)}$, and $\theta^{(3)}$ are individually disordered. However,
the differences $\theta^{(1)} - \theta^{(2)}$ and $\theta^{(2)} -\theta^{(3)}$ (and therefore also 
$\theta^{(1)} - \theta^{(3)}$) feature long-range order. This is therefore a state which is normal 
metallic, but nevertheless features two neutral superfluid modes. The bottom left panel illustrates 
a state where all of the phases $\theta^{(1)}$,  $\theta^{(2)}$, and $\theta^{(3)}$ are individually 
disordered. Only the phase difference  $\theta^{(2)} - \theta^{(3)}$ exhibits long-range order. This
is a normal metallic state featuring one neutral superfluid mode. The bottom right panel illustrates
a state where all of the phases $\theta^{(1)}$,  $\theta^{(2)}$, and $\theta^{(3)}$ are individually 
disordered and where none of the phase differences $\theta^{(1)} - \theta^{(2)}$ and $\theta^{(2)}-\theta^{(3)}$ 
and  $\theta^{(1)} -\theta^{(2)}$ feature long-range order. This is therefore a state which is normal 
metallic and normal fluid (no neutral superfluid modes). The states illustrated in the upper left, 
middle left, and lower right panel exist at zero as well as finite magnetic fields. The states 
illustrated in the middle right and lower left panels only exist at finite magnetic fields. } 
\end{figure}

In Fig.  \ref{ph1}, we illustrate graphically the various  phase disordering transitions and 
partial symmetry restorations discussed in the previous section.

\section{Applications}

In this Section, we will briefly mention some possible applications of the results obtained
in this paper. Emphasis will be on the case $N=2$, but some results for the cases
$N=3$ and even $N=4$ may find applications in  mixtures of
superconducting condensates in the not too distant future.  

\subsection{Applications of results for $N=2$}

The  results  for $N=2$ are expected to apply to two-component superconductivity which could be 
achieved in  metallic states of light atoms\cite{Ashcroft1999,Neilnew}, such as electronic 
and protonic condensates in liquid metallic hydrogen (LMH) under extreme pressure. Estimates 
exist for $T_{\rm c2}$ for such systems, $T_{\rm c2} \approx 160 {\rm K}$ 
\cite{Ashcroft1999}. A rough estimate for $T_{\rm c1}$ follows from the mass ratio of the electronic 
and protonic condensate, $T_{\rm c1} \approx 0.1 {\rm K}$. Hence,  at $T_{\rm c1}$ one should 
observe an extra low-temperature \xy specific heat anomaly, as well as an anomaly in the 
London penetration depth. An even more promising candidate is the the system $CH_x$  
\cite{Neilnew} where there are predictions of liquid metallic states at considerably lower 
pressures than those required to achieve LMH. In such multi-flavor 
superconductors it is the case $|\psi^{(1)}| \neq |\psi^{(2)}|$ which appears most 
naturally \cite{smiseth2004}. 

Here, it is appropriate to remark briefly on the microscopic origins of
superconductivity
in the projected liquid metallic phase of hydrogen
 \cite{Ashcroft1999,Neilnew}.
The proton is four times lighter than a ${}^4He$ atom.
It is well known that ${}^4He$
at normal conditions
is a classic {\it permanent liquid}, because of high zero-point energy
and  weak ordering energies.
Indeed zero-point energies of
protons in a dense environment
are also high, and at increasing compression
there is a shift of electron density from intra-molecular
regions to inter-molecular,
and with it a progressive decline in the effective inter-proton
attractions.
Because of this there is also a decline of ordering energies from
interactions relative to protonic zero-point energies.
The existence of {\it a melting point maximum}
as a function of pressure
in hydrogen, as well as
 a range of densities where hydrogen may take up \textit{a fluid
   phase} in its ground state  was
suggested in Ref. \onlinecite{Ashcroft1999}.
Another important circumstance
is that en route hydrogen should undergo  an insulator-metal
transition and therefore the resulting phase
should be the liquid metallic hydrogen, a
translationally invariant two-component fermionic liquid. There is
preliminary experimental evidence
that a melting point maximum may indeed exist \cite{Datchi2000} and it
has received recent powerful
backing in {\it ab initio}
calculations \cite{Bonev}. Experimentally a 12.4 fold compression of
hydrogen has already been achieved at
around 320 ${\rm GPa}$ \cite{Datchi2000,Bonev}. Estimates suggest that
LMH should appear at 13.6 fold
compression at pressure in the
vicinity of 400 ${\rm GPa}$  \cite{Bonev}, whereas hydrogen alloys may
exhibit metallic behavior at significantly
lower pressures \cite{Neilnew}. A predicted key feature of LMH at low
temperature is the {coexistence
of superconductivity of proton-proton and electron-electron Cooper
pairs} \cite{Ashcroft1999}.

The special point $|\psi^{(1)}| = |\psi^{(2)}|$  has a physical realization when Eq. (\ref{gl_action}) 
is viewed as an effective  field theory of a quantum antiferromagnet with easy plane anisotropy, which 
facilitates a suppression of topological defects in the form of "hedgehog"-configurations which appear 
in $O(3)$-symmetric models \cite{motrunich2004,sachdev2004}. (More generally, it may be viewed as a 
field theory of an $O(3)$ model where "hedgehogs" are suppressed by some unspecified mechanism, not 
necessarily limited to easy-plane anisotropy). 

We conclude this subsection on the $N=2$ case with a remark on how these results relate to 
multi-flavor {\it electronic} condensates. To describe this case, we need to include a 
Josephson coupling between the matter field species. The details of how to give a vortex 
representation of the $N$-flavor London model in the presence of inter-flavor Josephson 
coupling, is given in Appendix \ref{App:josephson}. Had a Josephson coupling term 
between condensate species been introduced in the theory for $N \geq 2$, this would 
have have altered the dual theory in a completely non-perturbative way, and would tend 
to lock the phases of individual condensate fields to each other. (For a dual representation 
of this case, where the non-perturbative character of the Josephson coupling is brought 
out in a particularly clear way, see Appendix \ref{App:josephson}). As a result, the transitions 
we describe here would collapse to one, namely the charged Higgs fixed point, which is
in the inverted \xy universality class.

\subsection{Applications of results for $N=3,4$}

Mixtures of superconducting condensates in LMH can be extended to include  also the 
hydrogen-isotopes deuterium and tritium \cite{thanksneil}. Tritium is a $S=1/2$ fermion, so this 
may give 
rise to a superconducting condensate via forming spin-singlet Cooper pairs, just as 
in the protonic case. Hence, our results for $N=3$ could be applicable to to the mixtures of 
liquid metallic hydrogen-tritium at extremely high pressures. Another possibility is to 
include  deuterium as a new component. Including  deuterium as a new component in 
addition to hydrogen means that we have an $N=3$-component mixture of superconducting 
condensates consisting of electrons, protons, 
and deuterons (deuterium nuclei), all of which in principle can undergo a metal-superconductor 
transition. Compared to the situation for $N=2$, the situation is complicated by at least 
two circumstances. Firstly, deuterons are bosons, and secondly they have spin $S=1$. This 
means that the electrons and protons become superconducting via forming Cooper pairs, while 
the deuterons undergo a metal-superconducting transition via Bose-Einstein Condensation.
Extending this to the case of having {both} tritium and deuterium in addition to
hydrogen, might provide a realization of the case $N=4$.


\section{\label{Concl} Summary }
We have  analyzed  the $N$-flavor London superconductor model
coupled to one gauge field with no Josephson coupling between the
matter field components. The dual theory is an $N$-flavor GL theory coupled to $N$
dual gauge fields, where the sum of all dual gauge fields is massive at all
couplings. There are $N-1$
charge-neutral superfluid modes and one charged superconducting mode
in this model. We have given a prescription for how to identify the
$N-1$ neutral modes for arbitrary  $N$.

For $N=2$, a case which should apply to a superconducting state of
liquid metallic hydrogen, as well as for the case $N=3$, we have
performed large scale MC simulations computing \textit{i)} critical
exponents $\alpha$ and $\nu$, \textit{ii)} gauge field and dual gauge
field correlators, \textit{iii)} the corresponding masses, and
\textit{iv)} critical couplings using FSS. For $N=2$ and $|\psi^{(1)}|\neq |\psi^{(2)}|$, 
we  find one low-temperature critical point in the \xy universality class at $T_{\rm c 1}$, 
and one critical point in the inverted \xy universality class at $T_{\rm c2} > T_{\rm c1}$.
For $N=2$ with $|\psi^{(1)}| = |\psi^{(2)}|$ we find one critical point with non-\xy 
values of $\alpha$ and $\nu$ \cite{motrunich2004}. We propose these to be critical exponents in the universality
class of a self-dual gauge theory. For $N=3$ and all $|\psi^{(\alpha)}|$ unequal, we find two  fixed points
in the \xy universality class at $T_{\rm c1}$ and $T_{\rm c2}$, and one
fixed point in the inverted \xy universality class at $T_{\rm c3} (> T_{\rm c2} >
T_{\rm c1})$. All critical points therefore exhibit \xy values
of $\alpha$ and $\nu$ when all bare phase stiffnesses
$|\psi^{(\alpha)}|$ are different. In the case $N=3$ with
$|\psi^{(1)}| = |\psi^{(2)}| < |\psi^{(3)}|$ we find two critical
points. The critical point at $T_{\rm c1} < T_{\rm c2}$ is found to be
in the \xy universality class, and the critical point at $T_{\rm c2}$
is in the inverted \xy universality class. For the case $N=3$ with
equal phase stiffnesses we find one critical point which is non-\xy.

In this context, we have also noted that collapsing two neutral critical points 
in the \xy universality class leads to a single critical point also in the 
\xy universality class. This follows from an argument implying that collapsing
any number of neutral critical points in the \xy universality class leads to a 
single critical point also in the \xy universality class. On the other hand, it 
appears that collapsing $N-1$ neutral critical points in the \xy universality 
class {\it and one charged fixed point in the inverted} \xy universality class 
leads to a single critical point in a universality class (which in principle 
could depend on $N$) which is not that of the \xy or inverted \xy type. For 
$N=2$, we may define the universality class as that of a 3D self-dual 
$U(1) \times U(1)$ 
gauge theory, while it is less clear what it is for other $N \geq 3$. The 
numerical values we have obtained for the critical exponents $\alpha$ and 
$\nu$ the two cases {\it{i)}} $N=2$, $|\psi^{(1)}| = |\psi^{(2)}|$ and 
{\it{ii)}} $N=3$, $|\psi^{(1)}| = |\psi^{(2)}| = |\psi^{(3)}|$ are 
remarkably similar, indicating that the values of the critical exponents 
are at most weakly dependent on $N$.

In an external magnetic field at low temperature, the ground state of
the system is an Abrikosov vortex lattice of composite vortices. 
However the effect of thermal fluctuations alters the physics significantly.
We discuss in detail that 
in the low-field regime and when the  bare stiffnesses of the condensates all
differ, we find that a \xy vortex sub-lattice melting transition takes place. Upon thermal
decomposition of field induced composite vortex lines, the
constituent vortices originating in the condensates with lowest bare stiffness disorder,
while the ones originating in the stiffer condensates remain arranged in an Abrikosov
vortex lattice. When  such a transition occurs, an $N=2$ system looses superfluid
properties, but remains superconducting. 
In contrast at high magnetic fields, the neutral mode 
disappears via the melting transition
of the lattice of composite vortices. 
This is  a transition from the superconducting state to a non-superconducting metallic superfluid state \cite{BSA}.
Inside this novel metallic superfluid phase, at a temperature much lower
than the zero-field metal-superconductor transition, we find  a
superfluid-normal fluid \xy
phase transition associated with a neutral mode-driven proliferation of
vortex loops nucleating
on field-induced vortex lines.
These various transitions  should in principle be detectable in flux-noise
experiments.
We extend the discussion for $N>2$, where we find phase transitions
associated with partial decomposition of composite vortices, yielding
several unusual states
of partial symmetry breakdown.
The universality class and partial symmetry breakdowns
are identified by mapping the system 
to an ensemble of electrically charged strings 
where for the $N$-component system there are $N(N-1)/2$ 
replicas of electric charges of different color.

The sublattice melting and partial decomposition transitions are 
of purely topological origin.
It involves what can be viewed as vortices and anti-vortices (positively 
and negatively charged objects). The existence of positively and negatively 
charged objects implies that the physics conceptually in some sense is 
similar to what occurs in a Kosterlitz-Thouless (KT) transition in two 
dimensions. In spite of being a decomposition of positively and negatively 
charged composite objects, {\it such a transition can  be mapped onto a 
proliferation of vortex loops in a vacuum}, {\it i.e} a phase transition 
in the \xy universality class. Another principal difference between this 
type of phase transition compared to the KT transition, is that in the 2D 
KT transition, vortices and anti-vortices (positively and negatively charged 
objects) are thermally generated. This can not happen in three dimensions
since any  vortex line  in 3D has an infinite energy in an infinite sample.
In the 3D transition which we consider, the neutral bound states of
the charged objects are introduced by an external magnetic field.
Thus we deal with a novel type of transition which in a very unusual 
form involves concepts both from two- and three- dimensional physics.

\begin{acknowledgments}
This work was funded by the Norwegian University of Science and Technology through the 
Norwegian High Performance Computing Program, by the Research Council of Norway, Grant
Nos. 157798/432, 158518/431, and 158547/431 (NANOMAT), by STINT and the Swedish Research 
Council, and by the US National Science Foundation DMR-0302347. We acknowledge helpful 
discussions and communications with  K. B\o rkje, H. Kleinert, O. Motrunich, F. S. Nogueira, 
S. Sachdev, Z. Tesanovic, and A. Vishwanath. E. B. thanks L. D. Faddeev and A. J. Niemi
for many discussions of multigap models. A. S. thanks H. Kleinert and the Institut f{\"u}r 
Theoretische Physik der Freie Universit{\"a}t Berlin, for hospitality while part of this 
work was being completed. In  particular, we wish to thank N. W. Ashcroft for many 
enlightening discussions and for collaboration on Ref. \onlinecite{BSA}. 
\end{acknowledgments}

\appendix

\begin{widetext}
\section{\label{App:charged_neutral} Identifying charged and neutral modes}
Here, we illustrate in detail the separation of variables in $N$-component London model and the 
extraction of the composite charged and neutral modes in the London limit. The general procedure 
beyond the London limit for the case $N=2$ can be found in \cite{egor2002}. The general 
$N$-component Ginzburg-Landau model is given by the action
\begin{equation}
S = \sum_{\ve r} \left \{ \sum_{\alpha=1}^N \frac{1}{2}{|(\nabla -\im
e\ve{A})\psi^{(\alpha)}|^2}
+ V(\{\psi^{(\alpha)}\}) + \frac{1}{2}(\nabla\times\ve{A})^2 \right \}.
\label{s1}
\end{equation}
Here, the masses $M^{(\alpha)}$  have been absorbed in the amplitudes $|\psi^{(\alpha)}|$ 
for notational simplicity.  Let us rewrite the first term in Eq. (\ref{s1}) as follows
\begin{eqnarray}
 \sum_{\alpha=1}^N \frac{1}{2}{|(\nabla -\im
e\ve{A})\psi^{(\alpha)}|^2}
& = & \sum_{\alpha=1}^N  \frac{1}{2}
|\psi^{(\alpha)}|^2(\nabla \theta^{(\alpha)})^2
-e\ve{A}\cdot(\nabla\theta^{(\alpha)})|\psi^{(\alpha)}|^2 
+ \frac{1}{2}e^2\ve{A}^2|\psi^{(\alpha)}|^2.
\label{s3}
\end{eqnarray}
The idea now is to first extract the charged mode of the system, and then identify the remaining terms 
as the neutral modes. The charged modes is the (only) linear combination of phase gradients
$\nabla \theta^{(\alpha)}$ that  couples to the gauge field $\ve A$.  We first introduce the quantity
\begin{equation}
\ve \Theta \equiv  \sum_{\alpha = 1}^N |\psi^{(\alpha)}|^2 \nabla \theta^{(\alpha)}.
\end{equation}
Using this, we form the combination
\begin{eqnarray}
\left( \sum_{\alpha=1}^N |\psi^{(\alpha)}|^2 \left[ \nabla \theta^{(\alpha)} - e \ve A \right] \right)^2
= (\ve \Theta - e \Psi^2 \ve A)^2,
\end{eqnarray}
where $\Psi^2$ is defined in \eq{defpsi2}. One can check that this combination is  gauge-invariant. By adding and subtracting ${\ve \Theta}^2/{2 \Psi^2}$, 
we now express Eq. (\ref{s3}) as follows
\begin{eqnarray}
&&\sum_{\alpha=1}^N  \frac{1}{2}
|\psi^{(\alpha)}|^2(\nabla \theta^{(\alpha)})^2
-e\ve{A}\cdot(\nabla\theta^{(\alpha)})|\psi^{(\alpha)}|^2 
+ \frac{1}{2} e^2\ve{A}^2|\psi^{(\alpha)}|^2 = \nonumber \\ &&
\frac{1}{2} 
\left[
\sum_{\alpha=1}^N 
|\psi^{(\alpha)}|^2(\nabla \theta^{(\alpha)})^2  - \frac{\ve \Theta^2}{\Psi^2} 
\right]  + \frac{1}{2 \Psi^2} (\ve \Theta - e \Psi^2 \ve A)^2.
\label{s4}
\end{eqnarray}
The last term is identified as the charged mode. The first term on the right hand side
of Eq. (\ref{s4}) can be rewritten as
follows
\begin{eqnarray}
\frac{1}{2} 
\left[
\sum_{\alpha=1}^N 
|\psi^{(\alpha)}|^2(\nabla \theta^{(\alpha)})^2  - \frac{\ve \Theta^2}{\Psi^2} 
\right]
& = & \frac{1}{2 \Psi^2} 
\left[
\sum_{\alpha,\beta=1}^N |\psi^{(\alpha)}|^2  |\psi^{(\beta)}|^2 
\nabla \theta^{(\alpha)} (\nabla \theta^{(\alpha)}- \nabla \theta^{(\beta)})
\right] \nonumber \\
& = & \frac{1}{4 \Psi^2} 
\left[
\sum_{\alpha,\beta=1}^N |\psi^{(\alpha)}|^2  |\psi^{(\beta)}|^2 
 (\nabla \theta^{(\alpha)}- \nabla \theta^{(\beta)})^2
\right].
\end{eqnarray}
Therefore, the action may be expressed as

\begin{eqnarray}
S = \sum_{\ve r} \left \{   \left( 
\frac{1}{2 \Psi^2}  \sum_{\alpha = 1}^N |\psi^{(\alpha)}|^2 \nabla \theta^{(\alpha)}  - e \Psi^2 \ve A 
\right)^2 + \frac{1}{2}(\nabla\times\ve{A})^2
+ \frac{1}{4 \Psi^2} \left[ \sum_{\alpha,\beta=1}^N |\psi^{(\alpha)}|^2  |\psi^{(\beta)}|^2 
 \left( \nabla (\theta^{(\alpha)}- \theta^{(\beta)}) \right)^2
\right]  \right \},
\label{action_separated}
\end{eqnarray}
which is now a sum of {\it one} charged mode and $N-1$  neutral modes.  The last term 
in Eq. (\ref{action_separated}) is a sum consisting of $N(N-1)/2$ terms,  involving all 
the color charges $ (\theta^{(\alpha)}- \theta^{(\beta)})$ (see Section \ref{N>2_extfield})  
which are needed to account for all the possible ways neutral modes can be excited in the 
system as a consequence of multiple connectedness of physical space in the presence
of vortices. Note how this term vanishes when $N=1$. 

\section{Derivation of Eqs. (\ref{vortex_action}) and (\ref{potential})}
\label{App:vortex}
Starting from the model in the Villain approximation Eq. (\ref{villain})
we linearize the kinetic energy terms by introducing $N$ auxiliary
fields $\ve v^{(\alpha)}$ where $\alpha \in 1,\dots,N$ with a
partition function
\begin{equation}
\label{lin}
\begin{split}
Z =& \int_{-\infty}^{\infty}\mathcal{D}\ve A\prod_{\gamma=1}^N\int_{-\pi}^{\pi}\mathcal{D}\theta^{(\gamma)}\int_{-\infty}^{\infty}\mathcal{D}\ve v^{(\gamma)}\sum_{\ \ve n^{(\gamma)}}\exp(-S), \\
S =& \sum_{\ve r}\left(\sum_{\alpha=1}^N\frac{1}{2\beta|\psi^{(\alpha)}|^2}(\ve v^{(\alpha)})^2 + \sum_{\alpha=1}^N\im(\Delta\theta^{(\alpha)} - e\ve{A} +2\pi \ve n^{(\alpha)})\cdot\ve v^{(\alpha)} +\frac{\beta}{2}(\Delta\times\ve{A})^2\right),
\end{split}
\end{equation}
\end{widetext}
where $|\psi^{(\alpha)}|^2 =|\Psi^{(\alpha)}_0|^2/ M^{(\alpha)}$. The
Poisson summation formula reads 
\begin{equation}
\label{poisson}
\sum_{n=-\infty}^{\infty}{\rm e}^{2\pi inB} = \sum_{m=-\infty}^\infty\delta(m-B).
\end{equation}
Here, $n,m \in \mathbb{Z}$ and $B \in \mathbb{R}$. We apply
Eq. (\ref{poisson}) to Eq. (\ref{lin}) and integrate out the integer fields
$\ve n^{(\alpha)}$ so that the fields $\ve v^{(\alpha)}$ take only integer
values which we denote $\ve{\hat v}^{(\alpha)}$. After a partial summation of $\sum_{\ve r}\sum_{\alpha=1}^N
\im\Delta\theta^{(\alpha)}\cdot\ve{\hat v}^{(\alpha)} = -\sum_{\ve r}\sum_{\alpha=1}^N
\im \theta^{(\alpha)}\Delta\cdot\ve{\hat v}^{(\alpha)}$, where the surface terms are omitted, we may 
integrate out the phase fields $\theta^{(\alpha)}$. This integration produces the local constraints
\begin{equation}
\Delta\cdot\ve{\hat v}^{(\alpha)} = 0.
\end{equation}
To fulfill this constraint we let $\ve{\hat v}^{(\alpha)}
= \Delta\times\ve{\hat h}^{(\alpha)}$ where $\ve{\hat h}^{(\alpha)}$
is an integer-valued vector field. At this point the theory reads 
\begin{widetext}
\begin{equation}
\begin{split}
\label{dualaction1}
Z =& \int_{-\infty}^{\infty}\mathcal{D}\ve A\prod_{\gamma=1}^N\sum_{\ \ve{\hat h}^{(\gamma)}}\exp(-S), \\
S =& \sum_{\ve r}\left[\sum_{\alpha=1}^N\frac{1}{2\beta|\psi^{(\alpha)}|^2}(\Delta\times \ve{\hat h}^{(\alpha)})^2 - \im e\ve{A}\cdot\left(\sum_{\alpha=1}^N\Delta\times\ve{\hat h}^{(\alpha)} \right)  +\frac{\beta}{2}(\Delta\times\ve{A})^2\right].
\end{split}
\end{equation}
Again, we apply the Poisson summation formula Eq. (\ref{poisson}) and the integer fields $\ve{\hat h}^{(\alpha)}$ 
are replaced by continuous dual gauge fields $\ve h^{(\alpha)}$ at the cost of introducing the
term $2\pi\im\sum_\alpha\ve h^{(\alpha)}\cdot\ve m^{(\alpha)}$ in the action and $\ve m^{(\alpha)}$ are 
integer vortex fields. Then the partition function reads
\begin{equation}
\begin{split}
\label{dualaction2}
Z =& \int_{-\infty}^{\infty}\mathcal{D}\ve A\prod_{\gamma=1}^N\int_{-\infty}^{\infty}\mathcal{D}\ve h^{(\gamma)}\sum_{\ \ve m^{(\gamma)}}\delta_{\Delta\cdot\ve m^{(\gamma)},0}\exp(-S), \\
S =& \sum_{\ve r}\left[\sum_{\alpha=1}^N\frac{1}{2\beta|\psi^{(\alpha)}|^2}(\Delta\times \ve h^{(\alpha)})^2 - \im e\ve{A}\cdot\left(\sum_{\alpha=1}^N\Delta\times\ve h^{(\alpha)} \right) + 2\pi\im\sum_{\alpha=1}^N\ve h^{(\alpha)}\cdot\ve m^{(\alpha)} +\frac{\beta}{2}(\Delta\times\ve{A})^2\right],
\end{split}
\end{equation}
where $\delta_{n,m}$ is the Kronecker delta. The gauge symmetry of the integer valued fields $\ve{\hat
  h}^{(\alpha)}$ in Eq. (\ref{dualaction1}) must be preserved through this transformation. Hence,
  the action Eq. (\ref{dualaction2}) must be invariant under the gauge transformation $\ve
  h^{(\alpha)} \to \ve h^{(\alpha)} + \Delta\chi$. The transformation
  produces the term $\sum_{\ve r}\sum_\alpha 2\pi\im\Delta\chi\cdot\ve
  m^{(\alpha)}$ which must be zero to preserve gauge
  symmetry. A partial summation therefore produces the  constraint
  $\Delta\cdot\ve m^{(\alpha)} = 0$ on each flavor of the vortex fields. 
At this point it is useful to write the theory in the Fourier
representation and the action becomes
\begin{equation}
\begin{split}
S =& \sum_{\ve q}\left\{\sum_{\alpha=1}^N\frac{1}{2\beta|\psi^{(\alpha)}|^2}(\ve Q_{\ve q}\times \ve h_{\ve q}^{(\alpha)})\cdot(\ve Q_{-\ve{q}}\times \ve h_{-\ve{q}}^{(\alpha)})  + \pi\im\sum_{\alpha=1}^N\left[\ve h^{(\alpha)}_{\ve q}\cdot\ve m^{(\alpha)}_{-\ve q} + \ve h^{(\alpha)}_{-\ve q}\cdot\ve m^{(\alpha)}_{\ve q}\right]  \right.
\\ &\left.- \frac{\im e}{2}\left[\ve{A_{\ve q}}\cdot\left(\sum_{\alpha=1}^N\ve Q_{-\ve q}\times\ve h^{(\alpha)}_{-\ve q} \right) + \ve{A_{-\ve q}}\cdot\left(\sum_{\alpha=1}^N\ve Q_{\ve q}\times\ve h^{(\alpha)}_{\ve q} \right)\right] + \frac{\beta}{2}(\ve Q_{\ve q}\times\ve{A}_{\ve q})\cdot(\ve Q_{-\ve q}\times\ve{A}_{-\ve q})\right\},
\end{split}
\end{equation}
where $\ve Q_{\ve q}$ is the Fourier representation of the lattice
difference operator $\Delta$. We choose the gauge $\Delta\cdot\ve A=0$
and $\Delta\cdot\ve h^{(\alpha)}=0$ which in the Fourier representation is
$\ve Q_{\ve q}\cdot\ve A_{\ve q}=0$ and $\ve Q_{\ve q}\cdot\ve
h^{(\alpha)}_{\ve q} = 0$. We complete the squares in $\ve A_{\ve q}$
and get the action
\begin{equation}
\begin{split}
S =& \sum_{\ve q}\left\{\sum_{\alpha=1}^N\frac{1}{2\beta|\psi^{(\alpha)}|^2}(\ve Q_{\ve q}\cdot\ve Q_{-\ve{q}})(\ve h_{\ve q}^{(\alpha)}\cdot\ve h_{-\ve{q}}^{(\alpha)}) + \pi\im\sum_{\alpha=1}^N\left[\ve h^{(\alpha)}_{\ve q}\cdot\ve m^{(\alpha)}_{-\ve q} + \ve h^{(\alpha)}_{-\ve q}\cdot\ve m^{(\alpha)}_{\ve q}\right] \right.
\\ &+\left.\left[\ve A_{\ve q} - \frac{\im e}{2}\left(\sum_{\alpha=1}^N\ve Q_{\ve q}\times\ve h^{(\alpha)}_{\ve q} \right)D_{\ve q}^{-1}\right]D_{\ve q}\left[\ve A_{-\ve q} - \frac{\im e}{2}\left(\sum_{\alpha=1}^N\ve Q_{-\ve q}\times\ve h^{(\alpha)}_{-\ve q} \right)D_{\ve q}^{-1}\right]\right.\\
 &+ \left. \frac{e^2}{4}\left(\sum_{\alpha=1}^N\ve Q_{\ve q}\times\ve h^{(\alpha)}_{\ve q}\right)D_{\ve q}^{-1}\left(\sum_{\alpha=1}^N\ve Q_{-\ve q}\times\ve h^{(\alpha)}_{-\ve q}\right)\right\},
\end{split}
\end{equation}
where $D_{\ve q} = \beta\ve Q_{\ve q}\cdot\ve Q_{-\ve
  q}/2$. After  performing the gaussian integral in $\ve A_{\ve
  q}$, the action reads 
\begin{equation}
\begin{split}
S =& \sum_{\ve q}\left\{\sum_{\alpha=1}^N\frac{1}{2\beta|\psi^{(\alpha)}|^2}(\ve Q_{\ve q}\cdot\ve Q_{-\ve{q}})(\ve h_{\ve q}^{(\alpha)}\cdot\ve h_{-\ve{q}}^{(\alpha)}) + \pi\im\sum_{\alpha=1}^N\left[\ve h^{(\alpha)}_{\ve q}\cdot\ve m^{(\alpha)}_{-\ve q} + \ve h^{(\alpha)}_{-\ve q}\cdot\ve m^{(\alpha)}_{\ve q}\right]  \right.\\
&+\left. \frac{e^2}{2\beta}\left(\sum_{\alpha=1}^N\ve h^{(\alpha)}_{\ve q}\right)\cdot\left(\sum_{\alpha=1}^N\ve h^{(\alpha)}_{-\ve q}\right)\right\}.
\end{split}
\end{equation}
At this point it is useful to introduce matrices and vectors in
\textit{flavor indices}. We write the action as
\begin{equation}
S = \sum_{\ve q}\left[ H^{\rm T}_{\ve q}  G_{\ve q}  H_{-\ve q} + \im\pi  M^{\rm T}_{\ve q}  H_{-\ve q} + 
\im\pi  H^{\rm T}_{\ve q}  M_{-\ve q}\right],
\end{equation}
where $H^{\rm T}_{\ve q} = (\ve h^{(1)}_{\ve q},\ve h^{(2)}_{\ve
  q},\dots,\ve h^{(N)}_{\ve q})$ and $M^{\rm T}_{\ve q} = (\ve
  m^{(1)}_{\ve q},\ve m^{(2)}_{\ve q},\dots,\ve m^{(N)}_{\ve q})$
are vectors in flavor indices, and the matrix $G_{\ve q}$ is given by  
\begin{equation}
G_{\ve q} = \left(
\begin{array}{ccccc}
\frac{|\ve Q_{\ve q}|^2}{2\beta|\psi^{(1)}|^2} + \frac{e^2}{2\beta}&
\frac{e^2}{2\beta} & \cdots & \frac{e^2}{2\beta} &
\frac{e^2}{2\beta}
\\
\frac{e^2}{2\beta} & \frac{|\ve Q_{\ve q}|^2}{2\beta|\psi^{(2)}|^2} +
\frac{e^2}{2\beta} & \cdots & \frac{e^2}{2\beta} & \frac{e^2}{2\beta}
\\
\vdots & \ddots & \ddots & \ddots & \vdots
\\

\frac{e^2}{2\beta} & \frac{e^2}{2\beta} &  \cdots &
\frac{|\ve Q_{\ve q}|^2}{2\beta|\psi^{(N-1)}|^2} + \frac{e^2}{2\beta} & \frac{e^2}{2\beta}
\\
\frac{e^2}{2\beta} & \frac{e^2}{2\beta} & \cdots & \frac{e^2}{2\beta} & \frac{|\ve Q_{\ve q}|^2}{2\beta|\psi^{(N)}|^2} +\frac{e^2}{2\beta}
\end{array}
\right),
\end{equation}
where $|\ve Q_{\ve q}|^2  =\ve Q_{\ve q}\cdot\ve Q_{-\ve q}$. In flavor indices $\eta,\lambda$ we write 
the matrix as $G_{\ve q}^{(\eta,\lambda)}
= d^{(\eta)}\delta_{\eta,\lambda} + c$, where $d^{(\eta)} = \frac{|\ve Q_{\ve q}|^2}{2\beta|\psi^{(\eta)}|^2}$ and $c = \frac{e^2}{2\beta}$. We
complete the square and obtain the expression
\begin{equation}
S = \sum_{\ve q}\left[ (H^{\rm T}_{\ve q} + \im\pi M^{\rm T}_{\ve q}G_{\ve q}^{-1})G_{\ve q}(H_{-\ve q} 
+ \im\pi M_{-\ve q} G_{\ve q}^{-1}) + \pi^2 M^{\rm T}_{\ve q}G_{\ve q}^{-1}  M_{-\ve q} \right],
\end{equation}
\end{widetext}
where $G_{\ve q}^{-1}$ is the solution of $G_{\ve q}G_{\ve q}^{-1} =
I$ and $I$ is the $N\times N$ unit matrix.
Gaussian integration of $H_{\ve q}$ yields an action expressed in vortex variables of different flavors
\begin{equation}
\label{vortexaction}
S = \sum_{\ve q} \pi^2 M^{\rm T}_{\ve q}  G_{\ve q}^{-1}  M_{-\ve q}.
\end{equation}

The matrix $G_{\ve q}^{-1}$ is found to have the following diagonal elements
\begin{equation}
  (G_{\ve q}^{-1})^{(\eta,\eta)} = \frac{\prod_{\alpha\neq\eta}d^{(\alpha)} + c\sum_{\alpha\neq\eta}\prod_{\gamma\neq\eta,\alpha}d^{(\gamma)}}{\prod_{\alpha=1}^N d^{(\alpha)} + c\sum_{\alpha=1}^N\prod_{\gamma\neq\alpha}d^{(\gamma)}},
\end{equation}
and
\begin{equation}
  (G_{\ve q}^{-1})^{(\eta,\lambda)} = -\frac{c\prod_{\alpha\neq\eta,\lambda}d^{(\alpha)}}{\prod_{\alpha=1}^N d^{(\alpha)} + c\sum_{\alpha=1}^N\prod_{\gamma\neq\alpha}d^{(\gamma)}},
\end{equation}
as off-diagonal elements ($\eta\neq\lambda$), where
$\prod_{\alpha=\varnothing}d^{(\alpha)}\equiv 1$. By dividing the
numerator and denominator by $\prod_{\alpha=1}^N d^{(\alpha)}$, we
obtain the diagonal elements
\begin{equation}
  (G_{\ve q}^{-1})^{(\eta,\eta)} = \frac{\frac{1}{d^{(\eta)}} 
  + \sum_{\alpha\neq\eta}\frac{c}{d^{(\eta)}d^{(\alpha)}}}{1 + \sum_{\alpha=1}^N\frac{c}{d^{(\alpha)}}},
\end{equation}
and the off-diagonal elements
\begin{equation}
  (G_{\ve q}^{-1})^{(\eta,\lambda)} = -\frac{\frac{c}{d^{(\eta)}d^{(\lambda)}}}{1 + \sum_{\alpha=1}^N\frac{c}{d^{(\alpha)}}},
\end{equation}
where $\eta\neq\lambda$. In total, the matrix $(G_{\ve
  q}^{-1})^{(\eta,\lambda)}$ reads
\begin{widetext}
\begin{equation}
  (G_{\ve q}^{-1})^{(\eta,\lambda)} = \frac{(\frac{1}{d^{(\eta)}} + \sum_{\alpha=1}^N\frac{c}{d^{(\eta)}d^{(\alpha)}})\delta_{\eta,\lambda} - \frac{c}{d^{(\eta)}d^{(\lambda)}}}{1 + \sum_{\alpha=1}^N\frac{c}{d^{(\alpha)}}}.
\end{equation}
Inserting the expressions for $d^{(\eta)}$ and $c$ and multiplying by $|\ve Q_{\ve q}|^4$ in the denominator and numerator, we obtain
\begin{equation}
  (G_{\ve q}^{-1})^{(\eta,\lambda)} = 2\beta\frac{(|\psi^{(\eta)}|^2|\ve Q_{\ve q}|^2 + e^2|\psi^{(\eta)}|^2\sum_{\alpha=1}^N|\psi^{(\alpha)}|^2)\delta_{\eta,\lambda} - e^2|\psi^{(\eta)}|^2|\psi^{(\lambda)}|^2}{|\ve Q_{\ve q}|^2(|\ve Q_{\ve q}|^2 + e^2\sum_{\alpha=1}^N|\psi^{(\alpha)}|^2)}.
\end{equation}
We introduce $\Psi^2 = \sum_{\alpha=1}^N|\psi^{(\alpha)}|^2$
and split the expression by partial fractioning and obtain the matrix
\begin{equation}
  (G_{\ve q}^{-1})^{(\eta,\lambda)} = \frac{2\beta|\psi^{(\eta)}|^2}{\Psi^2}\left[\frac{\Psi^2\delta_{\eta,\lambda} - |\psi^{(\lambda)}|^2}{|\ve Q_{\ve q}|^2} + \frac{|\psi^{(\lambda)}|^2}{|\ve Q_{\ve q}|^2 + e^2\Psi^2}\right].
\end{equation}
This is the vortex interaction matrix given in Eq. (\ref{potential}). Inserting 
this into Eq. (\ref{vortexaction}), the partition function of the system is
\begin{equation}
\begin{split}
Z =& \prod_{\alpha}  \sum_{ \ve m^{(\alpha)} }  \delta_{ \Delta\cdot\ve m^{(\alpha) },0}\exp(-S_{\rm V}), \\
S_{\rm V} =& \sum_{\ve q}\sum_{\eta,\lambda=1}^N\frac{2\pi^2\beta}{\Psi^2}\ve m^{(\eta)}_{\ve q}\cdot\ve m^{(\lambda)}_{-\ve q}|\psi^{(\eta)}|^2\left[\frac{\Psi^2\delta_{\eta,\lambda} - |\psi^{(\lambda)}|^2}{|\ve Q_{\ve q}|^2} + \frac{|\psi^{(\lambda)}|^2}{|\ve Q_{\ve q}|^2 + e^2\Psi^2}\right],
\end{split}
\end{equation}
which gives the action in the partition function in Eq. (\ref{vortex_action}). The above 
vortex action may be written in terms of charged and neutral vortex modes in a manner 
analogous to that of Eq. (\ref{action_separated}), as follows
\begin{equation}
\begin{split}
\frac{S_{\rm V}}{2 \pi^2 \beta/\Psi^2} = &  \sum_{\ve q} \left\{ 
\frac{
(\sum_{\alpha} |\psi^{(\alpha)}|^2  \ve m^{(\alpha)}_{\ve q} ) \cdot
(\sum_{\beta}  |\psi^{(\beta)} |^2  \ve m^{(\beta)}_{-\ve q} )
}
{|\ve Q_{\ve q}|^2 + m_0^2} 
 + 
\sum_{\alpha,\beta} 
\frac{
|\psi^{(\alpha)}|^2  |\psi^{(\beta)} |^2 (\ve m^{(\alpha)}_{\ve q} - \ve m^{(\beta)}_{\ve q})
\cdot (\ve m^{(\alpha)}_{-\ve q} - \ve m^{(\beta)}_{-\ve q})
}
{2 |\ve Q_{\ve q}|^2}  \right\}.
\end{split}
\end{equation}
Here, $m_0^2 = e^2 \Psi^2$. While the first, screened, term in general is present 
for all $N \geq 1$, it is clear from the above formulation that the second, 
unscreened, term is only present provided $N \geq 2$. 
The factor $2$ in the denominator in the unscreened terms is essential in order for 
the interaction terms between different vortex species to cancel out when $m_0^2=0$.

\section{\label{App:gauge} Gauge field correlator}
Introducing Fourier-transformed variables in Eq. (\ref{dual2}), 
{\it prior to integrating out the gauge field $\ve A$}, and adding
source terms in the form of electric currents $\ve J$ coupling 
linearly to $\ve A$, we obtain the action (note that for convenience,
we have redefined the $\ve A$-field in this Appendix by a factor
$\sqrt{\beta}$ as follows: $\ve A \to \tilde{\ve A} = \sqrt{\beta} \ve A$,
and then renaming $\tilde{\ve A} \to \ve A$. Thus, here we also
redefine the charge as follows: $e \to \tilde{e} = e/\sqrt{\beta}$
which we then rename  $ \tilde e \to e$.  In the end result we reinstate
the original definitions. We thus have the action Eq. (\ref{dual2}) on the 
form  
\begin{eqnarray}
S_J &=& \sum_{\ve q}  
\left[ \frac{ |\ve Q_{\ve q}|^2}{2 \beta |\psi^{(\alpha)}|^2}  \ve h^{(\alpha)}_{\ve q} \cdot \ve h^{(\alpha)}_{-\ve q}
+ \pi \im \left[ \ve m^{(\alpha)}_{\ve q} \cdot \ve h^{(\alpha)}_{-\ve q}
+\ve m^{(\alpha)}_{-\ve q} \cdot \ve h^{(\alpha)}_{\ve q} \right] 
-\frac{\im e}{2} 
\left[ \ve A_{\ve q} \cdot (\ve Q_{-\ve q} \times \ve h^{(\alpha)}_{-\ve q})
+  (\ve Q_{\ve q} \times \ve h^{(\alpha)}_{\ve q}) \cdot \ve A_{-\ve q}  \right] \nonumber \right. \\
 & + &  \left. \frac{1}{2} \left[ \ve J_{\ve q}  \cdot \ve A_{-\ve q}
                        +\ve J_{-\ve q} \cdot \ve A_{\ve q} \right] 
+ \frac{ |\ve Q_{\ve q}|^2 }{2}  \ve A_{\ve q} \cdot \ve A_{-\ve q} \right].
\label{dual_h_1_corr1}
\end{eqnarray}
Summations over indices $(\alpha,\beta) \in [1,\dots,N]$ is understood. Integrating out 
the gauge fields $\ve A_{\ve q}$, we get the action on the following form 
\begin{eqnarray}
\tilde S_J & = & 
\sum_{\ve q} \sum_{\alpha=1}^N  
\left[ \frac{ |\ve Q_{\ve q}|^2}{2 \beta |\psi^{(\alpha)}|^2}  \ve h^{(\alpha)}_{\ve q} \cdot \ve h^{(\alpha)}_{-\ve q}
+ \pi \im \left[ \ve m^{(\alpha)}_{\ve q} \cdot \ve h^{(\alpha)}_{-\ve q}
+\ve m^{(\alpha)}_{-\ve q} \cdot \ve h^{(\alpha)}_{\ve q} \right] 
-\frac{\ve D_{\ve q} \cdot \ve D_{-\ve q}}{2 |\ve Q_{\ve q}|^2}  \right], 
\label{dual_h_2_corr1a}
\end{eqnarray}
where we have defined $\ve D_{ \ve q} =  \ve J_{ \ve q} + \im e \ve
Q_{\ve q} \times \tilde{\ve h}_{ \ve q}$ and $\tilde{\ve h}_{\ve q} =  \sum_{\alpha=1}^{N} \ve h^{(\alpha)}_{\ve q}$.
Thus, the last term in Eq. (\ref{dual_h_2_corr1a}) may be written 
\begin{eqnarray}
\frac{\ve D_{\ve q} \cdot \ve D_{-\ve q}}{2 |\ve Q_{\ve q}|^2} & = & 
\frac{1}{2 |\ve Q_{\ve q}|^2} \ve J_{-\ve q}  \cdot \ve J_{\ve q}
+ \tilde{\ve h}_{\ve q} \cdot \ve \Lambda_{-\ve q} + \tilde{\ve h}_{-\ve q} \cdot \ve \Lambda_{\ve q}
-\frac{e^2}{2} \tilde{\ve h}_{\ve q} \cdot \tilde{\ve h}_{-\ve q},
\end{eqnarray}
where we have defined $\ve \Lambda_{ \ve q}  =  \frac{\im e}{2 |\ve Q_{\ve q}|^2}  \ve Q_{\ve q} \times \ve J_{ \ve q}$. Thus, the action may be written on the form
\begin{equation}
\begin{split}
\tilde S_J  =& 
\sum_{\ve q}  
\left[ \frac{ |\ve Q_{\ve q}|^2}{2 \beta |\psi^{(\alpha)}|^2}  \ve h^{(\alpha)}_{\ve q} \cdot \ve h^{(\alpha)}_{-\ve q}
+ \pi  \im  \left[ \ve m^{(\alpha)}_{\ve q} \cdot \ve h^{(\alpha)}_{-\ve q}
+\ve m^{(\alpha)}_{-\ve q} \cdot \ve h^{(\alpha)}_{\ve q} \right]
\right. \\ 
&- \left.\left[ \tilde{\ve h}_{\ve q} \cdot \ve \Lambda_{-\ve q} + \tilde{\ve h}_{-\ve q} \cdot \ve \Lambda_{\ve q}  \right]
+\frac{e^2}{2}  \tilde{\ve h}_{\ve q} \cdot \tilde{\ve h}_{-\ve q} - \frac{1}{2 |\ve Q_{\ve q}|^2} \ve J_{-\ve q}  \cdot \ve J_{\ve q}\right].
\label{dual_h_2_corr1b}
\end{split}
\end{equation}
In Eq. (\ref{dual_h_2_corr1b}), a summation over indices $(\alpha,\beta)$ is understood.
We can now integrate out the dual gauge fields $\ve h^{(\alpha)}_{\ve q}$ to obtain
\begin{eqnarray}
Z_J&= \prod_{\alpha=1}^N \sum_{\ve M^{(\alpha)}} \delta_{\Delta\cdot\ve M^{(\alpha)},0} \exp \left[-S_{J \rm{eff}} \right],
\label{gen_fun_Z1}
\end{eqnarray}
where
\begin{eqnarray}
S_{J \rm{eff}}& = & \sum_{\ve q}  \left[\pi^2  \ve m^{(\alpha)}_{\ve q} \widetilde{D}^{(\alpha,\beta)}(\ve q)
\ve m^{(\beta)}_{-\ve q} - F_{\ve A}(\ve J^{(\alpha)}_{\ve q},\ve J^{(\alpha)}_{-\ve q}) \right]. 
\label{dual_h_2_corr1}
\end{eqnarray}
Here, we have introduced
\begin{eqnarray}
F_{\ve A}(\{ \ve J_{\ve q},\ve J_{-\ve q} \}) 
& \equiv &  \frac{ J_{\ve q}^\mu  P_{\rm T}^{\mu \nu}  J_{-\ve q}^\nu}{2 |\ve Q_{\ve q}|^2 }
- \pi \im 
\left[
 \ve m^{(\alpha)}_{-\ve q} \widetilde{D}^{(\alpha,\beta)}(\ve q) \ve L^{(\beta)}_{\ve q} 
+\ve L_{-\ve q}^{(\alpha)} \widetilde{D}^{(\alpha,\beta)}(\ve q) \ve m^{(\beta)}_{\ve q} 
\right]
+\ve L^{(\alpha)}_{-\ve q} \widetilde{D}^{(\alpha,\beta)}(\ve q) \ve L^{(\beta)}_{\ve q}.
\label{Gen_functional_F1}
\end{eqnarray}

Here, the upper index denotes a "flavor" index $(\alpha,\beta) \in [1,\dots,N]$ indicating which 
matter field species the fields above correspond to, and we have introduced the vector 
$\ve L^{(\alpha)}_{\ve q} = \ve B_{\ve q}, \alpha \in [1,\dots,N]$.
Using this particular property of $ \ve L^{(\alpha)}_{\ve q}$, we may simplify 
$F_{\ve A}(\{ \ve J_{\ve q},\ve J_{-\ve q} \})$ somewhat, to obtain 
\begin{eqnarray}
F_{\ve A}(\{ \ve J_{\ve q},\ve J_{-\ve q} \}) 
& \equiv &   \frac{ J_{\ve q}^\mu  P_{\rm T}^{\mu \nu}  J_{-\ve q}^\nu}{2 |\ve Q_{\ve q}|^2 }
- \pi \im 
\left[ \ve m^{(\alpha)}_{-\ve q} \cdot \ve \Lambda_{\ve q} 
+\ve \Lambda_{-\ve q}  \cdot \ve m^{(\alpha)}{\ve q} \right]V^{(\alpha)}
+\ve \Lambda_{-\ve q} \cdot \ve \Lambda_{\ve q}  {\cal{S}}.  
\label{Gen_functional_F1a}
\end{eqnarray}
In Eq. (\ref{Gen_functional_F1a}), we have furthermore introduced
\begin{eqnarray}
V^{(\alpha)} & \equiv & \sum_{\beta = 1}^N         \widetilde{D}^{(\alpha,\beta)}(\ve q) 
=  \frac{2 \beta |\psi^{(\alpha)}|^2}{|\ve Q_{\ve q}|^2 + m_0^2}, \nonumber \\
{\cal{S}}            & \equiv & \sum_{\alpha,\beta = 1}^N  \widetilde{D}^{(\alpha,\beta)}(\ve q)
= \frac{2 \beta \psi^2}{|\ve Q_{\ve q}|^2 + m_0^2}. 
\label{VS_eqs}
\end{eqnarray}
The last equalities in Eqs. (\ref{VS_eqs}) are found from using the definition of 
$\widetilde{D}^{(\alpha,\beta)}(\ve q)$ given in Eq. (\ref{potential}), along with
the definition of $\psi^2$ given immediately after Eq. (\ref{potential}). We have  
also introduced the transverse projection operator
\begin{eqnarray}
P_{\rm T}^{\mu \nu} = \delta^{\mu \nu} - \frac{Q^\mu_{\ve
    q}Q^\nu_{-\ve q}}{|\ve Q_{\ve q}|^2},
\label{P_T}
\end{eqnarray}
which appears due to the transversality of the currents $\ve J$. Before doing functional 
derivations on $F_{\ve A}(\{ \ve J_{\ve q},\ve J_{-\ve q} \})$ it is useful to multiply 
out the term $\ve \Lambda_{-\ve q} \cdot \ve \Lambda_{\ve q} $ and explicitly use the constraint 
$\ve \nabla \cdot \ve J =0$ before derivation. We find
\begin{eqnarray}
&&\ve \Lambda_{-\ve q} \cdot \ve \Lambda_{\ve q}  =  
- \frac{e^2}{4|\ve Q_{\ve q}|^4} \varepsilon^{\alpha \nu \lambda}\varepsilon^{\alpha \rho \eta}
Q^{\nu}_{\ve q} Q^{\rho}_{-\ve q} J_{\ve q}^\lambda J_{- \ve q}^\eta  =  - \frac{e^2}{4|\ve Q_{\ve q}|^4} 
\left( 
\delta^{\nu \rho} \delta^{\lambda \eta} -\delta^{\nu \eta} \delta^{\lambda \rho} 
\right)
Q^{\nu}_{\ve q} Q^{\rho}_{-\ve q} J_{\ve q}^\lambda J_{- \ve q}^\eta \nonumber 
=  - \frac{e^2}{4|\ve Q_{\ve q}|^2}J_{\ve q}^\mu P_{\rm T}^{\mu \nu} J_{- \ve q}^\nu.
\end{eqnarray}
In the cross terms between vortex fields and current fields there is no need to introduce 
the transverse projection operator, since the inner product automatically projects out 
the transverse part of $\ve J$ since the vortices form closed loops. 

We may now express the gauge field correlators formally 
\begin{eqnarray}
\langle  A_{\ve q}^\mu  A_{-\ve q}^\nu \rangle
& = & \frac{1}{Z_0}  
\left. \frac{\delta^2 Z_J}{\delta  J_{-\ve q}^\mu \delta J_{\ve q}^\nu} 
\right|_{\ve J_{-\ve q} = \ve J_{\ve q} = 0 } \nonumber \\
& = & \frac{1}{Z_0}  
 \prod_{\alpha=1}^{N} \sum_{\ve M^{(\alpha)}} \delta_{\Delta\cdot\ve M^{(\alpha)},0}
    \left[ \frac{\delta F_{\ve A}}{\delta  J_{-\ve q}^\mu} 
\frac{\delta F_{\ve A}}{\delta  J_{\ve q}^\nu}
+ \frac{\delta^2 F_{\ve A}}{\delta  J_{-\ve q}^\mu \delta  J_{\ve q}^\nu} 
 \right]_{\ve J{-\ve q} = \ve J_{\ve q} = 0} \exp \left[ -S_{0 \rm{eff}} \right],
\label{A_corr_formal}
\end{eqnarray}
where  $S_{0 \rm{eff}} $ is Eq. (\ref{dual_h_2_corr1}) with all currents set to zero. We 
get for the required functional derivatives
\begin{eqnarray}
\frac{\delta F_{\ve A}}{\delta J_{\ve q}^\nu} & = &
\frac{P_{\rm T}^{\nu\lambda}J_{-\ve q}^\lambda}{|\ve Q_{\ve q}|^2}
- 2  \pi \im   \frac{\delta \ve \Lambda_{\ve q}}{\delta J_{\ve q}^\nu} \cdot \ve m^{(\alpha)}_{-\ve q}  V^{(\alpha)}
+ 2 \ve \Lambda_{-\ve q} \cdot \frac{\delta \ve \Lambda_{\ve q}}{\delta J_{\ve q}^\nu}  {\cal{S}}, \nonumber \\
\frac{\delta^2 F_{\ve A}}{\delta J_{\ve q}^\nu \delta J_{-\ve q}^\mu} & = &
\frac{P^{\mu \nu}_{\rm{T}}}{2|\ve Q_{\ve q}|^2} \left[ 2- e^2  {\cal{S}} \right].
\end{eqnarray}
Multiplying out everything, after setting the currents to zero, we obtain 
\begin{eqnarray}
\langle  A_{\ve q}^\mu  A_{-\ve q}^\nu \rangle =
\frac{P^{\mu \nu}_{\rm{T}}}{2 |\ve Q_{\ve q}|^2} \left[ 2 - e^2  {\cal{S}} \right] + 
\frac{\pi^2 e^2}{|\ve Q_{\ve q}|^2} V^{(\alpha)}V^{(\beta)}
\langle m^{\mu(\alpha)}_{\ve q} m^{\nu(\beta)}_{-\ve q} \rangle,
\end{eqnarray}
where a summation over $(\alpha,\beta) \in [1,\dots,N]$ is understood. Setting $\nu = \mu$ 
and summing over $\nu$, we get
\begin{eqnarray}
\langle  \ve A_{\ve q} \cdot  \ve A_{-\ve q} \rangle =
\frac{1}{|\ve Q_{\ve q}|^2} \left[ 2 - e^2  {\cal{S}} \right] + 
\frac{\pi^2 e^2}{|\ve Q_{\ve q}|^2} V^{(\alpha)}V^{(\beta)}
\langle \ve m^{(\alpha)}_{\ve q} \cdot  \ve m^{(\beta)}_{-\ve q} \rangle.
\end{eqnarray}
Using the definitions of $V^{(\alpha)}$ and ${\cal{S}}$ introduced above, and
moreover reintroducing the original gauge field $\ve A$ and the original
charge $e$ (see start of the Appendix), this becomes
\begin{eqnarray}
\langle  \ve A_{\ve q} \cdot  \ve A_{-\ve q} \rangle & = &
\frac{1}{\beta} \left\{ \
\frac{2}{|\ve Q_{\ve q}|^2 + m_0^2} + 
\frac{4\beta \pi^2 e^2}{|\ve Q_{\ve q}|^2} 
\frac{|\psi^{(\alpha)}|^2 |\psi^{(\beta)}|^2}{(|\ve Q_{\ve q}|^2 + m_0^2)^2}
\langle \ve m^{(\alpha)}_{\ve q} \cdot  \ve m^{(\beta)}_{-\ve q} \rangle  \right\} \nonumber \\
 & = &
\frac{2/\beta}{|\ve Q_{\ve q}|^2 + m_0^2} 
\left[1 + \frac{2 \beta \pi^2 e^2}{|\ve Q_{\ve q}|^2}  \frac{G^{(+)} (\ve q)}{|\ve Q_{\ve q}|^2 + m_0^2} \right],
\end{eqnarray}
when we introduce 
\begin{eqnarray}
G^{(+)}(\ve q) & = & 
\langle (\sum_{(\alpha)}|\psi^{(\alpha)}|^2\ve m^{(\alpha)}_{\ve q})
\cdot(\sum_{\beta}|\psi^{(\beta)}|^2\ve m^{(\beta)}_{-\ve q})\rangle,
\label{Gpluss_corr_app}
\end{eqnarray}
which are just Eqs. (\ref{A_gaugeprop})  and (\ref{Gpluss_corr}). Note how this 
has the structure of a "Dyson's equation", where $G^{(+)} (\ve q)$ plays the role
of a matter field loop or "polarizability", where the strength of the vertex is
given by $2 \beta \pi e/|\ve Q_{\ve q}|$. As we shall see below, a similar
statement holds for the dual gauge field correlators, but there
the vertex is a scalar of strength $\pi\im$. The difference lies in
the fact that while the dual gauge fields couple linearly to  
the vortex fields, it is the {\it curl} of the gauge field $\ve A$ 
that couples indirectly to the vortex fields  
(via the dual gauge fields).

\section{\label{App:dualgauge} Dual gauge field correlators}
The computation of correlators for the dual gauge fields $\ve h^{(\alpha)}$ proceeds along 
the same lines as for the gauge field $\ve A$, but sufficiently many details are different 
so we include it here for completeness.  Introducing Fourier-transformed variables in 
Eq. (\ref{dual2}), and after having added source terms in order to be able to compute 
correlators, the theory  may be written on the following form 
\begin{eqnarray}
S_J & = & \sum_{\ve q}  
\left[\frac{ |\ve Q_{\ve q}|^2}{2 \beta |\psi^{(\alpha)}|^2}  \ve h^{(\alpha)}_{\ve q} \cdot \ve h^{(\alpha)}_{-\ve q}
+ \pi \im  \left[ \ve m^{(\alpha)}_{\ve q} \cdot \ve h^{(\alpha)}_{-\ve q}
+\ve m^{(\alpha)}_{-\ve q} \cdot \ve h^{(\alpha)}_{\ve q} \right] 
+\frac{1}{2} \left[ \ve J^{(\alpha)}_{\ve q} \cdot \ve h^{(\alpha)}_{-\ve q}
                        +\ve J^{(\alpha)}_{-\ve q} \cdot \ve h^{(\alpha)}_{\ve q}     \right] 
+ \frac{e^2}{2 \beta} \ve h^{(\alpha)}_{\ve q} \cdot \ve h^{(\beta)}_{-\ve q} \right].
\label{dual_h_1_corr2}
\end{eqnarray}
In Eq. (\ref{dual_h_1_corr2}), summation over indices $(\alpha,\beta) \in [1,\dots,N]$ is 
understood and the last term appears after having integrated out the gauge field $\ve A$. 
Notice how we in this case have added $N$ source currents $\ve J^{(\alpha)}$, one for each
vortex field $\ve m^{(\alpha)}$.
Integrating out the dual gauge fields $\ve h^{(\alpha)}_{\ve q}$ we get the action on the 
following form 
\begin{eqnarray}
S_{J \rm{eff}} & = &   \sum_{\ve q} \left[\pi^2  \ve m^{(\alpha)}_{\ve q} \widetilde{D}^{(\alpha,\beta)}(\ve q)
\ve m^{(\beta)}_{-\ve q} - F_{\ve h}(\{ \ve J^{(\alpha)}_{\ve q},\ve J^{(\alpha)}_{-\ve q} \}) \right], 
\label{dual_h_2_corr2}
\end{eqnarray}
where we have defined
\begin{eqnarray}
F_{\ve h}(\{ \ve J^{(\alpha)}_{\ve q},\ve J^{(\alpha)}_{-\ve q} \}) & \equiv & 
\frac{1}{4} \ve J^{(\alpha)}_{\ve q}  
\widetilde{D}^{(\alpha,\beta)}(\ve q) \ve J^{(\beta)}_{-\ve q}) 
 +\frac{\pi \im}{2}  \ve J^{(\alpha)}_{-\ve q} \widetilde{D}^{(\alpha,\beta)}(\ve q) \ve m^{(\beta)}_{-\ve q}
+ \frac{\pi \im}{2}  \ve m^{(\alpha)}_{-\ve q} \widetilde{D}^{(\alpha,\beta)}(\ve q) \ve J^{(\beta)}_{-\ve q}.
\label{Gen_functional_F2}
\end{eqnarray}
Here, as in Appendix \ref{App:gauge}, the currents are divergence-free, 
$\nabla \cdot \ve J^{(\alpha)} = 0, \alpha \in [1,\dots,N]$, and the interaction matrix $
\widetilde{D}^{(\alpha,\beta)}(\ve q)$ is defined in Eq. (\ref{potential}). As was the 
case for the $\ve A$ field correlator, the constraints on the currents $\ve J^{(\alpha)}$ 
must be carefully kept track of when performing the necessary functional derivations in 
order to obtain the correlation functions. The generating functional is given by
\begin{eqnarray}
Z_J&= \prod_{\alpha=1}^N \sum_{\ve M^{(\alpha)}} \delta_{\Delta\cdot\ve M^{(\alpha)},0}
\exp \left[-S_{J \rm{eff}} \right].
\label{gen_fun_Z2}
\end{eqnarray}


Applying Eq. (\ref{A_corr_formal}) to Eqs. (\ref{dual_h_2_corr2}), (\ref{Gen_functional_F2}), 
and (\ref{gen_fun_Z2}), we find
\begin{eqnarray}
\langle h_{\ve q}^{\mu(\alpha)} h_{-\ve q}^{\nu(\beta)} \rangle
& = & \frac{1}{Z_0}  
\left. \frac{\delta^2 Z_J}{\delta  J_{-\ve q}^{\mu(\alpha)} \delta J_{\ve q}^{\nu(\beta)}} 
\right|_{\ve J^{(\alpha)}_{-\ve q} = \ve J^{(\beta)}_{\ve q} = 0 } \nonumber \\
& = & \frac{1}{Z_0}  
\prod_{\alpha=1}^N \sum_{\ve M^{(\alpha)}} \delta_{\Delta\cdot\ve M^{(\alpha)},0}
\left[ 
\frac{\delta F_{\ve h}}{\delta  J_{-\ve q}^{\mu(\alpha)}} 
\frac{\delta F_{\ve h}}{\delta  J_{\ve q}^{\nu(\beta)}}
+ \frac{\delta^2 F_{\ve h}}{\delta  J_{-\ve q}^{\mu(\alpha)} \delta  J_{\ve q}^{\nu(\beta)}} 
\right]_{\ve J^{(\alpha)}_{-\ve q} = \ve J^{(\beta)}_{\ve q} = 0} \exp \left[-S_{0J \rm{eff}} \right],
\end{eqnarray}
where $S_{0J \rm{eff}}$ is Eq. (\ref{dual_h_2_corr2}) in the absence of source terms. We obtain
from Eq. (\ref{Gen_functional_F2}) 
\begin{eqnarray}
\frac{\delta F_{\ve h}}{\delta  J_{-\ve q}^{\mu(\alpha)}}
& = & \frac{1}{2}  \widetilde{D}^{(\alpha,\beta)}(\ve q)  J_{\ve q}^{\mu(\beta)}(\ve q) 
- \pi \im \widetilde{D}^{(\alpha,\beta)}(\ve q)   m_{\ve q}^{\mu(\beta)}, \nonumber \\
\frac{\delta^2 F_{\ve h}}{\delta J_{-\ve q}^{\mu(\alpha)} \delta  J_{\ve q}^{\nu(\beta)}}
& = & \frac{P_{\rm T}^{\mu\nu}}{2} \widetilde{D}^{(\beta,\alpha)}(\ve q) .
\end{eqnarray}
Here, we must keep track of two indices on the "magnetic" currents $\ve J^{(\alpha)}$, since we 
have one current (source term) coupling to each of the $N$ dual gauge fields 
$\ve h^{(\alpha)}, \alpha \in [1,\dots,N]$. 
The notation we use is that $J_{\ve q}^{\mu(\alpha)}$ is the $\mu$ Cartesian component of the 
current coupling to dual gauge field of flavor $\alpha [1,\dots,N]$. 
We use a corresponding notation for $m^{\mu(\alpha)}_{\ve q}$.
Moreover, 
$P_{\rm T}^{\mu \nu}$ again is the transverse projection operator defined in Eq. (\ref{P_T}), 
appearing due to the transversality of the currents $\ve J^{(\alpha)}$. Multiplying all 
of this together, we find
\begin{eqnarray}
\langle  h_{\ve q}^{\mu(\alpha)}  h_{-\ve q}^{\nu(\beta)} \rangle
 = \frac{1}{2} \widetilde{D}^{(\alpha,\beta)} (\ve q)  P_{\rm T}^{\mu \nu}
- \pi^2   \widetilde{D}^{(\alpha,\eta)} (\ve q)  \widetilde{D}^{(\beta,\kappa)} (\ve q) 
\langle m_{\ve q}^{\mu (\eta)}  m_{-\ve q}^{\nu(\kappa)} \rangle.
\label{H_corr_final1}
\end{eqnarray}
Moreover, setting $\nu = \mu$ and summing over Cartesian components of the dual gauge 
fields, we find
\begin{eqnarray}
\langle  \ve h^{(\alpha)}_{\ve q}  \cdot \ve h^{(\beta)}_{-\ve q} \rangle
 = \widetilde{D}^{(\alpha,\beta)} (\ve q) 
- \pi^2   \widetilde{D}^{(\alpha,\eta)} (\ve q) \widetilde{D}^{(\beta,\kappa)} (\ve q) 
\langle \ve m^{(\eta)}_{-\ve q} \cdot  \ve m^{(\kappa)}_{\ve q} \rangle, 
\label{dual_h_individual_app}
\end{eqnarray}
where we have used the fact that the trace of the projection operator is equal to $2$. 
In Eqs. (\ref{H_corr_final1}) and (\ref{dual_h_individual_app}), a summation over the 
indices $(\kappa,\eta) \in [1,\dots,N]$ is understood. Note how this, as for
the $\ve A$-field correlator, has the structure
of a "Dyson's equation", where the vortex correlator plays a role analogous
to a matter field loop, or "polarizability", with a scalar vertex (charge)
of strength $\pi\im$. This is simply a reflection of the fact that
dual gauge fields $\ve h^{(\alpha)}$ couple linearly to the vortex fields
$\ve m^{(\alpha)}$ ("magnetic currents"). The factor $\im$ in the strength
of the vertex appearing in Eqs. (\ref{dual2}) and (\ref{dual_h_1_corr2})
gives rise to an "anti"-Biot Savart law between vortex
segments mediated by the dual gauge fields.  

\section{Generalization of Eq. (\ref{dual2}) in the presence of inter flavor Josephson coupling}
\label{App:josephson}
In this Appendix, we consider the $N$-flavor London superconductor model Eq. (\ref{villain}) 
in the dual representation, including Josephson  couplings between matter fields of different 
flavors. The Josephson coupling between $\theta^{(\alpha)}$ and $\theta^{(\eta)}$ is 
\textit{local in space-time}, represented in the Euclidean action by the terms
$g^{(\alpha,\eta)}\cos[\theta^{(\alpha)}(\ve r)-\theta^{(\eta)}(\ve r)]$. With $N$
matter fields there will be $N (N-1)/2$ such terms. (Note how there are no such terms when $N=1$).
However, since these terms act as ferromagnetic couplings between the phase fields of different 
flavors, the critical properties of the model are preserved if we only include the terms that are 
``nearest neighbors'' in flavor indices. This is precisely analogous to including only 
nearest-neighbor Josephson coupling in a Josephson junction array, but where the "lattice sites"
now are represented by flavor indices. In this case, we have $N-1$ Josephson terms.
Therefore, we consider the action
\begin{equation}
  \label{}
  \begin{split}
  S = -\sum_{\alpha=1}^N\beta|\psi^{(\alpha)}|^2\cos(\Delta\theta^{(\alpha)} - e\ve A) - \sum_{\eta=1}^{N-1}\beta g^{(\eta)}\cos(\theta^{(\eta)} - \theta^{(\eta+1)}) + \frac{\beta}{2}(\Delta\times\ve A)^2,
  \end{split}
\end{equation} 
where $g^{(\eta)}$ is the Josephson coupling. In the Villain
approximation the model reads
\begin{equation}
  \label{}
  \begin{split}
  Z =& \int\mathcal{D}\ve A\prod_{\alpha=1}^N\int\mathcal{D}\theta^{(\alpha)}\sum_{\ve n^{(\alpha)}}\prod_{\eta=1}^{N-1}\sum_{m^{(\eta)}}\exp(-S), \\
  S =& \sum_{\ve r} \left[\sum_{\alpha=1}^N\frac{\beta|\psi^{(\alpha)}|^2}{2}(\Delta\theta^{(\alpha)} -e\ve A + 2\pi\ve n^{(\alpha)})^2  
    + \sum_{\eta=1}^{N-1}\frac{\beta g^{(\eta)}}{2}(\theta^{(\eta)} - \theta^{(\eta+1)} + 2\pi  m^{(\eta)})^2 + \frac{\beta}{2}(\Delta\times\ve A)^2\right].
  \end{split}
\end{equation}
Here, $\ve n^{(\alpha)}$ are integer vector fields where $\alpha \in
[1,\dots,N]$ and $m^{(\eta)}$ are integer scalar fields with $\eta \in
[1,\dots,N-1]$ which take care of $2\pi$ periodicity. We introduce the Hubbard
Stratonovich fields $\ve v^{(\alpha)}$ and $q^{(\eta)}$, and apply
the Poisson summation formula so that they become integer fields 
\begin{equation}
  \label{hsjoseph}
  \begin{split}
  Z =& \int\mathcal{D}\ve A\prod_{\alpha=1}^N\int\mathcal{D}\theta^{(\alpha)}\sum_{\ve v^{(\alpha)}}\prod_{\eta=1}^{N-1}\sum_{q^{(\eta)}}\exp(-S), \\
  S =& \sum_{\ve r} \left\{\sum_{\alpha=1}^N\left[\frac{(\ve v^{(\alpha)})^2}{2\beta|\psi^{(\alpha)}|^2} + \im\ve v^{(\alpha)}\cdot(\Delta\theta^{(\alpha)} -e\ve A ) \right]
   +\sum_{\eta=1}^{N-1}\left[\frac{(q^{(\eta)})^2}{2\beta g^{(\eta)}} + \im q^{(\eta)}(\theta^{(\eta)} - \theta^{(\eta+1)})\right]+ \frac{\eta}{2}(\Delta\times\ve A)^2\right\}.
  \end{split}
\end{equation}
\end{widetext}
At this point we organize the phase fields and perform partial
summations so that they can be integrated out. This gives the
following constraints on the integer fields
\begin{equation}
  \label{}
  \begin{split}
  \Delta\cdot\ve v^{(1)} &= q^{(1)}, \\
  \Delta\cdot\ve v^{(\eta)} &= q^{(\eta)} - q^{(\eta-1)}, \\
  \Delta\cdot\ve v^{(N)} &= -q^{(N-1)}.
  \end{split}
\end{equation}
To enforce these constraints we introduce the non-compact gauge fields $\ve
h^{(\alpha)}$ and the integer fields
$\ve B^{(\eta)}$ where $\alpha\in[1,\dots,N]$ and $\eta\in[1,\dots,N-1]$ such that
\begin{equation}
  \label{}
  \begin{split}
  \ve v^{(1)} &= \ve B^{(1)} + \Delta\times\ve h^{(1)}, \\
  \ve v^{(\eta)} &= \ve B^{(\eta)} - \ve B^{(\eta-1)} + \Delta\times\ve h^{(\eta)}, \\
  \ve v^{(N)} &= -\ve B^{(N-1)} + \Delta\times\ve h^{(N)},
  \end{split}
\end{equation}
where $\eta\in[2,\dots,N-1]$, and $q^{(\eta)} = \Delta \cdot \ve
B^{(\eta)}$ are instantons. 
These expressions may be
simplified by introducing the dummy fields $q^{(0)}=q^{(N)}=0$ and
$\ve B^{(0)}=\ve B^{(N)}=0$, so that the constraints become slightly
more symmetric,   given by
\begin{eqnarray}
\Delta\cdot\ve v^{(\alpha)} & = & q^{(\alpha)} - q^{(\alpha-1)}, \nonumber \\
\ve v^{(\alpha)} & = &\ve B^{(\alpha)} - \ve B^{(\alpha-1)} + \Delta\times\ve h^{(\alpha)},
\end{eqnarray}
where $\alpha\in[1,\dots,N]$. Including inter-flavor Josephson couplings beyond
"nearest-neighbor" would merely have led to redundant additional constraints
in the problem.  Expressed in the new fields, the partition function reads
\begin{widetext}
\begin{equation}
  \label{}
  \begin{split}
  Z =& \left.\int\mathcal{D}\ve A\prod_{\alpha=1}^N\sum_{\ve h^{(\alpha)}}\prod_{\eta=1}^{N-1}\sum_{\ve B^{(\eta)}}\exp(-S)\right|_{\ve B^{(0)}=\ve B^{(N)}=0}, \\
  S =& \sum_{\ve r} \left\{  
   \sum_{\alpha=1}^{N}\frac{(\Delta\times\ve h^{(\alpha)} + \ve B^{(\alpha)} - \ve B^{(\alpha-1)})^2}{2\beta|\psi^{(\alpha)}|^2} 
   -\im e\ve A\cdot(\sum_{\alpha=1}^N\Delta\times\ve h^{(\alpha)}) 
   +\sum_{\eta=1}^{N-1}\frac{(\Delta\cdot\ve B^{(\eta)})^2}{2\beta g^{(\eta)}} + \frac{\beta}{2}(\Delta\times\ve A)^2\right\}.
  \end{split}
\end{equation}
The appearance of {\it instantons} and effectively {\it compact} dual gauge 
fields in the dual description when Josephson couplings are included, serves 
to illustrate what a non-perturbative effect this is. Instantons are singular
objects and cannot possibly be introduced perturbatively. 
Moreover, once instantons are required, a compactification
of the dual gauge fields is also required, a highly 
non-perturbative step. Therefore,
when we consider models where Josephson coupling is absent, it is 
essential to be able to rule out completely inter-flavor Josephson coupling
on symmetry grounds {\it a priori}, and at the level of the bare action.
(In  systems with multicomponent {\it electronic condensates}, 
with $N=2$ and where a weak inter-flavor Josephson coupling must be included, it 
may be possible to see the  resemblance of one phase transition for a neutral
mode and one phase transition for a charged mode, such as we have presented  
for the zero-Josephson coupling case. This would be so in small enough systems with 
linear extent smaller than the Josephson length given by
$\lambda_J^{(\alpha)} = \sqrt{|\psi^{(\alpha)}|^2/g^{(\alpha)}}$, which in this
context may be viewed as setting the scale of the inter-instanton separation. This 
would be a finite-size effect. For bulk systems, the apparent neutral mode will 
eventually be suppressed, leaving one phase transition in the universality class 
of the inverted \xy model.)

We proceed by integrating out the original gauge field and apply the Poisson 
summation formula to introduce the integer vortex fields $\ve m^{(\alpha)}$
and the integer fields $\ve J^{(\eta)}$, where $\alpha\in[1,\dots,N]$ and 
$\eta\in[1,\dots,N-1]$. The resulting theory is the generalization of 
Eq. (\ref{dual2})
\begin{equation}
  \label{gendual2}
  \begin{split}
  Z =& \left.\prod_{\alpha=1}^N\int\mathcal{D}\ve h^{(\alpha)}\sum_{\ve m^{(\alpha)}}\prod_{\eta=1}^{N-1}\int\mathcal{D}\ve B^{(\eta)}\sum_{\ve J^{(\eta)}}\exp(-S)\right|_{\ve B^{(0)}=\ve B^{(N)}=0},  \\
  S =& \sum_{\ve r} \left\{  
   \sum_{\alpha=1}^{N}\frac{(\Delta\times\ve h^{(\alpha)} + \ve B^{(\alpha)} - \ve B^{(\alpha-1)})^2}{2\beta|\psi^{(\alpha)}|^2}  + \frac{e^2}{2\beta}(\sum_{\alpha=1}^N\ve h^{(\alpha)})^2
   +\sum_{\eta=1}^{N-1}\frac{(\Delta\cdot\ve B^{(\eta)})^2}{2\beta g^{(\eta)}} \right.\\
   &+ \left.2\pi\im\left(\sum_{\alpha=1}^N\ve h^{(\alpha)}\cdot\ve m^{(\alpha)} + \sum_{\eta=1}^{N-1}\ve B^{(\eta)}\cdot\ve J^{(\eta)}\right)\right\}.
  \end{split}
\end{equation}
\end{widetext}
First we note that like in Eq. (\ref{dual2}) the integration of the
gauge field $\ve A$ makes the algebraic sum of the dual gauge fields
massive. This reflects the fact that the gauge field $\ve A$ cannot screen
instantons, it can only screen vortices. Furthermore, in the
limit $g^{(\eta)}\to0$ for all $\eta\in[1,\dots,N-1]$ there are no
Josephson coupling terms and each field $\ve B^{(\eta)}$ is
constrained locally so that $\Delta\cdot\ve B^{(\eta)}=0$. The
representation $\ve B^{(\eta)} \to \Delta\times\ve b^{(\eta)}$ takes
care of the constraint, and the substitution $\ve h^{(\alpha)} + \ve
b^{(\alpha)} - \ve b^{(\alpha-1)} \to \tilde{\ve h}^{(\alpha)}$
reduces Eq. (\ref{gendual2}) to Eq. (\ref{dual2}). 
Finally we consider the
uncharged case, $e\to 0$ for which it is useful to return to
Eq. (\ref{hsjoseph}), integrate out the phase fields, and write the 
theory in terms of the integer fields $\ve v^{(\alpha)}$ and 
$q^{(\eta)}$  
\begin{equation}
  \label{}
  \begin{split}
  Z =& \prod_{\alpha=1}^N\sum_{\ve v^{(\alpha)}}\prod_{\eta=1}^{N-1}\sum_{q^{(\eta)}}\delta_{\Delta\cdot\ve v^{(\alpha)},q^{(\alpha)} - q^{(\alpha-1)}}\exp(-S), \\
  S =&\sum_{\ve r} \left\{\sum_{\alpha=1}^N\frac{(\ve v^{(\alpha)})^2}{2\beta|\psi^{(\alpha)}|^2}
   +\sum_{\eta=1}^{N-1}\frac{(q^{(\eta)})^2}{2\beta g^{(\eta)}}\right\},
  \end{split}
\end{equation}
where $\delta_{x,y}$ is the Kronecker delta. We sum over the fields
$\ve q^{(\eta)}$ for all $\eta\in[1,\dots,N-1]$ and are left with the
partition function
\begin{equation}
  \label{}
  \begin{split}
  Z =& \prod_{\alpha=1}^N\sum_{\ve v^{(\alpha)}}\delta_{\sum_{\alpha=1}^N\Delta\cdot\ve v^{(\alpha)},0}\exp(-S), \\
  S =&\sum_{\ve r} \left\{\sum_{\alpha=1}^N\frac{(\ve v^{(\alpha)})^2}{2\beta|\psi^{(\alpha)}|^2}
   +\sum_{\eta=1}^{N-1}\frac{(\sum_{\gamma=1}^\eta\Delta\cdot\ve v^{(\gamma)})^2}{2\beta g^{(\eta)}}\right\}.
  \end{split}
\end{equation}
This is the theory of $N$ current fields $\ve v^{(\alpha)}$ which
individually can form closed loops, or dumbbells starting and ending
on instantons. There is only {\it one} remaining constraint on the
$N$ matter fields in the problem, after the $N-1$ instantons have been 
summed out. (If we had had $M < N-1$ Josephson coupling between flavors to
begin with, we would have had $N-M$ remaining constraints in the problem 
after summing out the instantons).  The  one remaining constraint leaves 
only one phase transition in the problem in the universality class of the 
\xy model, in contrast to the $N$ phase transitions we have in the complete 
absence of inter-flavor Josephson couplings. The local constraint 
$\sum_{\alpha=1}^N\Delta\cdot\ve v^{(\alpha)}=0$ forces each dumbbell
to form a closed loop with one or more dumbbells of any flavor. 
The instantons have been summed out of the problem, 
leaving as their only trace the possibility of having supercurrents change 
flavor, precisely what the inter-flavor Josephson coupling does when expressed 
in terms of the fields describing the supercurrents! 

It is the possibility of being able to "chop" closed current loops of a given flavor into 
dumbbell pieces starting and ending on instantons, and then joining  together dumbbell 
configurations of different flavors, that facilitates this. (It has already been noted 
\cite{sachdev2004} that the dual field theory of a gauge theory with two complex scalar  
matter fields minimally coupled to a compact gauge field,  features two complex scalar dual 
matter fields coupled to one non-compact dual gauge field and an inter-flavor  Josephson 
coupling between the two matter field components. This is the reverse of what we have shown 
in this Appendix for the case $N=2$, and nicely demonstrates that "duality squared equals 
unity".)
   
Taking the limit $g^{(\eta)}\to 0$ for all $\eta\in[1,\dots,N-1]$ constraints $\ve v^{(\alpha)}$ 
to be divergence free for all $\alpha\in[1,\dots,N]$, and the model  thus reverts back to the 
loop-gas representation of $N$ decoupled \xy models.

\section{\label{App:bkt} Kosterlitz-Thouless transitions in N-flavor superconductor in 
two spatial dimensions at finite temperatures}
In $2+1$ dimensions at finite temperature, the classical critical behavior of the N-flavor 
superconductor is very different from the true $2+1$-dimensional case, i.e. the quantum critical
behavior taking place in two spatial dimensions at zero temperature. Let us first recall some features
of planar superconductivity. It is well known that in two dimensional
models with a $U(1)$ gauge symmetry 
there is no quasi-long range order at any finite temperatures because a gauge field coupling makes 
the interaction between topological defects exponentially screened \cite{minnhagen}. The situation is 
however different if one takes into account the ``out of plane" magnetic field.
That is, taking into account a third  dimension, a vortex in a thin superconducting film produces 
a ``mushroom"-like magnetic field outside the plane, which as shown by Pearl \cite{pearl} gives rise 
to a logarithmic inter-vortex interaction at distances smaller than 
$[ penetration \  length]^2/ [film  \ thickness]$, while at a distance larger  than that
the vortices interact via a $1/r$-law. Thus, for a thin film with a penetration length which is 
significantly larger than the sample size,  a  Kosterlitz-Thouless (KT) transition/crossover 
should be observable \cite{pearl,minnhagen}. The same effect is also the reason for the appearance 
of  vortices with long-range interactions in layered systems
making them being essentially coupled $U(1)$ models,
where various KT transitions and 
crossovers were studied in numerous works \cite{Blatter1994,clem,layered}. 

Here, we are interested in KT phase transitions in the N-flavor London superconductor in 
$2+1$-dimensions in the regime when ({\it i}) the effect of  ``out of plane" Pearl field 
can be neglected (short penetration length limit or alternatively a planar field theory 
without a third dimension), and when ({\it ii}) all components have different stiffnesses.
In Ref. \onlinecite{npb},  the case $N=2$ in such a regime was considered. It was 
shown 
that the system has a KT transition into a state where quasi-long range  order is established 
only in phase difference which produces a quasi-superfluid state.
This state, however, is principally different from e.g. {\bf SSF} to {\bf MSF} transition 
considered in \cite{BSA} and in  this paper, because  ({\it i}) the quasi-superfluid
state considered in \cite{npb} is a purely two-dimensional phenomenon ({\it ii}) in this 
state there is no true off-diagonal long-range order, ({\it iii}) there is no phase transitions from superconductivity 
to superfluidity in a planar system at finite temperature \cite{npb}.

Let us now consider Eq. (\ref{charge_neutral}) for the case $N=3$, when 
$|\psi^{(1)}| \ll |\psi^{(2)}| \ll |\psi^{(3)}|$. In the most interesting 
case of finite  penetration length,  the charged mode
formally  can never
develop quasi-long range order. That is because the composite 
single-quantum vortices $(\Delta\theta^{(1)}=2\pi,\Delta\theta^{(2)}=2\pi,\Delta\theta^{(3)}=2\pi)$
and $(\Delta\theta^{(1)}=-2\pi,\Delta\theta^{(2)}=-2\pi,\Delta\theta^{(3)}=-2\pi)$
have finite energy and have only screened short range interaction. Thus, in the
limit where the magnetic penetration length is short, such vortices are
always  unbound at any finite temperature.
We do not consider here the possibility of
a ``would be" KT crossover 
which is possible in a charged system with significantly large penetration length 
\cite{minnhagen}. The absence of superconductivity means that  {individually} 
all phases are disordered and the system is not superconducting. However, considering 
quasi-long range order in phase differences in $d=2$, several interesting possibilities
arise. Composite  one flux quantum vortices have short range interactions. { On the other
hand, vortices with windings only in one or two phases  excite neutral modes and thus can 
undergo a true KT transition.} This opens up the possibility for a KT phase transitions  
associated with  establishing quasi-long-range order in phase differences
 \cite{npb}. The key feature 
of the $N>2$ system where  bare stiffnesses are different, is that, as discussed in 
Appendix \ref{App:charged_neutral}, the neutral modes have 
also different stiffnesses 
[see Eqs. (\ref{neutral}), (\ref{stiffnesses})]. 

Let us consider first the low temperature regime. Then the vortices with short-ranged 
interactions, namely 
$(\Delta\theta^{(1)}=2\pi,\Delta\theta^{(2)}=2\pi,\Delta\theta^{(3)}=2\pi)$
and 
$(\Delta\theta^{(1)}=-2\pi,\Delta\theta^{(2)}=-2\pi,\Delta\theta^{(3)}=-2\pi)$ 
are liberated, while vortices with phase windings only in one or two phases
are bound into pairs of vortices and anti-vortices. In this state there is
{\it quasi-long range order in the phase differences} 
$\theta^{(1)}-\theta^{(2)}$, $\theta^{(1)}-\theta^{(3)}$, and  $\theta^{(2)}-\theta^{(3)}$.
Recall that the gradient terms  of neutral modes
which follow from separating of variable in GL functional
and dropping terms describing charged modes in Eq. (\ref{neutral}) are
\begin{eqnarray}
H_{\rm neutral}&=&  
\frac{1}{2}\frac{|\psi^{(1)}|^2  |\psi^{(2)}|^2}
{\Psi^2}(\nabla (\theta^{(1)}-\theta^{(2)}))^2 + \nonumber \\
&&\frac{1}{2}\frac{|\psi^{(1)}|^2 |\psi^{(3)}|^2}
{\Psi^2}(\nabla (\theta^{(1)}-\theta^{(3)}))^2 + \nonumber \\
&&\frac{1}{2}\frac{|\psi^{(2)}|^2 |\psi^{(3)}|^2 }
{\Psi^2}(\nabla (\theta^{(2)}-\theta^{(3)}))^2.
\label{n2}
\end{eqnarray}
From this, we have the following stiffnesses of neutral modes
\begin{eqnarray}
J^{12}&=&
\frac{|\psi^{(1)}|^2 |\psi^{(2)}|^2 }
{\Psi^2}, \nonumber \\
J^{23}&=&\frac{|\psi^{(1)}|^2 |\psi^{(3)}|^2}
{\Psi^2}, \nonumber \\
J^{13}&=&\frac{|\psi^{(2)}|^2  |\psi^{(3)}|^2}
{\Psi^2}.
\label{s2}
\end{eqnarray}
Consider now what will happen as the temperature is increased. Pictorially, 
we may illustrate this by considering Fig. \ref{3spl_2}
by slicing through the pictures at a given coordinate along the vortex 
lines, considering typical cross sections.
Upon increasing the temperature, first there will take 
place a deconfinement of vortex pairs 
$(\Delta\theta^{(1)}=2\pi,\Delta\theta^{(2)}=0,\Delta\theta^{(3)}=0)$+
$(\Delta\theta^{(1)}=-2\pi,\Delta\theta^{(2)}=0,\Delta\theta^{(3)}=0)$
because vortices in such a pair are bound by the two weakest neutral
modes $J^{12}$  and $J^{13}$ [$J^{12} <J^{13}< |\psi^{(1)}|^2/2$].
This will be accompanied by partial decomposition 
of deconfined {\it thermally created} composite vortices
$(\Delta\theta^{(1)}=2\pi,\Delta\theta^{(2)}=2\pi,\Delta\theta^{(3)}=2\pi) \to
(\Delta\theta^{(1)}=2\pi,\Delta\theta^{(2)}=0,\Delta\theta^{(3)}=0) +
(\Delta\theta^{(1)}=0,\Delta\theta^{(2)}=2\pi,\Delta\theta^{(3)}=2\pi)$, because a vortex 
$(\Delta\theta^{(1)}=0,\Delta\theta^{(2)}=2\pi,\Delta\theta^{(3)}=2\pi)$
has the same neutral vorticity as a vortex 
$(\Delta\theta^{(1)}=-2\pi,\Delta\theta^{(2)}=0,\Delta\theta^{(3)}=0)$
namely $-2\pi$ windings in neutral modes $\theta^{(1)}-\theta^{(2)}$
and  $\theta^{(1)}-\theta^{(3)}$.
This transition takes place at
\begin{eqnarray}
T^{(1)}_{\rm KT}=\frac{\pi}{2}[J^{12}+J^{13}]= 
\frac{\pi}{2} \frac{|\psi^{(1)}|^2 }{\Psi^2}[{|\psi^{(2)}|^2 }+ {|\psi^{(3)}|^2}].
\end{eqnarray}
This phase transition disorders the variable $\theta^{(1)}$
and correspondingly eliminates quasi-long-range order
in phase differences $\theta^{(1)}-\theta^{(2)}$
and $\theta^{(1)}-\theta^{(3)}$. Consequently above 
$T^{(1)}_{\rm KT}$ the only surviving neutral mode
is associated with $\theta^{(2)}-\theta^{(3)}$. The remaining phase
transition can be mapped onto that in $N=2$ system \cite{npb}.
Thus the second phase transition takes place at 
\begin{eqnarray}
T^{(2)}_{\rm KT}=\frac{\pi}{2}
\frac{|\psi^{(2)}|^2 |\psi^{(3)}|^2 }{\Psi^2 }. 
\end{eqnarray}
We can now solve the general problem of KT transitions
in a system of $N$ planar condensates with all different bare
stiffnesses. 
In the general case of $N$-flavor London model the temperatures
of the lowest KT transition is given by
\begin{eqnarray}
T^{(1)}_{\rm KT}& = & \frac{\pi}{2}\sum_{\alpha=2}^NJ^{1\alpha}=\frac{\pi}{2}\frac{|\psi^{(1)}|^2}{\Psi^2}
\sum_{\alpha=2}^N|\psi^{(\alpha)}|^2 \nonumber \\
& = &\frac{\pi}{2}{|\psi^{(1)}|^2}\frac{\Psi^2-|\psi^{(1)}|^2}{\Psi^2}.
\label{tbkt}
\end{eqnarray}
The subsequent KT transitions at higher temperatures are mapped onto
the $N-1$, $N-2$, $\dots$ cases.
Taking the $N\to \infty$  limit in Eq. (\ref{tbkt}) and provided that $|\psi^{(1)}|^2\ll \Psi^2$
we obtain
\begin{eqnarray}
T^{(1) \ [{N \to \infty}]}_{\rm KT}\to
\frac{\pi}{2}{|\psi^{(1)}|^2}.
\label{tbkt2}
\end{eqnarray}
This  expression  quite remarkably shows that in the limit 
$N \to \infty$, even in the system with short penetration length,
$T^{(1)}_{\rm KT}$ 
tends to the value in a {\it neutral} system with the bare stiffness $|\psi^{(1)}|$.
In contrast in the one component case with short penetration length the system does 
not exhibit a KT transition.

In conclusion, we note that the KT transitions considered in this Appendix are 
still significantly simpler than the situation arising in this model in three 
dimensions because, {\it as  we have considered in previous sections, in 
three dimensions the charged mode plays an extremely important role.} It is precisely the 
interplay between neutral and charged modes which is particularly important in three 
dimensions and which gives the model a variety  of different phases and phase transitions. Also, 
we note that this situation is quite different from KT transitions that are known to 
exist in the $2+1$-dimensional N-component Chiral Gross-Neveu model \cite{pgNJL}
where there is only one KT transition which occurs at finite temperature (when the system 
is effectively two-dimensional through dimensional compactification).

\end{document}